\documentclass[10pt]{article}
\pdfoutput=1

\input{preamb.sty}

\pgfdeclarelayer{back}
\pgfdeclarelayer{front}
\pgfsetlayers{back,main,front}

\tikzset{->-/.style={
        decoration={
            markings,
            mark=at position #1 with {
                \pgftransformxshift{2.5pt}
                \arrow{Triangle[width=3.5pt, length=3.5pt]}
            }
        },
        postaction={decorate}
    }
}

\tikzset{-<-/.style={
        decoration={
            markings,
            mark=at position #1 with {
                \pgftransformxshift{-2.5pt}
                \arrowreversed{Triangle[width=3.5pt, length=3.5pt]}
            }
        },
        postaction={decorate}
    }
}

\tikzset{
    every picture/.append style={
        line join=round,
        line width = 0.7pt,
        line cap=round
    }
}

\tikzset{
    mathpoly/.style={
        baseline=0ex,        
        x=1.3ex, y=1.3ex,      
        line width=0.11ex,      
        line join=round        
    }
}

\tikzset{mpoint/.style={ transform shape, sloped, 
minimum width=20pt, minimum height=4pt, inner sep=0pt, ultra thin, font=\tiny}}

\tikzset{
    clip even odd rule/.code={\pgfseteorule},
    invclip/.style={
        clip,
        insert path=[clip even odd rule]{
            [reset cm](-20cm,-20cm) rectangle (20cm,20cm)
        }
    }
}
  
\newcommand\clipIntersection[1]{
    (#1.north west) -- (#1.north east) -- (#1.south east) -- (#1.south west) -- (#1.north west)
}

\tikzset{every label/.style={label distance=-1pt}}

\tikzstyle{dot}=[dash pattern= on 0.6pt off 2.1pt]
\tikzstyle{mor}=[circle, minimum size=6pt, inner sep=0pt, fill=gray!20, draw]
\tikzset{gmor/.style={mor, draw=none, fill=none, overlay}}
\tikzset{fake/.style={inner sep=0pt, draw=none, fill=none}}
\tikzstyle{obj}=[draw=black]
\tikzstyle{cobj}=[draw=CornflowerBlue]
\tikzstyle{kir}=[line width=.8pt, draw=black, dash pattern= on 0.8pt off 2.1pt]
\tikzstyle{onL}=[fill=white,rectangle,rounded corners=2pt,inner sep=1pt]

\newcommand\clipAroundNodes[2]{
    \begin{pgfscope}
        \begin{scope}[overlay]
            \path[invclip] \foreach \n in {#1} { \clipIntersection{\n} };
            #2
        \end{scope}
    \end{pgfscope}
}

\newcommand{\mtriangle}{
    \mathord{
        \tikz[mathpoly] 
            \draw[shift={(0, {cos(60)/(1+cos(60))})}] 
                ({-150}:{1/(1+cos(60))}) 
             -- ({-30}:{1/(1+cos(60))}) 
             -- ({90}:{1/(1+cos(60))}) 
             -- cycle;
    }
}

\newcommand{\msquare}{
    \mathord{
        \tikz[mathpoly] 
            \draw[shift={(0, 1/2)}] 
                ({-135}:{1/(2*cos(45))}) 
             -- ({-45}:{1/(2*cos(45))}) 
             -- ({45}:{1/(2*cos(45))}) 
             -- ({135}:{1/(2*cos(45))}) 
             -- cycle;
    }
}

\newcommand{\mpentagon}{
    \mathord{
        \tikz[mathpoly] 
            \draw[scale=1.1,shift={(0, {cos(36)/(1+cos(36))})}] 
                ({-126}:{1/(1+cos(36))}) 
             -- ({-54}:{1/(1+cos(36))}) 
             -- ({18}:{1/(1+cos(36))}) 
             -- ({90}:{1/(1+cos(36))}) 
             -- ({162}:{1/(1+cos(36))}) 
             -- cycle;
    }
}

\newcommand{\mhexagon}{
    \mathord{
        \tikz[mathpoly] 
            \draw[shift={(0, 1/2)}] 
                ({-120}:{1/(2*cos(30))}) 
             -- ({-60}:{1/(2*cos(30))}) 
             -- ({0}:{1/(2*cos(30))}) 
             -- ({60}:{1/(2*cos(30))}) 
             -- ({120}:{1/(2*cos(30))}) 
             -- ({180}:{1/(2*cos(30))}) 
             -- cycle;
    }
}

\newcommand{\lineT}[2]{
    \begin{tikzpicture}[baseline={([yshift=-.6ex]0,0)}]
        \def\l{.9};
        \node[gmor] (g1) at (0,0) {};
        \node[gmor] (g2) at (-\l, 0) {}; 
        
        \ifthenelse{\equal{#1}{kir}}{
            \draw[kir] (g1) --node[pos=.5,above]{${\sss #2}$} (g2);
        }{
            \draw[#1,->-=.5] (g1) --node[pos=.5,above]{${\sss #2}$} (g2);
        }
    \end{tikzpicture}
}

\newcommand{\lattice}[1]{
    \begin{tikzpicture}[baseline={([yshift=-.6ex]1.3,1.3)}]
        \def\l{.9};
        \ifthenelse{#1=1}{
            \path[pattern=north east lines, rounded corners=4pt, line width=.4pt, pattern color=black!60] 
                (.5*\l, .5*\l) rectangle (2.5*\l, 2.5*\l);
            \draw[step=\l, obj] (-.5*\l, -.5*\l) grid (3.5*\l, 3.5*\l);
            \foreach \x in {0, 1, 2, 3} {
                \foreach \y in {0, 1, 2, 3} {
                    \node[mor] at (\x*\l,\y*\l) {};
                }
            }
            \node at (2.6*\l,.4*\l) {${\sss \mathbb D}$};
        }{
            \draw[step=\l, obj] (-.5*\l, -.5*\l) grid (3.5*\l, 3.5*\l);
            \foreach \x in {0, 1, 2, 3} {
                \foreach \y in {0, 1, 2, 3} {
                    \node[mor] at (\x*\l,\y*\l) {};
                }
            }
            \fill[white] (.5*\l, .5*\l) rectangle (2.5*\l, 2.5*\l);
        }
    \end{tikzpicture}
}

\newcommand{\halfBraid}[2]{
    \begin{tikzpicture}[baseline={([yshift=-.6ex]0,0)}]
        \def\l{.9};
        \node[gmor] (g) at (0,0) {};
        \node[gmor] (g1) at (-\l,0) {};
        \node[gmor] (g2) at (\l,0) {};
        \node[gmor] (g3) at (0,-\l) {};
        \node[gmor] (g4) at (0,\l) {};
        \draw[obj,-<-=.25,-<-=.75] (g4) --node[pos=.5, mpoint](m){} node[pos=.25,right]{${\sss #2}$} node[pos=.75,left]{${\sss #2}$} (g3);
        \clipAroundNodes{m}{
            \draw[cobj,-<-=.25,-<-=.75] (g1) --node[pos=.25,above]{${\sss #1}$} node[pos=.75,below]{${\sss #1}$} node[pos=0](f1){} node[pos=1](f2){} (g2);
        }
        \node[fake] at (f1) {};
        \node[fake] at (f2) {};
    \end{tikzpicture}
}

\newcommand{\modT}[3]{
    \begin{tikzpicture}[baseline={([yshift=-.6ex]0,0)}]
        \def\l{.9};
        \node[mor,label=below:{${\sss #3}$}] (n1) at (0,0) {};
        \node[gmor] (g2) at (\l, 0) {};     
        \node[gmor] (g1) at (-\l,0) {};
        \draw[obj, -<-=.5] (n1) --node[pos=.5,above]{${\sss #1}$} (g2);
        \draw[cobj, ->-=.5] (n1) --node[pos=.5,above]{${\sss #2}$} (g1);
    \end{tikzpicture}
}

\newcommand{\innerModT}[5]{
    \begin{tikzpicture}[baseline={([yshift=-.6ex]0,0)}]
        \def\l{.9}
        \node[mor,label=below:{${\sss #2}$}] (n1) at (0,0) {};
        \node[mor,label=below:{${\sss #4}$}] (n2) at (\l, 0) {}; 
        \node[gmor] (g1) at (-\l,0) {};
        \node[gmor] (g2) at (2*\l,0) {};
        \draw[cobj, ->-=.5] (n2) --node[pos=.5,above]{${\sss #5}$} (n1);
        \draw[obj,->-=.5] (n1) --node[pos=.5,above]{${\sss #1}$} (g1);
        \draw[obj,->-=.5] (g2) --node[pos=.5,above]{${\sss #3}$} (n2);
    \end{tikzpicture}
}

\newcommand{\orthoLemma}[5]{
    \begin{tikzpicture}[baseline={([yshift=-.6ex]0,0)}]
        \def\l{.9}
        \node[mor,label=below:{${\sss #4}$}] (n1) at (0,0) {};
        \node[gmor] (g1) at (2*\l, 0) {};
        \node[mor,label=below:{${\sss #2}$}] (n2) at (-\l,0) {};
        \node[gmor] (g2) at (-3*\l, 0) {};
        \draw[kir] (0,-\l) arc(-90:90:\l) node[pos=.5, mpoint](m1){} -- (-\l,\l) arc(90:270:\l) node[pos=.5, mpoint](m2){} -- cycle;
        \draw[obj,->-=.5] (n1) --node[pos=.5,above]{${\sss #5}$} (n2);
        \clipAroundNodes{m1,m2}{
            \draw[cobj, -<-=.25,-<-=.75] (n1) --node[pos=.25,above]{${\sss #3}$} node[pos=.75,above]{${\sss #3}$} node[pos=1](f1){} (g1);
            \draw[cobj, ->-=.25,->-=.75] (n2) --node[pos=.25,above]{${\sss #1}$} node[pos=.75,above]{${\sss #1}$} node[pos=1](f2){} (g2);
        }
        \node[fake] at (f1) {};
        \node[fake] at (f2) {};
    \end{tikzpicture}
}

\newcommand{\tubeT}[4]{
    \begin{tikzpicture}[baseline={([yshift=-.6ex]0,0)}]
        \def\l{.9}
        \node[gmor] (g) at (0,0) {};
        \node[gmor] (g1) at (-\l, 0) {};
        \node[gmor] (g2) at (\l, 0) {};
        \draw[obj,-<-=.25,-<-=.75] (g) arc(0:360:\l) node[pos=.25,below]{${\sss #1}$} node[pos=.75,above]{${\sss #1}$};
        \draw[obj, -<-=.5] (g1) --node[pos=.5,above]{${\sss #2}$} (g);
        \draw[obj, -<-=.5] (g) --node[pos=.5,above]{${\sss #3}$} (g2);
        \node[mor,label=below right:{${\sss #4}$}] (n1) at (0,0) {};
    \end{tikzpicture}
}

\newcommand{\cylTube}[4]{
    \begin{tikzpicture}[baseline={([yshift=-.6ex]0,0)}]
        \def\l{.9}
        \coordinate (g1) at (0,-\l);
        \coordinate (g2) at (2*\l,-\l);
        \coordinate (g3) at (0,\l);
        \coordinate (g4) at (2*\l,\l);
        \draw[line width=.5pt] (g1) to[out=0, in=0, looseness=.7] coordinate[pos=.5](g5) (g3);
        \draw[line width=.5pt] (g1) to[out=180, in=180, looseness=.7] (g3);
        \draw[line width=.5pt] (g2) to[out=0, in=0, looseness=.7] coordinate[pos=.5](g6) (g4);
        \draw[line width=.5pt] (g1) --coordinate[pos=.5](g7) (g2);
        \draw[line width=.5pt] (g3) --coordinate[pos=.5](g8) (g4);
        \draw[obj,-<-=.25, -<-=.75] (g5) --coordinate[pos=.5](n) node[pos=.25,above]{${\sss #2}$} node[pos=.75,above]{${\sss #3}$} node[pos=.1,mpoint](m){} (g6);
        \clipAroundNodes{m}{
            \draw[obj,-<-=.125,-<-=.625] (g7) to[out=0,in=0,looseness=.7] node[pos=.25,right]{${\sss #1}$} (g8) to[out=180,in=180,looseness=.7] node[pos=.25,right]{${\sss #1}$}   (g7);
        }
        \node[mor] at (n) {};
    \end{tikzpicture}
}

\newcommand{\cylTubeAction}[1]{
    \begin{tikzpicture}[baseline={([yshift=-.6ex]0,0)}]
        \def\l{.9}
        \coordinate (g1) at (0,-\l);
        \coordinate (g2) at (3*\l,-\l);
        \coordinate (g3) at (0,\l);
        \coordinate (g4) at (3*\l,\l);
        \draw[line width=.5pt] (g1) to[out=0, in=0, looseness=.7] coordinate[pos=.5](g5) (g3);
        \draw[line width=.5pt] (g1) to[out=180, in=180, looseness=.7] (g3);
        \draw[line width=.5pt] (g2) to[out=0, in=0, looseness=.7] coordinate[pos=.5](g6) (g4);
        \draw[line width=.5pt] (g1) --coordinate[pos=1/3](g7) coordinate[pos=2/3](g9)  (g2);
        \draw[line width=.5pt] (g3) --coordinate[pos=1/3](g8) coordinate[pos=2/3](g10)  (g4);
        \ifthenelse{#1=1}{
            \path[pattern=north east lines, line width=.4pt, pattern color=black!60,opacity=.5] 
                (g7) to[out=180,in=180,looseness=.7] (g8) -- (g10) to[out=180,in=180,looseness=.7] (g9) -- (g7);
            \path (g7) to[out=0,in=0,looseness=.7] coordinate[pos=.45](g11) coordinate[pos=.55](g12) (g8);
            \path (g9) to[out=0,in=0,looseness=.7] coordinate[pos=.45,label=below right:{${\sss \mathbb B_\Omega^{[k]}}$}](g13) coordinate[pos=.55](g14) (g10);
            \path[pattern=north east lines, rounded corners=4pt, line width=.4pt, pattern color=black!60] 
                (g7) to[out=0,in=-100,looseness=.7] (g11) -- (g13) to[out=-100,in=0,looseness=.7] (g9);
            \path[pattern=north east lines, rounded corners=4pt, line width=.4pt, pattern color=black!60] 
                (g8) to[out=0,in=100,looseness=.7] (g12) -- (g14) to[out=100,in=0,looseness=.7] (g10);
            \draw[obj,-<-=1/2] (g5) --coordinate[pos=1/3](n1) coordinate[pos=2/3](n2) node[pos=5/6,above]{${\sss x}$}  (g6);
        }{}

        \ifthenelse{#1=2}{
            \draw[obj,-<-=1/2] (g5) --node[pos=.1,mpoint](m1){} node[pos=.4,mpoint](m2){} node[pos=5/6,above]{${\sss x}$} (g6);
            \path (g7) to[out=0,in=0,looseness=.7] coordinate[pos=.45](g11) coordinate[pos=.55](g12) (g8);
            \path (g9) to[out=0,in=0,looseness=.7] coordinate[pos=.45](g13) coordinate[pos=.55](g14) (g10);
            \draw[obj, rounded corners=4pt,-<-=1/6,-<-=5/6] (g7) to[out=0,in=-100,looseness=.7] (g11) --node[pos=.2,mpoint](m3){} (g13) to[out=-100,in=0,looseness=.7] (g9);
            \draw[obj, rounded corners=4pt] (g10) to[out=0,in=100,looseness=.7] (g14) --node[pos=.8,mpoint](m4){} (g12) to[out=100,in=0,looseness=.7] (g8); 
            \clipAroundNodes{m1,m2,m3,m4}{
                \draw[obj,-<-=3/4] (g9) to[out=180,in=180,looseness=.7] (g10);
                \draw[obj,-<-=1/4] (g8) to[out=180,in=180,looseness=.7] node[pos=.25,right]{${\sss g}$} (g7);
            }
        }{}
        
        \ifthenelse{#1=3}{
            \draw[obj,-<-=1/6,-<-=1/2,-<-=5/6] (g5) --coordinate[pos=1/3](n1) coordinate[pos=2/3](n2) node[pos=1/6,above]{${\sss x}$} node[pos=1/2,above]{${\sss ^g x}$} node[pos=5/6,above]{${\sss x}$} node[pos=.1,mpoint](m1){} node[pos=.4,mpoint](m2){} (g6);
            \clipAroundNodes{m1,m2}{
                \draw[obj,-<-=1/8,-<-=5/8] (g7) to[out=0,in=0,looseness=.7] node[pos=.25,left]{${\sss g}$} (g8) to[out=180,in=180,looseness=.7] node[pos=.25,right]{${\sss g}$}   (g7);
                \draw[obj,->-=1/8,->-=5/8] (g9) to[out=0,in=0,looseness=.7] node[pos=.25,left]{${\sss g}$} (g10) to[out=180,in=180,looseness=.7] node[pos=.25,right]{${\sss g}$}   (g9);
            }
            \node[mor] at (n1) {};
            \node[mor] at (n2) {};
        }{}
    \end{tikzpicture}
}

\newcommand{\tubeTExp}[4]{
    \begin{tikzpicture}[baseline={([yshift=-.6ex]0,0)}]
        \def\l{.9}
        \node[gmor] (g) at (0,0) {};
        \node[gmor] (g1) at (-\l, 0) {};
        \node[gmor] (g2) at (\l, 0) {};
        \node[gmor] (g3) at (2*\l, 0) {};
        \draw[obj,->-=.5] (g) arc(0:-180:\l) node[pos=.5,above]{${\sss #1}$};
        \draw[obj,->-=.5] (-2*\l, 0) to[out=90, in=90, looseness=1.45] node[pos=.5,below]{${\sss #1}$} (g2);
        \draw[obj, -<-=.5] (g1) --node[pos=.5,above]{${\sss #2}$} (g);
        \draw[obj, -<-=.5] (g) --node[pos=.5,above]{${\sss #3}$} (g2);
        \draw[obj, ->-=.5] (g3) --node[pos=.5,above]{${\sss #4}$} (g2);
        \node[mor] (n1) at (0,0) {};
        \node[mor] (n2) at (\l,0) {};
    \end{tikzpicture}
}

\newcommand{\tubeActionT}[6]{
    \begin{tikzpicture}[baseline={([yshift=-.6ex]0,0)}]
        \def\l{.9}
        \node[gmor] (g) at (0,0) {};
        \node[gmor] (g1) at (-3*\l, 0) {};
        \node[gmor] (g3) at (\l, 0) {};
        \node[mor,label=below:{${\sss #6}$}] (n2) at (-\l, 0) {};
        \draw[obj,-<-=.25,-<-=.75] (g) arc(0:360:\l) node[pos=.25,below]{${\sss #1}$} node[pos=.75,above]{${\sss #1}$} node[pos=.5,mpoint](m){};
        \node[mor,label=below right:{${\sss #4}$}] (n1) at (0,0) {};
        \draw[obj, ->-=.5] (g3) --node[pos=.5,above]{${\sss #3}$} (n1);
        \draw[obj, ->-=.5] (n1) --node[pos=.5,above]{${\sss #2}$} (n2);
        \clipAroundNodes{m}{
            \draw[cobj, ->-=.25,->-=.75] (n2) --node[pos=.25,above]{${\sss #5}$} node[pos=.75,above]{${\sss #5}$} node[pos=1](f1){} (g1);
        }
        \node[fake] at (f1) {};
    \end{tikzpicture}
}

\newcommand{\tubeActionTMod}[7]{
    \begin{tikzpicture}[baseline={([yshift=-.6ex]0,0)}]
        \def\l{.9}
        \node[gmor] (g) at (0,0) {};
        \node[gmor] (g0) at (-4*\l, 0) {};
        \node[gmor] (g1) at (-3*\l, 0) {};
        \node[gmor] (g3) at (\l, 0) {};
        \node[gmor] (g4) at (2*\l,0) {};
        \node[mor,label=below:{${\sss #7}$}] (n2) at (-\l, 0) {};
        \draw[obj,-<-=.25,-<-=.75] (g) arc(0:360:\l) node[pos=.25,below]{${\sss #1}$} node[pos=.75,above]{${\sss #1}$} node[pos=.5,mpoint](m1){};
        \draw[obj,-<-=.25,-<-=.75] (g3) arc(0:360:2*\l) node[pos=.25,below]{${\sss #2}$} node[pos=.75,above]{${\sss #2}$} node[pos=.5,mpoint](m2){};
        \node[mor] (n1) at (0,0) {};
        \node[mor] (n3) at (\l,0) {};
        \draw[obj, ->-=.5] (g3) --node[pos=.5,above]{${\sss #4}$} (n1);
        \draw[obj, ->-=.5] (n1) --node[pos=.5,above]{${\sss #3}$} (n2);
        \draw[obj, ->-=.5] (g4) --node[pos=.5,above]{${\sss #5}$} (g3);
        \clipAroundNodes{m1,m2}{
            \draw[cobj, ->-=.166,->-=.5,->-=.833] (n2) --node[pos=.166,above]{${\sss #6}$} node[pos=.5,above]{${\sss #6}$} node[pos=.833,above]{${\sss #6}$} node[pos=1](f1){} (g0);
        }
        \node[fake] at (f1) {};
    \end{tikzpicture}
}

\newcommand{\matrixUnits}[5]{
    \begin{tikzpicture}[baseline={([yshift=-.6ex]0,0)}]
        \def\l{.9}
        \node[gmor] (g1) at (-2*\l,0) {};
        \node[gmor] (g2) at (2*\l,0) {};
        \node[mor,label=below:{${\sss #3}$}] (n1) at (-\l, 0) {};
        \node[mor,label=below:{${\sss #5}$}] (n2) at (\l, 0) {};
        \draw[obj, ->-=.5] (n1) --node[pos=.5,above]{${\sss #2}$} (g1);
        \draw[obj, ->-=.5] (g2) --node[pos=.5,above]{${\sss #4}$} (n2);
        \draw[kir] (-\l,-\l) arc(-90:90:\l) node[pos=.5, mpoint](m){} -- (-2*\l,\l) arc(90:270:\l) -- cycle;
        \clipAroundNodes{m}{
            \draw[cobj,->-=.25,->-=.75] (n2) --node[pos=.25,above]{${\sss #1}$} node[pos=.75,above]{${\sss #1}$} (n1);
        }
    \end{tikzpicture} 
}

\newcommand{\tubeMult}[6]{
    \begin{tikzpicture}[baseline={([yshift=-.6ex]0,0)}]
        \def\l{.9}
        \node[gmor] (g) at (0,0) {};
        \node[gmor] (g1) at (-\l, 0) {};
        \node[gmor] (g2) at (\l, 0) {};
        \node[gmor] (g3) at (2*\l,0) {};
        \draw[obj,-<-=.25,-<-=.75] (g) arc(0:360:\l) node[pos=.25,below]{${\sss #1}$} node[pos=.75,above]{${\sss #1}$};
        \draw[obj,-<-=.25,-<-=.75] (g2) arc(0:360:2*\l) node[pos=.25,below]{${\sss #2}$} node[pos=.75,above]{${\sss #2}$};
        \draw[obj, -<-=.5] (g1) --node[pos=.5,above]{${\sss #3}$} (n1);
        \draw[obj, -<-=.5] (g) --node[pos=.5,above]{${\sss #4}$} (g2);
        \draw[obj,-<-=.5] (g2) --node[pos=.5,above]{${\sss #5}$} (g3);
        \node[mor,label=below right:{${\sss #6}$}] (n1) at (0,0) {};
        \node[mor,label=below right:{${\sss #6}$}] (n2) at (\l,0) {};
    \end{tikzpicture}
}

\newcommand{\vertexOp}[4]{
    \begin{tikzpicture}[baseline={([yshift=-.6ex]0,0)}]
        \def\l{.9}
        \node[mor] (n) at (0,0) {};
        \coordinate (n1) at (-\l,0);
        \coordinate (n2) at (0,\l);
        \coordinate (n3) at (\l,0);
        \coordinate (n4) at (0,-\l);
        \draw[obj] (n) --node[onL, pos=.5]{${\sss #1}$} (n1);
        \draw[obj] (n) --node[onL, pos=.5]{${\sss #2}$} (n2);
        \draw[obj] (n) --node[onL, pos=.5]{${\sss #3}$} (n3);
        \draw[obj] (n) --node[onL, pos=.5]{${\sss #4}$} (n4);
    \end{tikzpicture} 
}

\newcommand{\plaquetteOp}[4]{
    \begin{tikzpicture}[baseline={([yshift=-.6ex]0,0)}]
        \def\l{.9}
        \node[mor] (n1) at (-\l/2,-\l/2) {};
        \node[mor] (n2) at (\l/2,-\l/2) {};
        \node[mor] (n3) at (\l/2,\l/2) {};
        \node[mor] (n4) at (-\l/2,\l/2) {};
        \draw[obj] (n1) --node[onL, pos=.5]{${\sss #1}$} (n2);
        \draw[obj] (n2) --node[onL, pos=.5]{${\sss #2}$} (n3);
        \draw[obj] (n3) --node[onL, pos=.5]{${\sss #3}$} (n4);
        \draw[obj] (n1) --node[onL, pos=.5]{${\sss #4}$} (n4);
    \end{tikzpicture} 
}

\title{\boldmath Topological lattice gauge theory enriched by non-invertible symmetry}

\author[\mtriangle]{Lea E.\ Bottini,}
\author[\msquare]{Clement Delcamp,}
\author[\mpentagon]{Edmund Heng,}
\author[\mhexagon]{Campbell K. McLauchlan,}
\author[\, \mhexagon]{\par Dominic J.\ Williamson}

\affiliation[\normalfont \mtriangle]{Institut des Hautes \'Etudes Scientifiques, Bures-sur-Yvette, France}
\affiliation[\normalfont \msquare]{Laboratoire Alexander Grothendieck, Institut des Hautes \'Etudes Scientifiques \& CNRS, \\ Bures-sur-Yvette, France}
\affiliation[\normalfont \mpentagon]{School of Mathematics and Statistics, University of Sydney, NSW 2006, Australia}
\affiliation[\normalfont \mhexagon]{School of Physics, University of Sydney, NSW 2006, Australia}

\emailAdd{bottini@ihes.fr}
\emailAdd{delcamp@ihes.fr}

\abstract{\\~\\ We use finite group topological lattice gauge theory, also known as the quantum double model, as a lens to explore a notion of topological order enriched by a non-invertible symmetry. For invertible symmetry enriched topological order, there is an established axiomatisation in terms of a $G$-crossed braided fusion category.  We lay the foundations for a generalisation of this notion. By condensing an arbitrary algebra of charges in a quantum double model, we demonstrate that the category of localised excitations in the resulting theory forms a hypergroup-graded extension of the category of deconfined excitations. For every element in the hypergroup, the associated domain wall acts in a typically non-invertible way on these localised excitations. Both this action and the monoidal structure are compatible with the hypergroup grading. The actual categorical action is encoded in a Hopf monad on the category of localised excitations, and gauging the non-invertible symmetry amounts to computing the category of modules over this Hopf monad. Finally, we outline how this framework naturally extends to theories obtained by condensing algebras in a generic string-net model.

}

\begin{document} 
	\vspace*{-2em}
        \maketitle
	\flushbottom
	\newpage

\section{Introduction}

The interplay between  conventional global symmetries and (2+1)d topological orders gives rise to a class of gapped phases of matter known as \emph{symmetry enriched topological orders} (SETs). In addition to localised \emph{anyonic} excitations, which are characteristic of topological order, an SET enriched by an invertible symmetry $G$ supports a collection of codimension-one \emph{symmetry defects} labelled by group elements $g \in G$. 
The collection of localised excitations (encompassing both \emph{deconfined} and \emph{confined} excitations) then organises into \emph{twisted sectors}, one for each type of symmetry defect.  The trivial (untwisted) sector comprises the deconfined anyons of the underlying topological order, while the twisted sectors describe confined excitations that are attached to defect lines.
Since its introduction, this notion has received widespread attention; see for instance \cite{PhysRevB.87.155115,TEO2015349,Lan2017,PhysRevB.94.235136,Fidkowski:2016svr,Williamson:2017uzx,PhysRevB.100.115147,Christian:2023ekm}. A special class of symmetry enriched topological order, obtained by choosing the underlying topological order to be trivial, is that of \emph{symmetry protected topological phases} \cite{PhysRevB.86.115109,PhysRevB.84.235141,PhysRevB.87.155114}.

Beyond the fusion and braiding data of the underlying anyon theory, the basic physical data characterising such a phase consists of the fusion of symmetry defects, together with a global symmetry action that simultaneously permutes the twisted sectors  and transforms the corresponding localised excitations residing within them. In particular, as an anyon is dragged across a defect line, it may emerge as a different anyon species. Moreover, although the group $G$ acts linearly on the physical system, it may act \emph{projectively} on the anyon theory, giving rise to \emph{fractionalised quantum numbers}---one of the hallmarks of symmetry enriched topological order. This symmetry action further interacts with the fusion and the braiding of the anyons, yielding in particular a notion of \emph{$G$-crossed braiding}, which involves an application of the global symmetry. 

As suggested by the above remarks, the symmetry action on the anyon theory is not merely a group action. Formally, an anyon theory is specified by a choice of \emph{unitary modular tensor category} $\mc B$. It encodes the quasi-particle excitations as well as fusion and braiding data. In the presence of a symmetry $G$, this modular tensor category may be extended to a \emph{$G$-crossed braided fusion category} \cite{etingof2010fusion, etingof2016tensor, PhysRevB.100.115147}.
The symmetry action, which is a \emph{categorical action} rather than a mere group action, is encoded in a choice of \emph{monoidal functor} $\underline G \to \underline{\text{Aut}}_\otimes^{\rm br}(\mc B)$ with target the \emph{categorical group} of braided tensor autoequivalences of $\mc B$ \cite{etingof2010fusion,etingof2016tensor, PhysRevB.100.115147}.

\bigskip \noindent
Models of topological order are known to possess non-invertible 1-form symmetries generated by the line operators that create the anyonic excitations \cite{Gaiotto:2014kfa}.
In addition to these topological lines, it has been recently understood that such models host topological surfaces known as \emph{condensation defects} \cite{Roumpedakis:2022aik,Lin:2022xod,Choi:2022zal,Delcamp:2023kew,Choi:2024rjm,Inamura:2023qzl,Eck:2025ldx,Vancraeynest-DeCuiper:2025wkh}.
Together, they form a symmetry structure encoded in a \emph{fusion 2-category} \cite{Delcamp:2021szr,Delcamp:2023kew,Bhardwaj:2022lsg,Bhardwaj:2022maz,Bartsch:2022mpm}. While condensation defects act trivially on local operators, they do interact non-trivially with the line operators. The invertible condensation defects precisely implement the type of symmetry action that is a candidate for enriching the underlying topological order. In that spirit, the presence of \emph{non-invertible} condensation defects suggests the possibility of enriching a topological order by a non-invertible symmetry \cite{KNBalasubramanian:2025vum}. The goal of this manuscript is to take substantive steps towards formalising such a notion. 

Another hint at this notion of topological order enriched by a non-invertible symmetry comes from the mechanism by which ordinary symmetry-enriched topological orders are typically constructed, namely \emph{anyon condensation}. Given a topological order and its corresponding unitary modular tensor category $\mc B$, patterns of anyon condensation are classified by \emph{connected étale algebras} in $\mc B$ \cite{Bais:2008ni,Kong:2013aya}. Specifically, whenever $\Rep(G)$ for some group $G$ embeds in $\mc B$ as a \emph{Tannakian} fusion subcategory, there is a canonical connected étale algebra in $\mc B$, whose condensation yields a $G$-crossed braided fusion category describing a $G$-enriched topological order \cite{Drinfeld2010}. The \emph{$G$-equivariantisation} of the latter category recovers $\mc B$. Similarly, one expects any  condensation to yield a topological order enriched by a (possibly non-invertible) symmetry, generated by the corresponding twist defects.  In general, the data that should characterise such a theory is not explicitly known. However, it is clear that it should contain a generalisation of each piece of data entering the definition of a $G$-crossed braided fusion category.

The literature contains multiple speculations about potential generalisations of the $G$-crossed braided fusion category formalism by replacing the group $G$ with a \emph{fusion ring}. Following recent developments in the theory of \emph{vertex operator algebras} \cite{Bischoff:2016jmy,Riesen2025,Dong:2025ttr}, we instead propose an axiomatisation in terms of \emph{hypergroups}. Although every fusion ring gives rise to a hypergroup, the notion of hypergroup offers several advantages. The main one is that it is always possible to take \emph{quotients} of hypergroups---a mechanism that arises naturally when identifying twisted sectors of the condensed theory. This motivates a notion of hypergroup-graded fusion category equipped with a hypergroup action that is compatible with the grading. In particular, we explicitly spell out the compatibility conditions between the grading and the monoidal structure, and between the grading and the action. We also verify a generalisation of the \emph{fixed point theorem} \cite{PhysRevB.100.115147,Bischoff2020,JONES2021} first suggested in ref.~\cite{KNBalasubramanian:2025vum}.

The same way an invertible symmetry is not merely a group action, a hypergroup action is not enough to fully specify the non-invertible symmetry. As was suggested in ref.~\cite{Cui:2018hxz}, the complete description requires instead the notion of  \emph{Hopf monad} \cite{Bruguieres2000}. Concretely, the individual hypergroup actions organise into a Hopf monad, defined as an endofunctor of the category of localised excitations equipped with various structures. In particular, the multiplication of this Hopf monad encodes a choice of (generalised) symmetry \emph{fractionalisation}, with inequivalent such choices distinguishing  the various hypergroup-graded extensions of a given topological order. Extensions sharing the same fractionalisation may still differ by a choice of \emph{discrete torsion}, encoded in the monoidal associator.
Crucially, this gadget allows us to equivariantise, which physically amounts to gauging the non-invertible symmetry.    

Focusing on topological lattice gauge theory, one can describe how to extract the relevant data from a given model following the \emph{tube algebra approach} \cite{ocneanu1994chirality,Izumi2000,ocneanu2001operator,bockenhauer2001longo,MUGER2003159,Neshveyev_Yamashita_2018}. Starting from a parent theory whose topological order is encoded by the \emph{Drinfel'd center} $\mc Z(\Vect_G)$ for a finite group $G$,
we consider condensed phases associated with subgroups $H$ of $G$. The unconfined localised excitations of such a phase are encoded in $\mc Z(\Vect_H)$ with $H \leq G$, while the full category of localised excitations is provided by the \emph{relative} Drinfel’d centre $\mathcal Z_{\Vect_H} (\Vect_G)$ of $\Vect_G$ over $\Vect_H$ \cite{Majid1991}. Our construction implicitly relies on $\Vect_H$ being included in $\Vect_G$ as a (spherical) fusion subcategory. Despite focusing on the group case, we discuss generalisations to an arbitrary string-net model with input a (spherical) fusion category $\mc C$ \cite{Levin:2004mi},  related by anyon condensation to a model with localised excitations captured by $\mc Z_{\mc D} (\mc C)$ for some fusion subcategory $\mc D$ of $\mc C$.
Up to Morita equivalence, such a representation can be found for an arbitrary anyon condensation between two anyon theories that admit gapped boundaries. This follows by considering the fusion category of defects on the gapped boundary defined by extending the algebra in $\mc C$ whose condensate is $\mc Z_{\mc D} (\mc C)$ to a Lagrangian algebra. Finally, since any chiral phase can be mapped into a doubled system by stacking it with its orientation-reversed partner, this construction should be sufficiently general to cover such chiral models as well.

\bigskip \noindent
\textbf{Organisation of the manuscript:} The rest of the paper is structured in the following way. In sec.~\ref{sec2:lattice} we establish the physical framework by considering  a lattice model representing a topologically ordered phase enriched by a non-invertible symmetry. The model is obtained by condensing charges in a topological lattice gauge theory. Specifically, we introduce the quantum double model and describe its excitations via a tube algebra approach. We then discuss anyon condensation and the resulting theory, where we identify the presence of a non-invertible symmetry and compute its action on the excitations of the model. We conclude this section by elucidating the construction with the help of a simple example based on the dihedral group of order 6. While sec.~\ref{sec2:lattice} focuses on the physical picture, in sec.~\ref{sec:Maths} we propose a mathematical formalisation. Here, we cast the non-invertible symmetry appearing above in terms of a Hopf monad and we formalise the non-invertible symmetry enriched topological order in terms of a hypergroup-equivariant hypergroup-graded fusion category. We also describe the process of gauging this non-invertible symmetry to recover the original topological order. Then, we apply this  machinery to the same dihedral group example introduced in the previous section. Finally, in sec.~\ref{sec:outlook} we summarise our findings and sketch a generalisation of our construction that extends beyond the topological lattice gauge theory setting, which we illustrate with several further examples.

\bigskip \bigskip \noindent
\paragraph{Note added:} While preparing this manuscript, we became aware of independent, related work by L.~Eck, P.~Huston, K.~Kawagoe and D.~Penneys, which will appear in the same arXiv posting \cite{Eck:2026aiz}. We thank the authors for coordinating submission.  

\bigskip \noindent
\paragraph{Acknowledgements:} L.~B. and C.~D. are grateful to Peter Huston, Dmitri Nikshych and Brandon Rayhaun for illuminating exchanges. C.~D. and D.~W. are grateful to Kyle Kawagoe for insightful discussions. C.~D. is thankful to Mahesh Balasubramanian, Matthew Buican and  Rajath Radhakrishnan for collaboration on a related project. E.~H., C.~M., and D.~W. are grateful to Eugene Chon, Austin Lin, and Leighton Xia for useful discussions. The authors would like to thank the Isaac Newton Institute for Mathematical Sciences, Cambridge, for support and hospitality during the programme \emph{Quantum field theory with boundaries, impurities, and defects}, where work on this paper was undertaken. D.~W. is supported by the Australian Research Council Discovery Early Career Research Award (DE220100625).

\bigskip \bigskip \bigskip
\section{Anyon condensation in topological lattice gauge theory}\label{sec2:lattice}

\emph{In this section, we consider a lattice model representing a topologically ordered phase enriched by a non-invertible symmetry. The model is obtained by condensing charges in a topological lattice gauge theory. We focus here on the physical picture, with the mathematical formalism deferred to sec.~\ref{sec:Maths}.}

\subsection{Lattice model \label{sec:lattice}}

We begin by reviewing a lattice model of topological gauge theory introduced by Kitaev in ref.~\cite{KITAEV20032}, commonly referred to as the \emph{quantum double model}. The quantum double model---and its twisted variants---can be formulated as Hamiltonian realisations of \emph{Dijkgraaf--Witten theory} \cite{cmp/1104180750,Hu:2012wx}. These models form a special class of \emph{string-net} models \cite{Levin:2004mi}, which can themselves be understood as Hamiltonian realisations of \emph{Turaev--Viro--Barrett--Westbury theory} \cite{Turaev:1992hq,Barrett:1993ab}.
In this work, we also adopt the variant developed by Kirillov in ref.~\cite{Kirillov:2011mk}, which is well-suited to what follows. As the material is standard in the literature, we keep the exposition brief and refer the reader to the references above for further details.

Let $\Sigma$ be an oriented two-dimensional surface with possibly a non-empty boundary, and let $\Gamma_\triangle$ be the \emph{Poincaré dual} of an oriented cellular decomposition $\Sigma_\triangle$ of $\Sigma$. We denote the vertex set and the edge set of $\Gamma_\triangle$ by $\msf V$ and $\msf E$, respectively. For each $\msf e \in \msf E$, we write $-\msf e$ for the edge with the reversed orientation. A \emph{leaf} of $\Gamma_\triangle$ is a vertex with exactly one incident edge. The leaves form a subset $\msf L \subset \msf V$.  By construction, all the leaves lie on $\partial\Sigma$. The connected components of $\Sigma \,\setminus\, \Gamma_\triangle$ are called \emph{holes}, and we define a \emph{plaquette} as a hole that does not contain any leaf in its boundary. We write $\msf P$ for the collection of plaquettes, all of which we orient in the same way relative to the orientation of $\Sigma$. The microscopic Hilbert space of the system is\footnote{Notice that we conflate an assignment $\fr g \in G^\msf E$ and the corresponding state in $\mc H_G(\Gamma_\triangle)$.} 
\begin{equation}
    \mc H_{G}(\Gamma_\triangle)  := \mathbb C\{\, \fr g \, \}_{\fr g \in G^{\msf E}} \equiv \bigotimes_{\msf e \in \msf E} \mc H_\msf e,
\end{equation}
where we identified $\mc H_\msf e = \mathbb C[G] = \mathbb C\{g\}_{g \in G}$. By convention, the assignments $\fr g \in G^{\msf E}$ are subject to $\fr g(-\msf e) = \fr g(\msf e)^{-1}$. Given an assignment $\fr g \in G^\msf E$, we refer to its restriction to edges incident to a leaf as a choice of \emph{boundary condition}.

The dynamics of the system is governed by two families of \emph{mutually commuting local projectors}. To each vertex $\msf v \in \msf V \, \setminus \, \msf L$, we assign a projector $\mathbb A^G_\msf v$ enforcing the flatness condition $\prod_{\msf e \ni \msf v}\fr g(\msf e) = 1_G$, where the product is
over edges incident to $\msf v$ ordered \emph{counterclockwise} and assumed to be oriented in the \emph{inward} direction.\footnote{If an edge is oriented outward, then it contributes $\fr g(-\msf e) = \fr g(\msf e)^{-1}$ to the product.}
At every plaquette $\msf p \in \msf P$, one defines an action of the group $G$ on the microscopic Hilbert space as follows. For a basis state $\fr g \in \mc H_G(\Gamma_\triangle)$ and an edge $\msf e \subset \partial \msf p$, the group element $g \in G$ acts via $\mathbb B^g_\msf p$ mapping $\fr g(\msf e) \mapsto g \, \fr g(\msf e)$ or $\fr g(\msf e)\, g^{-1}$, depending on whether the orientation of $\msf e$ is induced by that of $\msf p$. To every plaquette $\msf p$, we then assign the projector $\mathbb B_\msf p^G := \frac{1}{|G|}\sum_{g \in G}\mathbb B_\msf p^g$ that  
averages over this action. 
The Hamiltonian then reads
\begin{equation}
    \label{eq:HGG}
    \mathbb H_G(\Gamma_\triangle) = -\! \sum_{\msf v \in \msf V \, \setminus \, \msf L} \!\mathbb A_\msf v^G -\sum_{\msf p \in \msf P} \, \mathbb B_\msf p^G \, .
\end{equation}
For concreteness, suppose $\Gamma_\triangle$ is a square lattice with every edge oriented from bottom-right to top-left. Introducing linear operators 
$L^g \colon \mc H_\msf e \to \mc H_\msf e$, $x \mapsto gx$, and $R^g \colon \mc H_\msf e \to \mc H_\msf e$, $x \mapsto xg^{-1}$, for every $\msf e\in \msf E$, the vertex and plaquette operators take the explicit form
\begin{equation}
    \mathbb A^G_\msf v \colon \vertexOp{g_3}{g_4}{g_2}{g_1} \, \mapsto \, \delta_{g_1 g_2,\, g_3 g_4}\; \vertexOp{g_3}{g_4}{g_2}{g_1}
    \q \text{and} \q
    \mathbb B^G_\msf p = \frac{1}{|G|} \sum_{g \in G}
    \plaquetteOp{R^g}{L^g}{L^g}{R^g} .
\end{equation}
Since $\mathbb H_G(\Gamma_\triangle)$ is a local commuting projector Hamiltonian, it is exactly solvable. We denote by $\mc H_{G}^0(\Gamma_\triangle)$ its ground state subspace, which is the subspace of $\mc H_G(\Gamma_\triangle)$ consisting of states $\psi \in \mc H_G(\Gamma_\triangle)$ such that $\mathbb A_\msf v^G(\psi) = +\psi$, for every $\msf v \in \msf V \, \setminus \, \msf L$, and $\mathbb B_{\msf p}^G(\psi) = +\psi$, for every $\msf p \in  \msf P$. Since boundary conditions are unconstrained by the Hamiltonian, the ground state subspace decomposes as
\begin{equation}
    \mc H_G^0(\Gamma_\triangle) = \bigoplus_{\fr b \in G^{\msf E_\msf L}} \mc H_G^0(\Gamma_\triangle, \fr b) ,
\end{equation}
into sectors $\mc H_G^0(\Gamma_\triangle, \fr b)$ of fixed boundary condition $\fr b \in G^{\msf E_\msf L}$, where $\msf E_\msf L := \{\msf e \in \msf E \, | \, \msf e \cap \msf L \neq \varnothing\}$. 

Whenever $\Sigma$ is closed the resulting ground state subspace is independent of the chosen cell decomposition up to canonical isomorphism. That is, $\mc H_G^0(\Gamma_\triangle) \cong \mc H_G^0(\Gamma_{\triangle'})$ for any two Poincaré duals $\Gamma_\triangle$ and $\Gamma_{\triangle'}$ associated with cell decompositions $\Sigma_\triangle$ and $\Sigma_{\triangle'}$ of $
\Sigma$ \cite{cmp/1104180750,Levin:2004mi,Hu:2012wx,Bullivant:2019fmk}.
This is a manifestation of the topological invariance of the theory. Accordingly, we denote the ground state subspace by $\mc H_G^0(\Sigma)$.

\bigskip \noindent
In practice, a more conceptual realisation of the ground state subspace is often useful \cite{Kirillov:2011mk}. Suppose for now that the two-dimensional surface $\Sigma$ is closed.
We consider (finite connected oriented) graphs $\Gamma = (\msf V, \msf E)$ embedded in $\Sigma$ whose leaf sets $\msf L$ are empty. A \emph{$G$-colouring} of a graph $\Gamma$ is an assignment $\fr g \in G^\msf E$ satisfying the flatness condition $\prod_{\msf e \ni \msf v} \fr g(\msf e) = 1_G$ at every vertex $\msf v \in \msf V$, where the product is again
over edges incident to $\msf v$ ordered counterclockwise and assumed to be oriented in the inward direction. We define a \emph{$G$-coloured graph} as a  pair $(\Gamma, \fr g)$ consisting of such a graph and $G$-colouring. We write $\mc V_G(\Sigma)$ for the complex vector space freely generated by the set of $G$-coloured graphs. We endow $\mc V_G(\Sigma)$ with the equivalence relation $\sim$ generated by the relation defined as follows. Two $G$-coloured graphs $(\Gamma_1,\fr g_1), (\Gamma_2,\fr g_2) \in \mc V_G(\Sigma)$  are said to be related whenever there exists an embedded disc $\mathbb D \subset \Sigma$ whose boundary $\partial \mathbb D$ is intersected \emph{transversely} by both $\Gamma_1$ and $\Gamma_2$, such that $\Gamma_1 \, \setminus \, {\rm int}(\mathbb D) = \Gamma_2 \, \setminus \, {\rm int}(\mathbb D)$ and $\fr g_1(\msf e_1) = \fr g_2(\msf e_2)$ for all edges $\msf e_1 \in \mathsf{E}_1$ and $\msf e_2 \in \mathsf{E}_2$ satisfying $\msf e_1 \setminus \, {\rm int}(\mathbb D) = \msf e_2 \setminus \, {\rm int}(\mathbb D) \neq \varnothing$. Introducing the subspace
\begin{equation}
    \mc V^{{\rm null}}_G(\Sigma) := \mathbb C\big\{(\Gamma_1,\fr g_1) - (\Gamma_2,\fr g_2) \, | \, (\Gamma_1,\fr g_1) \sim (\Gamma_2,\fr g_2)\big\},
\end{equation}
we define $\mc V^0_{G}(\Sigma) := \mc V_G(\Sigma)\, / \,  \mc V^{\rm null}_G(\Sigma)$.
We then have the isomorphism \cite{Kirillov:2011mk}
\begin{equation}
    \mc H_G^0(\Sigma) \cong \mc V_G^0(\Sigma).
\end{equation}
Colloquially, for every ground state---here realised as a superposition of $G$-coloured graphs---each coloured graph appearing in the superposition can be simplified within any embedded disc by substituting it with another one inducing the same assignment on the disc boundary. 

The construction naturally extends to surfaces with boundary. Fix a finite set $\msf B$ of points on $\partial \Sigma$ together with an assignment $\fr b \in G^\msf B$. 
We consider graphs $\Gamma = (\msf V, \msf E)$ embedded in $\Sigma$ whose leaf set $\msf L$ equals $\msf B$. A $G$-colouring with boundary condition $\fr b$ of such a graph $\Gamma$ is an assignment $\fr g \in G^\msf E$ such that $\prod_{\msf e \ni \msf v} \fr g(\msf e) = 1_G$ at every vertex $\msf v \in \msf V \, \setminus \, \msf L$, and $\fr g(\msf e_\msf v) = \fr b(\msf v)$ for every $\msf v \in \msf B$, with $\msf e_\msf v$ the unique edge incident to $\msf v$ assumed to be oriented in the inward direction.
We define a \emph{$G$-coloured graph} with boundary condition $\fr b$ as a  pair $(\Gamma, \fr g)$ consisting of such a graph and $G$-colouring. We write $\mc V_G(\Sigma, \fr b)$ for the complex vector space freely generated by the set of $G$-coloured graphs with boundary condition $\fr b$. Generalising the equivalence relation $\sim$ above to embedded discs $\mathbb D \subset {\rm int}(\Sigma)$, we define $\mc V^0_{G}(\Sigma,\fr b) := \mc V_G(\Sigma,\fr b) \, / \, \mc V^{\rm null}_G(\Sigma, \fr b)$. This space is canonically isomorphic to the ground-state subspace $\mc H_G^0(\Gamma_\triangle, \fr b)$ of the quantum double model defined on any Poincaré dual graph $\Gamma_\triangle$ of a cell decomposition $\Sigma_\triangle$ of $\Sigma$ whose boundary $1$-cells are dual to $\msf B$. Finally, taking the direct sum over all possible boundary conditions produce $\mc V^0_G(\Sigma)$.

\subsection{Localised excitations and tube algebra\label{sec:anyons}}

Let us review the excitation content of the quantum double model. We adopt the `tube algebra' approach of refs.~\cite{Koenig:2010uua,kongBdries,PhysRevB.90.115119,Hu:2015dga,Aasen:2017ubm,Bultinck:2015bot,Bullivant:2019fmk}, which builds upon the eponymous construction from subfactor theory \cite{ocneanu1994chirality,Izumi2000,ocneanu2001operator,bockenhauer2001longo,MUGER2003159,Neshveyev_Yamashita_2018}. Let us choose $\Sigma$ to be a closed two-dimensional surface and let $\Gamma_\triangle$ be the Poincaré dual of a cell decomposition $\Sigma_\triangle$. A low-lying \emph{excitation} of the quantum double model is defined as an eigenstate of the Hamiltonian $\mathbb H_G(\Gamma_\triangle)$ with an energy `slightly' above the ground state. Because the local operators $\mathbb A_\msf v^G$, where $\msf v \in \msf V$, and $\mathbb B_\msf p^G$, where $\msf p \in \msf P$, are simultaneously diagonalisable, a low-lying excitation $\psi$ violates only a small number of the defining ground state constraints $\mathbb A^G_\msf v(\psi) = + \psi$ and $\mathbb B_\msf p^G(\psi) = + \psi$.
This invites us to consider localised excitations that violate ground state constraints only within a small number of disconnected embedded discs $\mathbb D \subset \Sigma$.\footnote{\label{foot:cylinder} On a closed two-dimensional surface, it follows from global constraints that it is not possible to violate a single vertex constraint and/or a single plaquette constraint. Therefore, the minimal topology required to isolate an excitation is the twice-punctured two-sphere, which is homeomorphic to the cylinder. On the cylinder, the ground state subspace of the model decomposes into sectors corresponding to a quasi-particle at one end and its quasi-antiparticle partner at the other end.}  The fact that such excitations are supported in arbitrarily small regions of $\Sigma$ allows us to regard them as point-like \emph{quasi-particles} of the theory. Our task is to classify all possible types of such quasi-particle excitations.

Consider an embedded disc $\mathbb D \subset \Sigma$ whose boundary $\partial \mathbb D$ is intersected transversely by the graph $\Gamma_\triangle$, and suppose the energy of the configuration within $\mathbb D$ exceeds the ground state energy. Excising the interior of $\mathbb D$ from $\Sigma$ produces a new boundary component, carrying one marked point at each intersection of $\Gamma_\triangle$ with $\partial \mathbb D$. Marked points in $\msf B := \Gamma_\triangle \cap \partial \mathbb D$ coincide with the leaves of the resulting graph $\Gamma_\triangle \,\setminus\, \mathrm{int}(\mathbb D)$. Schematically, 
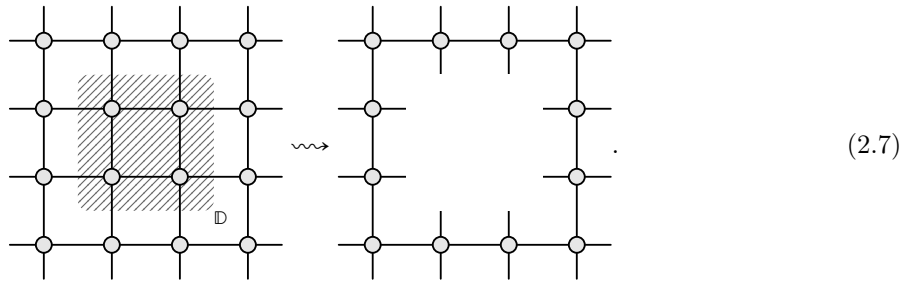
\begin{equation}
    \lattice{1} \rightsquigarrow \lattice{2}.
\end{equation}
The guiding philosophy is to view such localised excited states as elements of the ground state subspace of the model $\mathbb H_G(\Gamma_\triangle \,\setminus\, \mathrm{int}(\mathbb D))$. As mentioned in sec.~\ref{sec:lattice}, this Hamiltonian leaves boundary conditions unconstrained, so its ground state subspace decomposes into sectors of fixed boundary condition. Classifying the different types of localised excitation amounts to classifying  \emph{irreducible} sectors under a gauge action that we specify below.

Consider the oriented \emph{annulus} $(\partial \mathbb D)^\mathbb I \equiv \partial \mathbb D \times [0,1]$, viewed as a two-dimensional surface with two boundary components $\partial \mathbb D \times \{0\}$ and $\partial \mathbb D \times \{1\}$, each carrying a copy of the set $\msf B$ of marked points. Consider the space $\mc V^0_G((\partial \mathbb D)^\mathbb I)$ of equivalence classes of formal linear combinations of $G$-coloured graphs on the cylinder. It turns out that the space $\mc V^0_G((\partial \mathbb D)^\mathbb I)$ is endowed with the structure of a (finite semisimple) $*$-algebra. Briefly, given two elements of $\mc V^0_G((\partial \mathbb D)^\mathbb I)$, their product is obtained by gluing them along $\partial \mathbb D$ before invoking $\mc V^0_G(\partial \mathbb D \times [0,2]) \cong \mc V^0_G(\partial \mathbb D \times [0,1])$ (the precise definition is given below). Algebras constructed in this way are referred to as \emph{tube algebras}. Crucially, the ground state subspace $\mc H_G^0(\Gamma_\triangle \,\setminus\, \mathrm{int}(\mathbb D)) \cong \mc V_G^0(\Sigma \, \setminus \, {\rm int}(\mathbb D))$ is found to carry a (left) module structure over this tube algebra. The module action is also obtained by gluing elements of $\mc V^0_G((\partial \mathbb D)^\mathbb I)$ to states in $\mc V_G^0(\Sigma \, \setminus \, {\rm int}(\mathbb D))$ in such a way that the topology of the punctured surface with marked points is preserved. This is the gauge action mentioned above. Since the collar neighbourhood of $\partial\mathbb{D}$ in the surface $\Sigma \,\setminus\, {\rm int}(\mathbb{D})$ is homeomorphic to $(\partial\mathbb{D})^\mathbb{I}$, one can always reduce the module action of the tube algebra to multiplication within the tube algebra. It follows that localised excitations are classified by simple modules over the tube algebra. 

\bigskip \noindent
Topological invariance further implies that the \emph{Morita class} of tube algebras constructed in this way does not depend on the number of marked points on $\mathbb D$. Therefore, we pick from now on the Morita class representative obtained by setting $|\msf B|=1$. We denote the corresponding tube algebra by $\Tu_G^G$. Concretely, as a vector space,
\begin{equation}
    \label{eq:basisTubeGG}
    \Tu_G^G = \mathbb C \left\{ \, \tubeT{g}{x}{{}^gx}{}  \right\}_{g,x \in G} \!\!\!\!\!\! \equiv 
    \mathbb C \left\{ \, \cylTube{g}{x}{{}^gx}{}  \right\}_{g,x \in G} \!\!\!\!\!\! \equiv 
    \mathbb C \big\{\tub{x}{g} \! \big\}_{g,x \in G} \, , 
\end{equation}
where we introduced the shorthand ${}^gx \equiv gxg^{-1}$. The diagram in eq.~\eqref{eq:basisTubeGG} should be interpreted as a $G$-coloured graph on the cylinder. The multiplication rule is defined by   
\begin{align}
    \tub{x_1}{g_1} \cdot \tub{x_2}{g_2} 
    &\equiv 
    \tubeT{g_1}{x_1}{{}^{g_1}x_1}{}
    \cdot     
    \tubeT{g_2}{x_2}{{}^{g_2}x_2}{}
    := \delta_{x_1,{}^{g_2}x_2} \; 
    \tubeMult{g_2}{g_1}{x_2}{{}^{g_2}x_2}{{}^{g_1g_2}x_2}{}
    \nn \\[-1ex]
    &=
    \delta_{x_1,{}^{g_2}x_2} \; \tubeT{g_1g_2}{x_2}{{}^{g_1g_2}x_2}{}
    = \delta_{x_1,{}^{g_2}x_2} \, \tub{x_2}{g_1g_2},
\end{align}
for every $g_1,g_2,x_1,x_2 \in G$. To go from the first line to the second line, we simply replaced the $G$-coloured graph in $\mc V_G^0((\partial \mathbb D)^\mathbb I)$ by another one with the same boundary condition.  
The $*$-structure reads
\begin{equation}
    (\tub{x}{g})^* \equiv \left( \tubeT{g}{x}{{}^gx}{}\right)^* := \tubeT{g^{-1}}{{}^gx}{x}{} = \tub{gxg^{-1
    }}{g^{-1}},
\end{equation}
for every $g,x \in G$.

The modules over this tube algebra---which is isomorphic to the \emph{quantum double} of the group algebra $\mathbb C[G]$---are well understood  \cite{DIJKGRAAF199160}. In particular, simple objects in the category $\Mod(\Tu_G^G)$ are labelled by pairs $(\cl(f),\hat V)$ consisting of a conjugacy class $\cl(f) \in \cl(G)$, with representative $f \in G$, and an irreducible representation $\hat V \in \Irr(\Rep(Z_G(f)))$ of the centraliser $Z_G(f)$ of $f$ in $G$. For every simple module $(\cl(f),\hat V)$ over $\Tu_G^G$, the corresponding \emph{minimal central idempotent} reads 
\begin{equation}
    \mc E_{(\cl(f),\hat V)} = \frac{\dim_\mathbb C \hat V}{|Z_G(f)|} \sum_{x \in \cl(f)} \sum_{g \in Z_G(x)} \overline{\chi_{\hat V}(z_{g,x})} \; \tub{x}{g},
\end{equation}
where $\chi_{\hat V}$ denotes  the character of $\hat V$ and $z_{g,x}$ is an element in $Z_G(f)$ that is uniquely defined, up to conjugacy, given $x \in \cl(f)$ and $g \in Z_G(x)$ (see sec~\ref{sec:Maths_VGG} for details). Minimal central idempotents satisfy
\begin{align}
    \mc E_{(\cl(f_1),\hat V_1)} \cdot \mc E_{(\cl(f_2),\hat V_2)}
    = \delta_{\cl(f_1),\cl(f_2)} \; \delta_{\hat V_1,\hat V_2} \; \mc E_{(\cl(f_1),\hat V_1)} 
    \q \text{and} \q
    \sum_{\cl(f),\hat V}  \mc E_{(\cl(f),\hat V)}= 1_{\Tu_G^G}.
\end{align}
Physically, a localised excitation associated with a simple module of the form $(\cl(1_G),\hat V)$ is referred to as a (pure) \emph{charge}. These correspond to violations of the Gau{\ss} constraints $\mathbb B_\msf p^G(\psi) = + \psi$. On the other hand, a localised excitation associated with a simple module of the form $(\cl(f),\mathbb C)$ is referred to as a (pure) \emph{flux}. These correspond to violations of the flatness constraints $\mathbb A_\msf v(\psi) = + \psi$. 
The localised excitations can be fused and braided, with the property that no single excitation braids trivially with all others \cite{KITAEV20032}. Mathematically, this is captured by the fact that the category $\Mod(\Tu_G^G)$ carries the structure of a \emph{non-degenerate braided fusion category} (see sec.~\ref{sec:Maths_VGG}) \cite{DIJKGRAAF199160,Levin:2004mi,Kitaev:2005hzj}.
In general, the exchange statistics of these quasi-particles is neither bosonic nor fermionic, so that we refer to them collectively as \emph{anyons}.

\subsection{Anyon condensation\label{sec:condensation}}

Phase transitions between topologically ordered phases can be induced by the condensation of certain collections of localised excitations into the vacuum. Generically, the condensing object is a composite quasi-particle equipped with a structure that must satisfy strict consistency conditions. Notably, it must behaves like a \emph{boson}. All properties of the resulting child theory can be systematically derived from the knowledge of the parent theory and the chosen condensation pattern.
The fate of the parent excitations is dictated by their mutual braiding statistics with the condensate. Roughly, quasi-particles that braid trivially with the condensed composite  give rise to \emph{deconfined} localised excitations in the new vacuum. The remaining ones become dynamically \emph{confined} and are restricted to live at the endpoints of \emph{domain walls}.
Mathematically, the theory of anyon condensation is well established~\cite{Bais:2008ni,Kong:2013aya}. In particular, patterns of anyon condensation are classified by \emph{condensable algebras} in the (non-degenerate braided fusion) category of localised excitations (see sec.~\ref{sec:Maths_VGGA}).

While the quantum double model admits various condensation patterns, we focus on the condensation of pure charges. Given a subgroup $H \leq G$, we wish to condense the composite pure charge associated with the $\Tu_G^G$-module $(\cl(1_G),\mathbb C(G/H))$, where $\mathbb C(G/H) \cong \Ind_H^G(\mathbb C)$ is the permutation $G$-representation of the set $G/H$ of left cosets. As discussed in sec.~\ref{sec:Maths_VGGA}, these form a condensable algebra in $\Mod(\Tu_G^G)$ with product the pointwise multiplication of functions $G/H \to \mathbb C$. It follows from \emph{Frobenius reciprocity} that the charges that (at least partially) condense are precisely those labelled by $G$-representations whose restriction to $H$ contains the trivial representation.

Let us construct the Hamiltonian of the condensed theory. The setup is the same as for the initial quantum double model. We work on the Poincaré dual graph $\Gamma_\triangle$ of a cellular decomposition $\Sigma_\triangle$ of a surface $\Sigma$ with a possibly non-empty boundary. The microscopic Hilbert space is still provided by $\mc H_G(\Gamma_\triangle)$. 
Since pure charges correspond to violations of the Gau{\ss} constraints, the vertex terms need not be altered. 
However, the plaquette operators need to be modified to allow for the condensate to form. Allowing for charges in the condensed composite to freely proliferate in the ground state sector requires reducing the local gauge symmetry from $G$ to $H$. Therefore, to every plaquette $\msf p \in \msf P$, we now assign the projector $\mathbb B_\msf p^H$ that averages over the action of the subgroup $H$. 
Accordingly, at every edge $\msf e \in \msf E \,\setminus\, \msf E_\msf L$---where recall that $\msf E_\msf L = \{\msf e \in \msf E \mid \msf e \cap \msf L \neq \varnothing\}$ is the set of edges incident to the leaves in $\msf L$---one assigns a projector $\mathbb \Pi_\msf e^H$ that forces basis states $\fr g \in \mc H_G(\Gamma_\triangle)$ to satisfy $\fr g(\msf e) \in H$. 
This is necessary in order to prevent degrees of freedom to fluctuate freely in $G/H$ without violating the flatness constraints. The new Gau{\ss} constraints being blind to such fluctuations, this would lead to an unphysical extensive ground state degeneracy otherwise. Bringing everything together, we are led to consider the Hamiltonian 
\begin{equation}
    \label{eq:HGH}
    \mathbb H_{G,H}(\Gamma_\triangle) 
    = - \! \sum_{\msf v \in \msf V \, \setminus \, \msf L} 
    \! \mathbb A_\msf v^G -\sum_{\substack{\msf p \in \msf P}} \, \mathbb B_\msf p^H - \! \sum_{\msf e \in \msf E \, \setminus \, \msf E_\msf L} \! \mathbb \Pi_\msf e^H.
\end{equation}
This model was initially considered in ref.~\cite{Bombin:2007qv}, and we refer the reader to this reference for additional motivation. Lattice models associated with more general patterns of anyon condensation and for broader classes of topological orders were constructed in refs.~\cite{PhysRevB.87.155115,TEO2015349,PhysRevB.94.235136,Fidkowski:2016svr,Williamson:2017uzx,PhysRevB.100.115147,Christian:2023ekm}.

The ground state subspace $\mc H_{G,H}^0(\Gamma_\triangle)$ of \eqref{eq:HGH} can still be formulated in terms of equivalence classes of coloured graphs. 
Fix a finite set $\msf B$ of points on $\partial \Sigma$ together with an assignment $\fr b \in G^\msf B$. 
We consider graphs $\Gamma = (\msf V, \msf E)$ embedded in $\Sigma$ whose leaf set $\msf L$ equals $\msf B$. A $(G,H)$-colouring with boundary condition $\fr b$ of such a graph $\Gamma$ is an assignment $\fr g \in G^\msf E$ such that $\fr g(\msf e) \in H$ at every edge $\msf e \in \msf E \, \setminus \, \msf E_\msf L$, $\prod_{\msf e \ni \msf v} \fr g(\msf e) = 1_G$ at every vertex $\msf v \in \msf V \, \setminus \, \msf L$, and $\fr g(\msf e_\msf v) = \fr b(\msf v)$ for every $\msf v \in \msf B$.
Notice that in order for $\fr b(\msf v_1) \in G \setminus  H$ at some $\msf v_1 \in \msf L$, there must be at least another $\msf v_2 \in \msf L$ such that $\fr b(\msf v_2) \in G  \setminus  H$ and $\msf e_{\msf v_1} \cap \msf e_{\msf v_2} \neq \varnothing$.
We define a \emph{$(G,H)$-coloured graph} with boundary condition $\fr b$ as a  pair $(\Gamma, \fr g)$ consisting of such a graph and a  $(G,H)$-colouring. We write $\mc V_{G,H}(\Sigma, \fr b)$ for the complex vector space freely generated by the set of $(G,H)$-coloured graphs with boundary condition $\fr b$. Constructing $\mc V^{\rm null}_{G,H}(\Sigma,\fr b)$ as before, we ultimately define $\mc V^0_{G,H}(\Sigma,\fr b) := \mc V_{G,H}(\Sigma,\fr b) \, / \, \mc V^{\rm null}_{G,H}(\Sigma, \fr b)$. Taking the direct sum over all possible boundary conditions in $G^\msf B$ produces $\mc V^0_{G,H}(\Sigma)$, which is isomorphic to $\mc H^0_{G,H}(\Gamma_\triangle)$, defined on the Poincaré dual of a cell decomposition $\Sigma_\triangle$ whose boundary 1-cells are dual to $\msf B$.  

\bigskip \noindent
Equipped with the above parametrisation of the ground state subspace, the analysis of the excitation content of the theory proceeds as before. Specifically, localised excitations are classified by simple modules over 
a Morita class of tube algebras. We denote the Morita class representative obtained by fixing the number of marked points on the circle to one by $\Tu_G^H$.  As a vector space
\begin{equation}
    \label{eq:basisTubeGH}
    \Tu_G^H = \mathbb C \left\{ \, \tubeT{h}{x}{{}^hx}{}  \right\}_{x \in G, \, h \in H} \!\!\!\!\!\! \equiv 
    \mathbb C \left\{ \, \cylTube{h}{x}{{}^hx}{}  \right\}_{x \in G, \, h \in H} \!\!\!\!\!\! \equiv
    \mathbb C \big\{\tub{x}{h} \! \big\}_{x \in G, \, h \in H} \, .
\end{equation}
The diagram in eq.~\eqref{eq:basisTubeGH} should be interpreted as a $(G,H)$-coloured graph on the cylinder.
The multiplication rule is defined by   
\begin{align}
    \tub{x_1}{h_1} \cdot \tub{x_2}{h_2} 
    := \delta_{x_1,{}^{h_2}x_2} \, \tub{x_2}{h_1h_2},
\end{align}
for every $x_1,x_2 \in G$ and $h_1,h_2 \in H$, while the $*$-structure reads
\begin{equation}
    (\tub{x}{h})^* = \tub{hxh^{-1}}{h^{-1}},
\end{equation}
for every $x \in G$ and $h \in H$.
The representation of this tube algebra is also well understood. In particular, simple objects in $\Mod(\Tu_G^H)$ are labelled by pairs $(\orb(f),\hat W)$ consisting of an $H$-conjugacy class $\orb(f) \in \orb(G)$, with representative $f \in G$, and an irreducible representation $\hat W \in \Irr(\Rep(\Stab_H(f)))$ of the centraliser $\Stab_H(f)$ of $f$ in $H$. For every simple module $(\orb(f),\hat W)$ over $\Tu_G^H$, the corresponding minimal central idempotent reads 
\begin{equation}
    \label{eq:MCI}
    \mc E_{(\orb(f),\hat W)} = \frac{\dim_\mathbb C \hat W}{|\Stab_H(f)|} \sum_{x \in \orb(f)} \sum_{h \in \Stab_H(f)} \overline{\chi_{\hat W}(s_{h,x})} \; \tub{x}{h},
\end{equation}
where $\chi_{\hat W}$ denotes  the character of $\hat W$ and $s_{h,x}$ is an element in $\Stab_H(f)$ that is uniquely defined, up to $H$-conjugacy, given $x \in \orb(f)$ and $h \in \Stab_H(x)$. Explicit formulas for \emph{primitive idempotents} and \emph{nilpotents} can also be provided (see sec~\ref{sec:Maths_tube} for details). 

A few preliminary remarks about the excitation content are in order.
First of all, note that the algebra $\Tu_H^H$ is a subalgebra of $\Tu_G^H$. From the discussion above, we deduce that simple modules over $\Tu_H^H$ label the deconfined excitations of the condensed theory. They form the \emph{untwisted sector} of the theory. The remaining localised excitations---the confined excitations---organise into distinct sectors labelled by \emph{double cosets} $[k] \equiv HkH \in H\backslash G/H$. Each double coset corresponds to a topological domain wall, i.e., violations of the edge constraints $\mathbb \Pi_\msf e^H(\psi)= + \psi$. The collection of localised excitations confined to live at the endpoint of the domain wall labelled by $[k]$ form the \emph{$[k]$-twisted sector}.\footnote{The domain wall itself is not a localised excitation.} Algebraically, the simple modules of this sector are precisely those of $\Tu_G^H$ supported on the subspace spanned by tube elements $\mc T_x^h$ with $x \in [k]$. In sec.~\ref{sec:Maths_HyperGr}, we refine this structure further and show that the category $\Mod(\Tu_G^H)$ is a \emph{hypergroup} $H\setminus G/H$-\emph{graded extension} of $\Mod(\Tu_H^H)$. In particular, we discuss in which sense the monoidal structure of $\Mod(\Tu_G^H)$, which encodes the fusion of the localised excitations, is compatible with this hypergroup grading.

\subsection{Non-invertible symmetry\label{sec:symmetry}}

This brings us to the main construction of this section. Our goal is to identify the (non-anomalous) non-invertible symmetry of the condensed theory and compute the symmetry action on the localised excitations. From the invertible scenario \cite{PhysRevB.100.115147}---where the symmetry is implemented by unitary operators---one expects the non-invertible symmetry to be implemented by operators that, when restricted to a bounded region, produce domain wall excitations along its boundary. 

Let us begin by defining the symmetry action within the ground state subspace. Consider $\mathbb H_{G,H}(\Gamma_\triangle)$ for $\Gamma_\triangle$ the Poincaré dual graph of a cellular decomposition $\Sigma_\triangle$ of a \emph{closed} surface $\Sigma$. For every double coset $[k] \in H \setminus G / H$, we assign to every plaquette $\msf p \in \msf P$ the operator
\begin{equation}
    \mathbb B_\msf p^{[k]} = \frac{1}{|[k]|} \sum_{g \in [k]} \mathbb B_\msf p^g.
\end{equation}
Let us mention that these form a representation of the hypergroup $H \setminus G/H$ mentioned above such that 
\begin{equation}
    \mathbb B^{[k_1]}_\msf p \mathbb \cdot \mathbb B_\msf p^{[k_2]} = \sum_{[k_3] \in H \setminus G /H} C^{[k_1][k_2]}_{[k_3]} \mathbb B^{[k_3]}_\msf p,
\end{equation} 
for every $[k_1],[k_2] \in H \setminus G /H$, where $C^{[k_1][k_2]}_{[k_3]}$ are real non-negative constants (see sec.~\ref{sec:Maths_HyperGr} and \ref{sec:Maths_monad} for details).\footnote{Symmetries associated with double cosets have appeared previously in a different context, see e.g. ref.~\cite{Cao:2025qnc}.} Note that when choosing $[1_G] = H$, one recovers $\mathbb B_\msf p^H$, which satisfy
\begin{equation}
    \label{eq:commutTriv}
    \mathbb B_\msf p^H \cdot \mathbb B_\msf p^{[k]} = \mathbb B_\msf p^{[k]} = \mathbb B_\msf p^{[k]} \cdot \mathbb B_\msf p^H, 
\end{equation} 
for every $[k] \in H \setminus G /H$.
Moreover, let us remark that for neighbouring plaquettes $\msf p_1, \msf p_2 \in \msf P$, the operators $\mathbb B_{\msf p_1}^{[k]}$ and $\mathbb B_{\msf p_2}^{[k]}$ commute. Indeed, by convention, the orientation of any edge $\msf e$ cannot be induced by the orientation of both adjacent plaquettes. Therefore, the operators cannot simultaneously act by left or right action on the corresponding degree of freedom.

The symmetry operator associated with $[k] \in H \setminus G /H$ is then given by 
\begin{equation}
    \mathbb B^{[k]} := \bigg(\prod_{\msf e \in \msf E} \mathbb \Pi_\msf e^H\bigg) \cdot \bigg(\prod_{\msf p \in \msf P} \mathbb B_\msf p^{[k]} \bigg) \cdot \bigg( \prod_{\msf e \in \msf E} \mathbb \Pi_\msf e^H \bigg) .
\end{equation}
Physically, this amounts to `sweeping' a line defect from the $[k]$-twisted sector across the whole surface $\Sigma$. 
Notice that we have included projectors onto the trivial sector at every edge.\footnote{This additional operation is not necessary for invertible symmetries.} This is to prevent domain walls from proliferating during the process. One can verify that the symmetry operator $\mathbb B^{[k]}$ commutes with the ground state projector. We have already addressed the edge operators; it clearly commutes with the vertex operators for the same reason that $\mathbb H_{G,H}(\Gamma_\triangle)$ is a local commuting projector Hamiltonian; it commutes with plaquette operators thanks to eq.~\eqref{eq:commutTriv}.

Next, let $\Omega \subset \msf P$ be a simply connected bounded region of $\Sigma$ formed by a finite, contiguous subset of plaquettes in $\msf P$, and let $\msf E^0_\Omega$ denote the set of edges that are strictly in the interior of $\Omega$, i.e., edges that are shared by exactly two adjacent plaquettes in $\Omega$. Let us restrict the action of $\mathbb B^{[k]}$ to this region $\Omega$. Concretely, consider the operator
\begin{equation}
    \mathbb B^{[k]}_\Omega := \bigg(\prod_{\msf e \in \msf E^0_\Omega} \mathbb \Pi_\msf e^H\bigg) \cdot \bigg(\prod_{\msf p \in \Omega} \mathbb B_\msf p^{[k]} \bigg) \cdot \bigg( \prod_{\msf e \in \msf E^0_\Omega}\mathbb \Pi_\msf e^H \bigg) .
\end{equation}
It follows from the analysis above that the operator $\mathbb B_\Omega^{[k]}$ commutes with the ground state projector within the region $\Omega$. On its boundary, it creates a domain wall with degrees of freedom now valued in the double coset $[k] \in H \setminus G / H$. 

Let us now compute the action of the non-invertible symmetry on the localised excitations.
We exploit the fact that applying the symmetry to a bounded region creates a domain wall excitation along its boundary. Consider the twice-punctured two-sphere (or cylinder). As we mentioned in foot.~\ref{foot:cylinder}, this topology supports a quasi-particle and its quasi-antiparticle partner. Provided that these localised excitations are in some twisted sector, a non-trivial domain wall needs to be running along the cylinder between these two punctures.\footnote{Recall that domain walls continuously violate edge constraints along their paths, so that separating the excitations incurs an extensive energy cost. In particular, this implies that a state describing separated, localised excitations---in such a way that the graph is not the minimal one---in some twisted sector is not literally a ground state of the Hamiltonian $\mathbb H_{G,H}(\Gamma_\triangle)$ on the cylinder.}  In any local region away from the punctures, a state in this sector looks like a cylinder with a $G$-coloured edge running along it. Now, we can identify a disc-like region $\Omega$ wrapping around the cylinder, within which we can apply the non-invertible symmetry operator $\mathbb B_\Omega^{[k]}$. Graphically,\footnote{Crucially, the edge labelled by $x \in G$ is not considered to be in interior of $\Omega$ here.}  
\begin{equation*}
    \cylTubeAction{1} = \frac{1}{|[k]|} \sum_{g \in [k]}
    \cylTubeAction{2} = \frac{1}{|[k]|} \sum_{g \in [k]}
    \cylTubeAction{3},
\end{equation*}
where in the last step, we replace the $G$-coloured graph by another one with the same boundary condition. How should we interpret this operation? The action of $\mathbb B^{[k]}_\Omega$ translates into a simultaneous action on the quasi-particle and quasi-antiparticle partner, as well as on the domain wall. Concretely, this action is by multiplication in $\Tu_G^G$ by some tube elements of the form $\mc T^g_x$, where $x \in G$ and $g \in [k]$. Turning the argument around, consider the following space of tubes:  
\begin{equation}
    M_{[k]} = \mathbb C \left\{ \, \tubeT{g}{x}{{}^gx}{}  \right\}_{\substack{\!\! x \in G \\ g \in [k]}} \!\!\!\! \equiv 
    \mathbb C \left\{ \, \cylTube{g}{x}{{}^gx}{}  \right\}_{\substack{\!\! x \in G \\ g \in [k]}} \!\!\!\! \equiv 
    \mathbb C \big\{\tub{x}{g} \! \big\}_{\substack{\!\!x \in G \\ g \in [k]}} . 
\end{equation}
After embedding in $\Tu_G^G$, multiplying tube elements in $M_{[k]}$ by tube elements in $\Tu_G^H$ from the left and/or from the right results in tube elements in $M_{[k]}$. Therefore, $M_{[k]}$ carries the structure of a bimodule over $\Tu_G^H$. It follows there is a natural action of $M_{[k]}$ on modules over $\Tu_G^H$, which we formalise in sec.~\ref{sec:Maths_monad}. Consider a simple excitation and its corresponding simple module $(\cl_{H}(f_1),\hat W_1)$ over $\Tu_G^H$. We provided in eq.~\eqref{eq:MCI} the corresponding minimal central idempotent $\mc E_{(\cl_{H}(f_1),\hat W_1)}$, from which we can extract primitive idempotents. Let $(\mc E_{(\cl_{H}(f_1),\hat W_1)})_1^1$ be such a primitive idempotent. Then, the multiplicity of another simple module $(\cl_{H}(f_2),\hat W_2)$ in the image of the action of the hypergroup element $[k] \in H \setminus G /H$ is provided by the dimension of the subspace
$(\mc E_{(\cl_{H}(f_2),\hat W_2)})_1^1 \cdot M_{[k]} \cdot (\mc E_{(\cl_{H}(f_1),\hat W_1)})^1_1 \subset M_{[k]}$, where one recalls that the multiplication takes place in $\Tu_G^G$ (see sec.~\ref{sec:Maths_monad} for details). 

When a pair of localised excitations braid with each other, the above action on the domain wall defect line and localised excitation arise similarly, which gives rise to `defect-crossed' braiding matrix. Furthermore, by considering thrice-punctured two-spheres (aka pair of pants), one can define an action of the non-invertible symmetry of fusion spaces of the localised excitations.\footnote{The strategy here is similar to the one above. Starting from a pair of pants with some $G$-coloured graph capturing a configuration of domain walls, one applies the non-invertible symmetry operator to disc-like regions bounded by the edges of the coloured graph.} 

Starting from the condensed theory, we may want to gauge the non-invertible symmetry so as to recover the parent theory. In general, this amounts to freely proliferating domain wall of all possible types $[k] \in H \setminus G /H$ across the entire lattice. In the lattice model we consider, this operation is particularly straightforward. 
It is achieved by projecting onto the symmetric subspace of the hypergroup action on each plaquette. This projects onto the simultaneous $+1$-eigenspace of the plaquette operators $\mathbb B_\msf p^G$. In this case, the additional defect string types on each edge play the role of the gauge fields that allow the confined defects to fluctuate and become deconfined. 

\bigskip \noindent
Below, we illustrate this construction with a simple example. A related example, arising from gauging a non-normal subgroup $H\leq G$ was discussed in ref.~\cite{Hsin:2024aqb}, where the non-invertible symmetry is associated with the set $G /H$ of cosets; this is different from our setting, where the non-invertible symmetry is rather associated with double cosets in $H \backslash G / H$. 
All the properties that we motivated in this section are formalised and expanded upon in the next section.

\subsection{Example\label{sec:example}}

We choose $G$ to be the dihedral group $\mathbb D_6 = \la r,s \, | \, r^3=s^2=(sr)^2=1\ra $ of order 6. As a set, $\mathbb D_6$ is the disjoint union of conjugacy classes $\cl(1)=\{1\}$, $\cl(r)=\{r,r^2\}$, and $\cl(s) =\{s,sr,sr^2\}$. The centraliser subgroups of these conjugacy classes are isomorphic to $\mathbb D_6$, $\mathbb Z_3$, and $\mathbb Z_2$ respectively. Irreducible representations of $\mathbb Z_2$ are denoted $\{\pm\}$, irreducible representations of $\mathbb Z_3$ are denoted $\{1,\omega,\bar \omega\}$ such that $\omega \otimes \omega \cong \bar \omega$ and $\omega \otimes \bar \omega \cong 1$, while irreducible representations of $\mathbb D_6$ are denoted $\{1,e,\pi\}$, such that $e \otimes e \cong 1$, $e \otimes \pi \cong \pi$, and $\pi \otimes \pi \cong 1 \oplus e \oplus \pi$. 

Recall that simple objects in $\Mod(\Tu_{\mathbb D_6}^{\mathbb D_6})$ are labelled by pairs $(\cl(f),\hat V)$ consisting of $\cl(f) \in \cl(\mathbb D_6)$ and $\hat V \in \Irr(\Rep(Z_{\mathbb D_6}(f)))$. We obtain eight simple objects:  $(\cl(1),1), (\cl(1),e), (\cl(1),\pi)$, $(\cl(r),1)$, $(\cl(r),\omega)$, $(\cl(r),\bar \omega)$, $(\cl(s),+)$, and $(\cl(s),-)$. Note that $(\cl(1),\pi)$ and $(\cl(r),1/\omega/\bar \omega)$ are two-dimensional, while $(\cl(s),\pm)$ are three-dimensional. 
The corresponding minimal central idempotents are 
\begin{equation}
\begin{split}
    \mc E_{(\cl(1),1)} &= \frac{1}{6} \big( \mc T^1 + \mc T^r + \mc T^{r^2} + \mc T^s + \mc T^{rs} + \mc T^{r^2s}\big),
    \\
    \mc E_{(\cl(1),e)} &= \frac{1}{6} \big(\mc T^1_1 + \mc T^r_1 + \mc T^{r^2}_1 - \mc T^s_1 - \mc T^{sr}_1 - \mc T^{sr^2}_1 \big),
    \\
    \mc E_{(\cl(1),\pi)}&= \frac{1}{3} \big(2\mc T^1_1 - \mc T^r_1 - \mc T^{r^2}_1 \big),
    \\
    \mc E_{(\cl(r),1)} &= \frac{1}{3} \big(\mc T^1_r + \mc T^r_r + \mc T^{r^2}_r + \mc T^1_{r^2} + \mc T^r_{r^2} + \mc T^{r^2}_{r^2} \big),
    \\
    \mc E_{(\cl(r),\omega)} &= \frac{1}{3} \big(\mc T^1_r + \omega\mc T^r_r + \bar{\omega}\mc T^{r^2}_r + \mc T^1_{r^2} + \omega\mc T^r_{r^2} + \bar{\omega}\mc T^{r^2}_{r^2} \big),
    \\
    \mc E_{(\cl(r),\bar \omega)} &= \frac{1}{3} \big(\mc T^1_r + \bar{\omega}\mc T^r_r + \omega\mc T^{r^2}_r + \mc T^1_{r^2} + \bar{\omega}\mc T^r_{r^2} + \omega\mc T^{r^2}_{r^2} \big),
    \\
    \mc E_{(\cl(s),+)} &= \frac{1}{2} \big(\mc T^1_s + \mc T^s_s + \mc T^{1}_{r^2s} + \mc T^{r^2s}_{r^2s} + \mc T^{1}_{rs} + \mc T^{rs}_{rs} \big),
    \\
    \mc E_{(\cl(1),-)} &= \frac{1}{2} \big(\mc T^1_s - \mc T^s_s + \mc T^{1}_{r^2s} - \mc T^{r^2s}_{r^2s} + \mc T^{1}_{rs} - \mc T^{rs}_{rs} \big),
\end{split}
\end{equation}
where $\omega := \exp(\frac{2\pi i}{3})$.

Choose the non-normal subgroup $H = \mathbb Z_2 = \la s \ra < \mathbb D_6$. One condenses the composite quasi-particle $(\cl(1),\mathbb C(\mathbb D_6/\mathbb Z_2)) \cong (\cl(1),1) \oplus (\cl(1),\pi)$. The localised excitations of the condensed theory are classified by simple modules over the tube algebra $\TDZ$, similar to the approach in ref.~\cite{Williamson:2017uzx}. 
There are two double cosets in $\mathbb Z_2 \setminus \mathbb D_6/ \mathbb Z_2$: the trivial one $[1]=\mathbb Z_2 = \{1,s\}$, which labels the untwisted sector, and $[r]=\mathbb Z_2 r \mathbb Z_2 = \{r,r^2,rs,r^2s\}$, which labels the only twisted sector of the theory. The untwisted sector boils down to the topological order of the \emph{toric code} \cite{KITAEV20032}. The four quasi-particle excitations are $\bm{1} \equiv (\cl_{\mathbb Z_2}(1),+)$, $\bm{e} \equiv (\cl_{\mathbb Z_2}(1),-)$, $\bm{m} \equiv (\cl_{\mathbb Z_2}(s),+)$ and $\bm{f} \equiv (\cl_{\mathbb Z_2}(s),-)$, which are all one-dimensional. The $[r]$-twisted sector consists of two confined quasi-particles living at the endpoint of the domain wall, namely $\bm{c_1} \equiv (\cl_{\mathbb Z_2}(rs),+)$ and $\bm{c_2} \equiv (\cl_{\mathbb Z_2}(r),+)$, which are both two-dimensional. The corresponding minimal central idempotents are 
\begin{equation}
\begin{alignedat}{2}
    \mc E_{\bm 1} &= \frac{1}{2} \big(\mc T^1_1 + \mc T^s_1 \big),
    &\q\q 
    \mc E_{\bm{c_1}} &= \mc T^1_{rs} + \mc T^1_{r^2s},
    \\
    \mc E_{\bm e} &= \frac{1}{2} \big(\mc T^1_1 - \mc T^s_1 \big),
    &\q \q
    \mc E_{\bm{c_2}} &= \mc T^1_r + \mc T^1_{r^2}.
    \\
    \mc E_{\bm m} &= \frac{1}{2} \big(\mc T^1_s + \mc T^s_s \big), &
    \\
    \mc E_{\bm f} &= \frac{1}{2} \big(\mc T^1_s - \mc T^s_s \big), &
\end{alignedat}
\end{equation}
The expressions above also provide the decomposition of the minimal central idempotents $\mc E_{\bm{c_1}}$ and $\mc E_{\bm{c_2}}$ into sums of primitive idempotents. Concretely, minimal central idempotents of the one-dimensional modules are also the primitive idempotents, while  $\mc E_{\bm{c_1}} \equiv (\mc E_{\bm{c_1}})_1^1 + (\mc E_{\bm{c_1}})_2^2$ with $(\mc E_{\bm{c_1}})_1^1 = \mc T^1_{rs}$ and $(\mc E_{\bm{c_1}})_2^2 = \mc T^1_{r^2s}$,  $\mc E_{\bm{c_2}} \equiv (\mc E_{\bm{c_2}})_1^1 + (\mc E_{\bm{c_2}})_2^2$ with $(\mc E_{\bm{c_2}})_1^1 = \mc T^1_{r}$ and $(\mc E_{\bm{c_2}})_2^2 = \mc T^1_{r^2}$.

By inspection of the minimal central idempotents, one derives the image of the initial quasi-particles in the condensed theory: $(\cl(1),1) \mapsto \bm{1}$, $(\cl(1),e) \mapsto \bm{e}$, $(\cl(1),\pi) \mapsto \bm{1} \oplus \bm{e}$, $(\cl(s),+) \mapsto \bm{m} \oplus \bm{c_1}$, $(\cl(s),-) \mapsto \bm{f} \oplus \bm{c_1}$ and $(\cl(r),1/w/\bar w) \mapsto \bm{c_2}$. Indeed, recall that condensation is essentially accomplished by restricting the minimal central idempotents to tube elements $\mc T^h_x \in \TDZ$ such that $hx=xh$. For instance, 
\begin{equation}
\begin{split}
    \mc E_{(\cl(1),\pi)} &\mapsto \frac{2}{3} \mc T_1^1 = \frac{2}{3} \big(\mc E_{\bm 1} + \mc E_{\bm e} \big),
    \\
    \mc E_{(\cl(s),+)} &\mapsto \frac{1}{2} \big(\mc T^1_s + \mc T^s_s + \mc T^{1}_{r^2s}  + \mc T^{1}_{rs}\big) = \mc E_{\bm m} + \frac{1}{2} \mc E_{\bm{c_1}},
    \\ 
    \mc E_{(\cl(1),\omega)} &\mapsto \frac{1}{3} \big(\mc T^1_r + \mc T^1_{r^2}\big) = \frac{1}{3} \mc E_{\bm{c_2}}.     
\end{split}
\end{equation}
Let us now derive the action of the hypergroup $\mathbb Z_2 \setminus \mathbb D_6 / \mathbb Z_2$ on the localised excitations. Recall from sec.~\ref{sec:symmetry} that the hypergroup element $[r]$ acts on modules over the tube algebra $\TDZ$ via its corresponding $\TDZ$-bimodule $M_{[r]}$. We denote this action by $T_{[r]}$. As a vector space,
\begin{equation}
    M_{[r]} = \mathbb C\big\{\mc T^{g}_{x}\big\}_{g \in [r], \; x \in G}.
\end{equation}
The bimodule structure is obtained by embedding elements in $M_{[r]}$ and $\TDZ$ into $\Tu_{\mathbb D_6}^{\mathbb D_6}$ before using the multiplication rule of $\TDD$. Following the approach described in sec.~\ref{sec:symmetry}, given a simple module $(\cl_{\mathbb Z_2}(f),\hat W)$ over $\TDZ$, one obtains the decomposition of $T_{[r]}\big((\cl_{\mathbb Z_2}(f_1),\hat W_1)\big)$ into simple $\TDZ$-modules by computing the dimension of subspaces $(\mc E_{(\cl_{\mathbb Z_2}(f_2),\hat W_2)})_1^1 \cdot W_{[r]} \cdot (\mc E_{(\cl_{\mathbb Z_2}(f_1),\hat W_1)})^1_1$, for any choice of primitive idempotents. We list below all non-vanishing such subspaces. First of all, we have
\begin{equation}
\begin{split}
    \mc E_{\bm 1} \cdot M_{[r]} \cdot \mc E_{\bm 1} &= \mathbb C\big\{\frac{1}{4} \big( \mc T_1^r + \mc T_1^{rs}+\mc T_1^{r^2}+\mc T_1^{r^2s} \big)\big\},
    \\
    \mc E_{\bm 1} \cdot M_{[r]} \cdot \mc E_{\bm e} &= \mathbb C\big\{\frac{1}{4} \big(\mc T_1^r - \mc T_1^{rs} - \mc T_1^{r^2} + \mc T_1^{r^2s}\big)\big\},
    \\
    \mc E_{\bm e} \cdot M_{[r]} \cdot \mc E_{\bm 1} &= \mathbb C\big\{\frac{1}{4} \big(\mc T_1^r + \mc T_1^{rs} - \mc T_1^{r^2} - \mc T_1^{r^2s}\big)\big\},
    \\
    \mc E_{\bm e} \cdot M_{[r]} \cdot \mc E_{\bm e} &= \mathbb C\big\{\frac{1}{4} \big( \mc T_1^r - \mc T_1^{rs}+\mc T_1^{r^2} - \mc T_1^{r^2s}\big)\big\},
\end{split}
\end{equation}
from which we deduce that $T_{[r]}(\bm{1}) \cong \bm{1} \oplus 
\bm{e}$ and $T_{[r]}(\bm{e}) \cong \bm{1} \oplus 
\bm{e}$. Already, this indicates that the symmetry associated with $\mathbb Z_2 \setminus \mathbb D_6 / \mathbb Z_2$ is non-invertible. Similarly, one finds that 
\begin{equation}
\begin{split}
    (\mc E_{\bm{c_1}})^1_1 \cdot M_{[r]} \cdot \mc E_{\bm m} &= \mathbb C\big\{\frac{1}{2} \big( \mc T_s^{r^2} + \mc T_s^{r^2s} \big)\big\},
    \\
    (\mc E_{\bm{c_1}})^1_1 \cdot M_{[r]} \cdot \mc E_{\bm f} &= \mathbb C\big\{\frac{1}{2} \big(\mc T_s^{r^2} - \mc T_s^{r^2s}\big)\big\},
    \\
    (\mc E_{\bm{c_1}})^1_1 \cdot M_{[r]} \cdot (\mc E_{\bm{c_1}})_1^1 &= \mathbb C\big\{\mc T_{rs}^{rs} \big\},
    \\
    \mc E_{\boldsymbol m} \cdot M_{[r]} \cdot (\mc E_{\bm{c_1}})^1_1 &= \mathbb C\big\{\frac{1}{2} \big(\mc T_{rs}^{r} + \mc T_{rs}^{r^2s}\big)\big\},
    \\
    \mc E_{\bm f} \cdot M_{[r]} \cdot (\mc E_{\bm{c_1}})^1_1
    &= \mathbb C\big\{\frac{1}{2} \big(\mc T_{rs}^{r} - \mc T_{rs}^{r^2s}\big)\big\},
\end{split}
\end{equation}
which implies that $T_{[r]}(\bm{m}) \cong \bm{c_1}$, $T_{[r]}(\bm{f}) \cong \bm{c_1}$ and $T_{[r]}(\bm{c_1}) \cong \bm{m} \oplus \bm{f} \oplus \bm{c_1}$. Finally, the only two-dimensional such subspace is provided by
\begin{equation}
    (\mc E_{\bm{c_2}})_1^1 \cdot M_{[r]} \cdot (\mc E_{\bm{c_2}})^1_1 = \mathbb C\{\mc T^r_r,\mc T^{r^2}_r\},
\end{equation}
which implies that $T_{[r]}(\bm{c_2}) \cong \bm{c_2} \oplus \bm{c_2}$. 
The above defect tubes provide examples of the non-invertible-symmetry defects, their generalised permutation action that mixes anyons and defects, as well as their action on individual defects via algebras that are not necessarily equivalent to the underlying hypergroup, which is a generalisation of symmetry fractionalisation. 

We conclude by highlighting the following property. Let $\mc E_{[r]} := \frac{1}{4}(\mc T_1^r + \mc T_1^{rs} + \mc T_1^{r^2} + \mc T_1^{r^2s})$ be our choice of basis vector for $\mc E_{\bm 1} \cdot M_{[r]} \cdot \mc E_{\bm 1}$. We have
\begin{equation}
    \mc E_{[r]}^2 = \frac{1}{2}\mc E_{\bm 1} + \frac{1}{2} \mc E_{[r]},
\end{equation}
which, as we clarify in the next section, is another hint of the hypergroup structure of $\mathbb Z_2 \setminus \mathbb D_6 / \mathbb Z_2$.
In particular, we revisit the very same example in sec.~\ref{sec:Maths_example} from this vantage point. 

\section{Topological order enriched by a non-invertible symmetry\label{sec:Maths}}

\emph{Guided by the lattice picture of the previous section, we propose a formalisation of the non-invertible symmetry that appeared there in terms of a Hopf monad. Furthermore, we formalise the topological order enriched by such a non-invertible symmetry in terms of a hypergroup-equivariant hypergroup-graded fusion category. Some familiarity with the theory of braided fusion categories is assumed. We encourage the reader to consult \cite{etingof2016tensor} for background material.}

\subsection{Topological order enriched by an invertible symmetry}

We begin by reviewing the mathematical formalism used to describe a (2+1)d topological order enriched by an invertible symmetry group $G$, which will serve as the foundation for our subsequent generalisation to non-invertible symmetry enrichments. We follow references  \cite{KirillovJr2002,etingof2010fusion,etingof2016tensor,PhysRevB.100.115147}. 

Given a modular tensor category $\mc B$, a \emph{categorical $G$-action} is a \emph{monoidal functor} $T \colon \underline{G} \to \underline{\rm Aut}_\otimes^{\rm br}(\mc B)$ from the group $G$ treated as a monoidal category with object set $G$ to the \emph{categorical group} $\underline{\rm Aut}_\otimes^{\rm br}(\mc B)$ of braided tensor autoequivalences of $\mc B$.\footnote{Objects in $\underline{\rm Aut}_\otimes^{\rm br}(\mc B)$ are braided tensor autoequivalences, and morphisms are natural isomorphisms between them. The monoidal structure is provided by composition of functors.} Recall that a braided tensor autoequivalence of $\cB$ is a tensor functor $F \colon \cB \to \cB$ with natural isomorphisms $J_{X,Y}: F(X)\otimes F(Y)\xrightarrow{\sim} F(X\otimes Y)$ satisfying $F(R_{X,Y}) \circ J_{X,Y}=J_{Y,X} \circ R_{F(X),F(Y)}$, where $R_{X,Y} \colon X \otimes Y \xrightarrow{\sim} Y \otimes X$ is the braiding. Given a categorical action, we may consider a $G$-\emph{crossed braided extension} of $\mc B$. Such an extension is in general not unique. Once a categorical $G$-action is fixed, inequivalent extensions are classified by an additional piece of cohomological data, namely a class in $H^3(G,{\rm U}(1))$, subject to the vanishing of an \emph{obstruction} in $H^4(G,{\rm U}(1))$ (see ref.~\cite{etingof2010fusion,PhysRevB.100.115147} for details). Provided it exists, we denote such an extension by $\mc B_G^\times$.

As a $G$-crossed braided extension of $\mc B$, $\mc B_G^\times$ is equipped with various structures: (i) First of all, $\mc B_G^\times$ is $G$-graded, $\mc B^\times_G = \bigoplus_{g \in G}  (\mc B^\times_G)_g$, with $(\mc B^\times_G)_{1_G} \cong \mc B$. We refer to the component $(\mc B^\times_G)_g$ for $g\neq 1_G$ as the \emph{$g$-twisted sector}. We assume the $G$-grading to be \emph{faithful}, so none of the components $(\mc B^\times_G)_g$ are zero. Moreover, the tensor product $\otimes$ of $\mc B_G^\times$ maps $(\mc B^\times_G)_{g_1} \boxtimes (\mc B^\times_G)_{g_2} $ to $(\mc B^\times_G)_{g_1 g_2}$, for every $g_1,g_2\in G$, i.e., it is compatible with the $G$-grading. It follows that every $(\mc B^\times_G)_g$ defines an invertible bimodule category over the untwisted sector $\mathcal B$. (ii) There is a $G$-action $g \mapsto T_g$ on $\mc B_G^\times$ such that $T_g \big( (\mc B^\times_G)_x \big) \subset (\mc B^\times_G)_{gxg^{-1}}$ for every $g,x \in G$, i.e., it is compatible with the $G$-grading. (iii) The braiding is twisted by the $G$-action, i.e., for $X_1 \in (\mc B^\times_G)_g$ and $X_2\in \cB_G^\times$ we have natural isomorphisms $R_{X_1,X_2} \colon X_1 \otimes X_2 \xrightarrow{\sim} T_g (X_2) \otimes X_1$. This collection of natural isomorphisms is referred to as the \emph{$G$-crossed braiding}. These structures are subject to various coherence relations, which we omit here. Finally, let us mention two important properties. 
The first one---sometimes referred to as the \emph{fixed point theorem}---states  that the rank of $(\mc B^\times_G)_g$ is equal to the number of simple objects in $\cB$ left invariant by $T_g$. The second one is that the \emph{global quantum dimension} is the same in each component.

Once we have a $G$-crossed braided extension $\cB_G^\times$, we can consider its \emph{$G$-equivariantisation}. A $G$-equivariant object is a pair $(X,\{\lambda_g\}_{g\in G})$ consisting of an object $X \in \cB_G^\times$ and isomorphisms $\lambda_g \colon T_g(X) \cong X$ such that $\lambda_{g_1} \circ T_{g_1}(\lambda_{g_2})= \lambda_{g_1 g_2}\circ \gamma_{g_1,g_2}(X)$, where $\gamma_{g_1,g_2}$ is the isomorphism $T_{g_1} \circ T_{g_2} \cong T_{g_1 g_2} $ defining the monoidal structure of $T \colon \underline G \to \underline{\text{Aut}}_\otimes^{\rm br}(\mc B_G^\times)$. Equivariant objects form a modular tensor category $(\cB_G^\times)^G$ with braiding inherited from the $G$-crossed braiding. Importantly, this is equivalent to the category $\Mod_{\cB^\times_G}(T)$ of modules over the \textit{Hopf monad} $T = \bigoplus_{g\in G}T_g$ (see \cite{bruguieres2011exact}).  This is the viewpoint we generalise.

\bigskip \noindent
Given a non-degenerate braided fusion category $\mc C$, recall that condensation patterns are classified by \emph{condensable algebras} in $\mc C$ \cite{Kong:2013aya}. Formally, a condensable algebra is a \emph{connected étale} algebra $A$, i.e., an algebra in $\mc C$ that is connected, indecomposable, separable and commutative. Localised excitations in the condensed theory are then encoded in the category $\mc C_A$ of right $A$-modules in $\mc C$, the deconfined excitations corresponding to the \emph{local} $A$-modules.\footnote{Recall that a right $A$-module is a pair $(M,\rho)$ consisting of an object $M \in \mathcal C$ and $\rho \colon M \otimes A\to M$. A module $(M,\rho)$ is called local if $\rho \circ R_{A,M} \circ R_{M,A} = \rho$ where $R_{A,M}$ is the braiding between $A$ and $M$.} Importantly, $(\cB_G^\times)^G$ always contains $\Rep(G)$ as a Tannakian fusion subcategory so that the algebra $\mathbb C(G)$ of functions $\phi \colon G\to\mathbb C$ with pointwise multiplication gives rise to a condensable algebra  $A$ \cite{Drinfeld2010}. Then, one can show that $((\cB_G^\times)^G)_A \simeq \cB_G^\times$, with the local modules corresponding to $(\mc B^\times_G)_{1_G} \cong \cB$. This process is referred to as \emph{de-equivariantisation}. More generally, given a non-degenerate braided fusion category $\mathcal C$ containing $\Rep(G)$ as a Tannakian fusion subcategory,  one can show that $\mathcal C_A$, with $A=\mathbb C(G) = \mathbb C\{ \delta_g\}_{g\in G}$, is a $G$-crossed braided fusion category with $(\mathcal C_A)_{1_G} \cong \mathcal C_A^\text{loc}$, such that $(\mathcal C_A)^G \simeq \mathcal C$. There is an action  $\alpha_g \colon \mathbb C(G) \to \mathbb C(G)$ of $G$ on $A$ by automorphisms of $A$, which is given by $\alpha_g(\delta_x) = \delta_{xg^{-1}}$.
This induces a $G$-action on $\mathcal C_A$. In particular, we have that $T_g \big( (M,\rho)\big) = (M,g(\rho))$, i.e., the $G$-action does not change the underlying object, but only the module structure. In fact, $g(\rho) = \rho \circ (1_M \otimes \alpha_{g^{-1}} )$, i.e., $T_g$ twists the module action by precomposing it with the algebra automorphism associated to $g^{-1} \in G$ \cite{KirillovJr2002}.

As an illustrative example that we will generalise below, we consider $\mathcal C =\mathcal Z(\Vect_G)$ and a normal subgroup $N \lhd G$. The algebra $A(N)=\mathbb C(G/N)$ of functions $\phi:G/N \to \mathbb C$ with pointwise multiplication is a condensable algebra such that $\mathcal Z(\Vect_G)_{A(N)}^\text{loc} \simeq \mathcal Z(\Vect_N)$. Including the non-local modules, $\mathcal Z(\Vect_G)_{A(N)}$ has the structure of a $G/N$-crossed braided fusion category. Each component $(\mc Z(\Vect_G)_{A(N)})_g$ for $g \in G/N$ is an invertible bimodule category over $\mathcal Z(\Vect_N)$. Moreover, $\mathcal Z (\Vect_G)_{A(N)} \simeq \mathcal Z_{\Vect_N}(\Vect_G)$, where the latter is the relative centre of $\Vect_N$ in $\Vect_G$. Doing the equivariantisation with respect to the $G/N$-action recovers $\mathcal Z(\Vect_G)$, i.e.\ $(\mathcal Z(\Vect_G)_{A(N)})^{G/N} \simeq \mathcal Z(\Vect_G)$.

\subsection{Topological lattice gauge theory\label{sec:Maths_VGG}}

In the previous section, we reviewed that the topological order of the quantum double model with input datum a finite group $G$ is described by the category $\Mod(\Tu_G^G)$ of modules over the tube algebra $\Tu_G^G$. This is an example of a non-degenerate braided fusion category. There exist many alternative formulations of the same algebraic structure. For instance, $\Mod(\Tu_G^G)$ is equivalent, as a non-degenerate braided fusion category, to the \emph{Drinfel'd centre} $\mc Z(\Vect_G)$ of the category of $G$-graded vector spaces, but also to the category $\VGG$ of \emph{$G$-equivariant $G$-graded vector spaces}. The latter formulation is particularly well-suited to what follows. We review this construction in some detail below following ref.~\cite{DAVYDOV2017149}.

\bigskip\noindent
\paragraph{$G$-equivariant $G$-graded vector space:} Define a $G$-equivariant $G$-graded vector space as a $G$-graded vector space $V = \bigoplus_{x \in G}V_x$ together with a $G$-action such that $g \cdot V_x \subseteq V_{gxg^{-1}}$, for every $g,x \in G$. 
Consider the non-degenerate braided fusion category $\VGG$ whose objects are $G$-equivariant $G$-graded vector spaces and morphisms are $G$-grading preserving $G$-equivariant linear maps. 
The monoidal structure is given by the monoidal structure of $\Vect_G$ with the group $G$ acting diagonally. Finally, the braiding $R_{V_1,V_2} \colon V_1 \otimes V_2 \xrightarrow{\sim} V_2 \otimes V_1$ is given by $ v_1 \otimes v_2 \mapsto x \cdot v_2 \otimes v_1$,
for every $v_1 \in (V_1)_x$ and $v_2 \in V_2$.

Let $\cl(G)$ denote the set of conjugacy classes of $G$ and $\cl(f) \in \cl(G)$ the conjugacy class with representative $f \in G$. 
Since conjugacy classes partition the group, the support $\supp(V)$ of any object $V \in \VGG$ is a union of conjugacy classes in $\cl(G)$. Therefore, 
\begin{equation}
    \label{eq:VecGG_decomp}
    \VGG = \! \bigoplus_{\cl(f) \in \cl(G)} \! (\VGG)_{\cl(f)} ,
\end{equation}
where $(\VGG)_{\cl(f)}$ is the subcategory of $\VGG$ consisting of objects $V \in \VGG$ such that $\text{supp}(V) = \cl(f)$. Given $\cl(f) \in \cl(G)$, let $Z_G(f)$ denote the centraliser of $f$ in $G$. There is an equivalence
\begin{equation}
    \label{eq:VecGG_equiv}
    (\VGG)_{\cl(f)} \simeq \Rep(Z_G(f)) \simeq \Mod(\mathbb C[Z_G(f)]).
\end{equation}
On the one hand, given $V_{\cl(f)} \in (\VGG)_{\cl(f)}$, $V_{f}$ is an object in $\Rep(Z_G(f))$. On the other hand, given $\hat V \in \Rep(Z_G(f))$, let $V := \Ind_{Z_G(f)}^G(\hat V)$ be the induced representation of $\hat V$ in $G$. By definition,
\begin{equation}
    V = \CG \otimes_{\mathbb C[Z_G(f)]} \hat V , 
\end{equation}
where $\CG = \mathbb C\{g\}_{g \in G}$ is treated here as $(\CG,\mathbb C[Z_G(f)])$-bimodule. We define a $G$-grading on $V$ by declaring that $g \otimes \hat v \in V$ is homogeneous with degree $gfg^{-1} \in G$. The $G$-equivariant structure is provided by the left $\CG$-module structure on the first factor, and reads $g_1 \cdot (g_2 \otimes \hat v) = g_1g_2 \otimes \hat v$ for every $g_1,g_2 \in G$ and $v \in \hat V$. 
Since $|g_1g_2 \otimes \hat v| = g_1 |g_2 \otimes \hat v|g_1^{-1}$, we conclude that $V \in (\VGG)_{\cl(f)}$. 
It remains to check that the two functors thus constructed are inverse to each other up to natural isomorphisms. Let $\{p_x\}_{x \in \cl(f)}$ be a choice of representatives in $G$ of the left cosets in $G/Z_G(f)$ such that $p_{f} = 1_G$. Firstly, for $V_{\cl(f)} \in (\VGG)_{\cl(f)}$ and a homogeneous vector $v \in V_x$, $v \mapsto p_x \otimes p_x^{-1} \cdot v$ realises $V_{\cl(f)} \cong \CG \otimes_{\mathbb C[Z_G(f)]} V_{f}$. Secondly, for $\hat V \in \Rep(Z_G(f))$ and $\hat v \in \hat V$, $\hat v \mapsto 1_G \otimes \hat v$ realises $\hat V \cong (\CG \otimes_{\mathbb C[Z_G(f)]} \hat V)_{f}$.

\bigskip \noindent
It follows from equivalence \eqref{eq:VecGG_equiv} that simple objects in $\VGG$ are labelled by pairs $(\cl(f),\hat V)$ consisting of $\cl(f) \in \cl(G)$ and $\hat V \in \Irr(\Rep(Z_G(f)))$. As an object in $\Vect_G$, the simple object $(\cl(f),\hat V)$ is the $G$-graded vector space $V = \bigoplus_{x \in \cl(f)} V_x$ with $\supp(V)=\cl(f)$ such that $V_x \cong \hat V$, for every $x \in \cl(f)$. For each $g \in G$, there is a unique $x \in \cl(f)$ such that $g \in p_x Z_G(f)$. Thus, writing $\hat V = \mathbb C\{\hat v_b\}_b$, we have $V = \CG \otimes_{\mathbb C[Z_G(f)]} \hat V = \mathbb C\{p_x \otimes \hat v_b\}_{x \in \cl(f),b}$ such that $|p_x \otimes \hat v_b| = p_xfp_x^{-1} = x$. By definition, $g \cdot p_x Z_G(f) = p_{gxg^{-1}}Z_G(f)$. Moreover, for every $g \in G$ and $p_xZ_G(f) \in G/Z_G(f)$, there is a unique $z_{g,x} \in Z_G(f)$ such that $gp_x = p_{gxg^{-1}} z_{g,x}$. It follows that
\begin{equation}
    g \cdot (p_x \otimes \hat v_b) = gp_x \otimes \hat v_b = p_{gxg^{-1}}z_{g,x} \otimes \hat v_b = p_{gxg^{-1}} \otimes z_{g,x} \cdot \hat v_b ,
\end{equation}
for every $g \in G$, $x \in \cl(f)$ and $b \in \{1,\ldots,\dim_\mathbb C \hat V \}$. Finally, one verifies that $|g \cdot (p_x \otimes \hat v_b)| = gxg^{-1} = g |p_x \otimes \hat v_b|g^{-1}$, as required.

We have recovered the classification of simple modules over $\Tu_G^G$ in terms of pairs $(\cl(f),\hat V)$. The explicit equivalences $\VGG \simeq \mc Z(\Vect_G) \simeq \Mod(\Tu_G^G)$ are be established in sec.~\ref{sec:Maths_ZVGH} in a slightly more general context.

\subsection{Condensable algebra and condensed theory\label{sec:Maths_VGGA}}

As reviewed previously, condensation patterns in the quantum double model are encoded in condensable algebras in the non-degenerate braided fusion category $\VGG$. The classification of condensable algebras in $\VGG$ was carried out in ref.~\cite{DAVYDOV2017149}. Throughout this manuscript, we focus on a special family of condensable algebras. 

Let $H \leq G$ be a (possibly non-normal) subgroup of the finite group $G$.
Consider the algebra $\mathbb C(G/H)$ of functions $\phi : G/H \to \mathbb C$ with pointwise multiplication. It is $G$-equivariant with $(g \cdot \phi)(-) = \phi(g^{-1} - )$. Actually, as an object in $\Rep(G)$, we have $\Ind_H^G(\mathbb C) \cong \mathbb C(G/H)$. This makes $\mathbb C(G/H)$ a \emph{$G$-algebra}, i.e., an algebra object in $\Rep(G)$, which we denote by $A(H)$. Letting $\{r_a\}_{a \in \{1,\ldots,(G:H)\}}$ be a choice of representatives in $G$ for the set of left cosets $G/H$ such that $r_1 = 1_G$, a convenient choice of basis is $\mathbb C(G/H) = \mathbb C\{\delta_{r_aH}\}_{a}$, where $\delta_{r_aH}$ is the minimal idempotent in $\mathbb C(G/H)$ such that $\delta_{r_aH}(g)=1$ if $g \in r_aH$, and 0 otherwise. It follows from $\Rep(G)$ being a fusion subcategory $\VGG$ that the object $A(H)$ lifts to a connected étale (or condensable) algebra in $\VGG$.

Condensing $A(H)$ in $\VGG$ results in a theory whose defects are encoded in the category $(\VGG)_{A(H)}$ of right $A(H)$-modules in $\Vect_G^G$. A very useful observation is that $(\VGG)_{A(H)}$ is equivalent, as a fusion category, to the category $\VGH$ of \emph{$H$-equivariant $G$-graded vector spaces}. 
Before reviewing this equivalence, we first detail the definition of $\VGH$ to establish the necessary notation, even though this is somewhat redundant with the construction of $\VGG$.

\bigskip \noindent
\paragraph{$H$-equivariant $G$-graded vector spaces:} Given $H \leq G$, define a $H$-equivariant $G$-graded vector space as a $G$-graded vector space $W = \bigoplus_{x \in G} W_x$ together with an $H$-action such that $h \cdot W_x \subseteq W_{hxh^{-1}}$, for every $h \in H$ and $x \in G$. Let $\VGH$ be the fusion category whose objects are $H$-equivariant $G$-graded vector spaces and morphisms are $G$-grading preserving $H$-equivariant linear maps. The monoidal structure is given by the monoidal structure of $\Vect_G$ with the group $H$ acting diagonally.

Let $\orb(G)$ denote the set of $H$-conjugacy classes of $G$ and $\orb(f) \in \orb(G)$ the $H$-conjugacy class with representative $f \in G$. Since $H$-conjugacy classes in $\orb(G)$ partition the group, the support $\supp(W)$ of any object $W \in \VGH$ is a union of $H$-orbits in $\orb(G)$. Therefore, 
\begin{equation}
    \VGH = \! \bigoplus_{\orb(f) \in \orb(G)} \! (\VGH)_{\orb(f)} ,
\end{equation}
where $(\VGH)_{\orb(f)}$ is the subcategory of $\VGH$ consisting of objects $W \in \VGH$ such that $\text{supp}(W) = \orb(f)$. Given $\orb(f) \in \orb(G)$, let $\Stab_H(f)$ denote the stabiliser of $f$ in $H$. There is an equivalence
\begin{equation}
    \label{eq:VecGH_equiv}
    (\VGH)_{\orb(f)} \simeq \Rep(\Stab_H(f)) \simeq \Mod(\mathbb C[\Stab_H(f)]),
\end{equation}
which can be demonstrated following the same steps as for $\VGG$. In particular, given a representation $\hat W$ of $\Stab_H(f)$, the induced representation
\begin{equation}
    W := \Ind^H_{\Stab_H(f)}(\hat W) = \CH \otimes_{\mathbb C[\Stab_H(f)]} \hat W 
\end{equation}
of $\hat W$ in $H$ defines an object in $\VGH$, whereby $|h \otimes \hat w| := hfh^{-1}$ and $h_1 \cdot (h_2 \otimes \hat w) := h_1h_2 \otimes \hat w$, for every $h,h_1,h_2 \in H$ and $w \in \hat W$.

\bigskip\noindent
It follows from \eqref{eq:VecGH_equiv} that simple objects in $\VGH$ are labelled by pairs $(\orb(f),\hat W)$ consisting of $\orb(f) \in \orb(G)$ and $\hat W \in \Irr(\Rep(\Stab_H(f)))$. As an object in $\Vect_G$, the simple object $(\orb(f),\hat W)$ is the $G$-graded vector space $W = \bigoplus_{x \in \orb(f)} W_x$ with $\supp(W)=\orb(f)$ such that $W_x \cong \hat W$, for every $x \in \orb(f)$. Let $\{q_x\}_{x \in \orb(f)}$ be a choice of representatives in $H$ of the left cosets in $H/\Stab_H(f)$ such that $q_{f} = 1_H$. For each $h \in H$, there is a unique $x \in \cl(f)$ such that $h \in q_x \Stab_H(f)$. Thus, writing $\hat W = \mathbb C\{\hat w_c\}_c$, we have $W = \CH \otimes_{\mathbb C[\Stab_H(f)]} \hat W = \mathbb C\{q_x \otimes \hat w_c\}_{x \in \orb(f),c}$ such that $|q_x \otimes \hat w_c| = x$. By definition, $h \cdot q_x \Stab_H(f) = q_{hxh^{-1}}\Stab_H(f)$. Moreover, for every $h \in H$ and $q_x\Stab_H(f) \in H/\Stab_H(f)$, there is a unique $s_{h,x} \in \Stab_H(f)$ such that $hq_x = q_{hxh^{-1}} s_{h,x}$. It follows that
\begin{equation}
    h \cdot (q_x \otimes \hat w_c) = hq_x \otimes \hat w_c = q_{hxh^{-1}}s_{h,x} \otimes \hat w_c = q_{hxh^{-1}} \otimes s_{h,x} \cdot \hat w_c ,
\end{equation}
for every $h \in H$, $x \in \orb(f)$ and $c \in \{1,\ldots,\dim_\mathbb C \hat W \}$. Finally, one verifies that $|h \cdot (q_x \otimes \hat w_c)| = hxh^{-1} = h |q_x \otimes \hat w_c|h^{-1}$, as required.

\bigskip\noindent
\paragraph{$A(H)$-modules in $\VGG$:} We are now ready to verify $(\VGG)_{A(H)} \simeq \VGH$ following ref.~\cite{DAVYDOV2017149}.\footnote{Forgetting about the $G$-grading, we already know that $(\Vect^G)_{A(H)} \simeq \Rep(G)_{A(H)} \simeq \Rep(H) \simeq \Vect^H$.} We construct a monoidal functor $(\VGG)_{A(H)} \to \VGH$ as follows. For an object $(M,\rho) \in (\VGG)_{A(H)}$, with $\rho \colon M \otimes A(H) \to M$ the module structure,  consider the subspace $M \cdot \delta_H$. Since the underlying space $M = \bigoplus_{x \in G} M_x$ is an object in $\Vect_G$, $M \cdot \delta_H$ forms a $G$-graded subspace with homogeneous components $(M \cdot \delta_H)_x = M_x \cdot \delta_H$. Furthermore, the $G$-action on $M$ restricts to an $H$-action on $M \cdot \delta_H$, yielding a well-defined $H$-equivariant $G$-grading. Note that $M \cdot \delta_H$ inherits a trivial $(\delta_H \cdot A(H) \cdot \delta_H)$-module structure from the $A(H)$-module structure of $M$. For two objects $(M_1,\rho_{1})$ and $(M_2,\rho_{2})$ in $(\VGG)_{A(H)}$, we have
\begin{equation}
\begin{split}
    (M_1 \otimes_{A(H)} M_2) \cdot \delta_H
    &=
    (M_1 \otimes_{A(H)} M_2) \cdot (\delta_H \cdot \delta_H)
    \\
    &= (M_1 \cdot \delta_H) \otimes_{\delta_H \cdot A(H) \cdot \delta_H} (M_2 \cdot \delta_H)
    = (M_1 \cdot \delta_H) \otimes (M_2 \otimes \delta_H) ,
\end{split}
\end{equation}
where we used the commutativity of $A(H)$ inside the braided category $\VGG$. This makes $(\VGG)_{A(H)} \to \VGH, (M,\rho) \mapsto M \cdot \delta_H$ a tensor functor.

Consider the induction functor $\Ind_H^G : \VGH \to \VGG$. Given $W \in \VGH$, let $M := \Ind_H^G(W)$. By definition,
\begin{equation}
    M = \CG \otimes_{\CH} W ,
\end{equation}
where $\CG = \mathbb C\{g\}_{g \in G}$ is treated here as a $(\CG,\CH)$-bimodule. We define a $G$-grading on $M$ by declaring that, for every homogeneous $w \in W$ and $g \in G$, $g \otimes w \in M$ is homogeneous with degree $|g \otimes w | := g |w|g^{-1} \in G$. This is compatible with the $H$-equivariant $G$-graded structure of $W$ because, for every $h \in H$ and homogeneous $w \in W$, $h|w|h^{-1} = |h \otimes w| = |1_G \otimes h \cdot w| = |h \cdot w| = h|w|h^{-1}$. The $G$-equivariant structure is provided by the left $\CG$-module structure on the first factor, and reads $g_1 \cdot (g_2 \otimes w) = g_1g_2 \otimes w$, for every $g_1,g_2 \in G$ and $w \in W$. Since $|g_1g_2 \otimes w| = (g_1g_2)|w|(g_1g_2)^{-1} = g_1|g_2 \otimes w|g_1^{-1}$, we conclude that $M \in \VGG$. Moreover, $M$ possesses the structure of a right $A(H)$-module in $\VGG$ defined via 
\begin{equation}
    \begin{array}{rccl}
        \rho : & M \otimes A(H) & \longrightarrow & M
        \\
        {} & (g \otimes w) \otimes \phi & \longmapsto & (g \otimes w) \cdot \phi := \phi(gH) \, g \otimes w ,
    \end{array}
\end{equation}
for every $g \in G$, $w \in W$ and $\phi \in A(H)$. For two objects $W_1,W_2 \in \VGH$, let $M_1 := \Ind_H^G(W_1)$ and $M_2 := 
\Ind_H^G(W_2)$.  We have
\begin{equation}
    M_1 \otimes_{A(H)} M_2 = \Ind_H^G(W_1) \otimes_{\Ind_H^G(\mathbb C)} \Ind_H^G(W_2) \cong \Ind_H^G(W_1 \otimes W_2) .
\end{equation}
This makes $\Ind_H^G \colon \VGH \to (\VGG)_{A(H)}$ a tensor functor.

It remains to check that the two functors thus constructed are inverse to each other up to natural isomorphisms. Firstly, for $W \in \VGH$, $w \mapsto (1_G \otimes w)$
realises $W \cong \Ind_H^G(W) \cdot \delta_H$. Secondly, for $M \in (\VGG)_{A(H)}$ and a homogeneous vector $m \in M_x$, $m \mapsto x \otimes x^{-1} \cdot m$ realises $M \cong \Ind_H^G(M \cdot \delta_H)$. This completes the proof of the equivalence $(\VGG)_{A(H)} \simeq \VGH$.

Finally, given a simple object $W \equiv (\orb(f),\hat W) \in \VGH$, let us construct the corresponding object $M \equiv \Ind_H^G(W)$ in $(\VGG)_{A(H)}$. Recall that $W = \mathbb C\{q_x \otimes \hat w_c\}_{x \in \orb(f),c}$ such that $|q_x \otimes \hat w_c| = x$ and $h \cdot (q_x \otimes \hat w_c) = q_{hxh^{-1}} \otimes s_{h,x}\cdot \hat w_c$, for every $h \in H$, where $s_{h,x} = q_{hxh^{-1}}^{-1}h q_x \in \Stab_H(f)$.
Let $\{r_a\}_{a=1,\ldots,(G:H)}$ be a choice of representatives in $G$ of the left cosets in $G/H$ such that $r_1 = 1_G$. For each $g \in G$, there is a unique $a \in \{1,\ldots,(G:H)\}$ such that $g \in r_a H$. Thus, we have $M = \mathbb C\{r_a \otimes (q_x \otimes \hat w_c)\}_{a,x \in \orb(f),c}$ such that
\begin{equation}
    |r_a \otimes (q_x \otimes \hat w_c)| = r_a|q_x \otimes \hat w_c|r_a^{-1}= r_axr_a^{-1} \in G .
\end{equation}
By definition, for every $g \in G$ and $r_aH \in G/H$, there is a unique $g(a) \in \{1,\ldots,(G:H)\}$ such that $g \cdot r_aH = r_{g(a)}H$. Moreover, there is a unique $h_{g,a} \in H$ such that $gr_a = r_{g(a)}h_{g,a}$. It follows that
\begin{equation}
\begin{split}
    g \cdot \big(r_a \otimes (q_x \otimes \hat w_c) \big) 
    &= gr_a \otimes (q_x \otimes \hat w_c)
    \\
    &= r_{g(a)} \otimes h_{g,a} \cdot (q_x \otimes \hat w_c)
    = r_{g(a)} \otimes ( q_{h_{g,a}xh_{g,a}^{-1}} \otimes s_{h_{g,a},x} \cdot \hat w_c) ,
\end{split}
\end{equation}
for every $g \in G$, $a \in \{1,\ldots,(G:H)\}$, $x \in \orb(f)$ and $c \in \{1,\ldots,\dim_\mathbb C \hat W\}$. In particular, one verifies that 
\begin{equation}
\begin{split}
    \big|g \cdot \big( r_a \otimes (q_x \otimes \hat w_c)\big) \big|
    &= r_{g(a)} |q_{h_{g,a}xh_{g,a}^{-1}} \otimes s_{h_{g,a},x} \cdot \hat w_c|r_{g(a)}^{-1}
    \\
    &= r_{g(a)} h_{g,a}xh_{g,a}^{-1} r_{g(a)}^{-1} = g r_a x r_a^{-1} g^{-1} = g|r_a \otimes (q_x \otimes \hat w_c)|g^{-1} ,
\end{split}
\end{equation}
as required.

\bigskip \noindent
We have established that the fusion category $(\VGG)_{A(H)}$ of $A(H)$-modules in $\VGG$ is equivalent to $\VGH$. Now, let us show that the category $(\VGG)_{A(H)}^{\rm loc}$ of \emph{local} $A(H)$-modules is equivalent to $\VHH$. Let $(M,\rho)$ be an object in $(\VGG)_{A(H)}^{\rm loc}$. By definition, the module action is required to satisfy $\rho = \rho \circ R_{A(H),M}\circ R_{M,A(H)}$. Recall that $A(H)$ is concentrated purely in degree $1_G \in G$ so that, for $m \in M_x$, we have $(R_{A(H),M} \circ R_{M,A(H)})(m \otimes \delta_H) = m \otimes x \cdot \delta_H$. Therefore, $m \cdot \delta_H = m \cdot \delta_{xH}$. But $\delta_H$ is an idempotent so that $m \cdot \delta_H = m \cdot (\delta_H)^2 = m \cdot (\delta_{xH} \cdot \delta_H)$. If $x \notin H$, then $m \cdot \delta_H = 0$. 
Therefore, $\supp(M \cdot \delta_H) \subseteq H$. The full subcategory of $\VGH$ consisting of $G$-graded vector spaces with support in $H$ is equivalent to $\VHH$. Conversely, following the same steps as before, one can show that $\Ind_H^G \colon \VHH \to (\VGG)_{A(H)}^{\rm loc}$, and ultimately that $(\VGG)_{A(H)}^{\rm loc} \simeq \VHH$. This is an equivalence of (non-degenerate) braided fusion categories. 

Anticipating the results below, which imply the equivalence $\Vect_H^H \simeq \Mod(\Tu_H^H)$, one recovers the result of sec.~\ref{sec:condensation} that deconfined excitations in the condensed theory are encoded in $\Mod(\Tu_H^H)$. In particular, the locality condition of the $A(H)$-modules formalises the notion that deconfined excitations arise from excitations of the parent theory that braid trivially with the quasi-particles forming the condensable algebra.

\subsection{Relative centres\label{sec:Maths_ZVGH}}

As mentioned earlier, the category $\VGG$ of $G$-equivariant $G$-graded vector spaces---which we are using throughout this section to model the topological order of the quantum double---is equivalent to the category of modules over the tube algebra $\Tu_G^G$, which naturally arises in the lattice picture. Similarly, the category $\VGH$ of $H$-equivariant $G$-graded vector spaces is found to be equivalent to the category of modules over the tube algebra $\Tu_G^H$. 
A particularly enlightening way of establishing this equivalence is to first show that $\VGH$ is equivalent to the \emph{relative Drinfel'd centre} of $\Vect_G$ over $\Vect_H$. This serves two purposes. On the one hand, the relative Drinfel'd centre is the natural category-theoretic construction for describing localised excitations beyond the topological lattice gauge theory setting (see sec.~\ref{sec:outlook}). On the other hand, it readily justifies why modules over the tube algebra encodes both deconfined excitations and confined excitations living at the endpoints of domain walls (see comments below).

\bigskip\noindent
\paragraph{Relative centre over $\Vect_H$:}
Given a subgroup $H \leq G$, $\Vect_H$ forms a fusion subcategory of $\Vect_G$, which makes $\Vect_G$ a $\Vect_H$-bimodule category. Therefore, one can consider the relative Drinfel'd centre $\ZVHVG$ \cite{Majid1991}.
The relative Drinfel'd centre $\ZVHVG$ of $\Vect_G$ over $\Vect_H$ is equivalent, as a fusion category, to the category $\VGH$ of $H$-equivariant $G$-graded vector spaces. The proof goes as follows. Recall that objects in $\VGH$ consist of $G$-graded vector spaces $W = \bigoplus_{x \in G}W_x$ together with an $H$-action such that $h \cdot W_x \subseteq W_{hxh^{-1}}$, for every $h \in G$ and $x \in G$. Let $(W,R_{-,W})$ be an object in $\ZVHVG$. Unpacking the definition, it consists of a $G$-graded vector space $W= \bigoplus_{x \in G}W_x$ and a collection of $G$-grading preserving isomorphisms $R_{V,W} \colon V \act W\xrightarrow{\sim} W\cat V$ that are natural in $V$, for every $V \in \Vect_H$. In particular, for every $h \in H$, we have $R_{\mathbb C_h,W} \colon \mathbb C_h \act W\xrightarrow{\sim} W\cat \mathbb C_h$. 
By definition of the monoidal structure of $\Vect_G$, we have $\mathbb C_h \act W\cong \bigoplus_{x \in G} W_{h^{-1}x}$ and $W\cat \mathbb C_h \cong \bigoplus_{x \in G}W_{xh^{-1}}$, in such a way that $R_{\mathbb C_h,W}$ boils down to a collection of isomorphisms $W_x \xrightarrow{\sim}W_{hxh^{-1}}$, for every $h \in H$ and $x \in G$. Furthermore, the mixed half-braiding natural isomorphism $R_{-,W}$ is subject to a hexagon axiom, which imposes
\begin{equation}
    (R_{\mathbb C_{h_1},W} \otimes {\rm id}_{\mathbb C_{h_2}}) \circ ({\rm id}_{\mathbb C_{h_1}} \otimes R_{\mathbb C_{h_2},W})= R_{\mathbb C_{h_1h_2},W} ,
\end{equation}
for every $h_1,h_2 \in H$. Consequently, we find that the isomorphisms $W_x \xrightarrow{\sim}W_{hxh^{-1}}$ equip $W$ with a $H$-equivariant structure that is compatible with the $G$-grading. Conversely, given an object $W \in \VGH$, the corresponding object in $\ZVHVG$ has the same underlying $G$-graded vector space and the mixed half-braiding natural isomorphism reads
\begin{equation}
    R_{V,W}(v \act w) = h \cdot w \cat v ,
\end{equation}
for $V = \bigoplus_{h \in H}V_h \in \Vect_H$, $v \in V_h$ and $w \in W$. This is the essence of the equivalence $\ZVHVG \simeq \VGH$.

Given a simple object $W \equiv (\orb(f), \hat{W})$, let us construct the corresponding object in $\ZVHVG$. As an object in $\Vect_G$, $W$ decomposes into homogeneous components as $W = \bigoplus_{x \in \orb(f)} W_x$, where $W_x \cong \hat{W}$ for every $x \in \orb(f)$. As an object in $\Rep(H)$, $W$ is given by the induced representation $W \cong \Ind_{\Stab_H(f)}^H(\hat{W}) = \mathbb C\{q_x \otimes \hat w_c\}_{x \in \orb(f),c}$. Recall that the $H$-action on this basis is given by $h \cdot (q_x \otimes \hat{w}_c) = q_{hxh^{-1}} \otimes s_{h,x} \cdot \hat{w}_c$,
where $s_{h,x} \in \Stab_H(f)$ is defined by the relation $h q_x = q_{hxh^{-1}} s_{h,x}$. Therefore, the mixed half-braiding natural isomorphism of the corresponding object in $\ZVHVG$ is provided by
\begin{equation}
    \label{eq:ZVHVG_halfBraid}
    R_{\mathbb C_h,W}\big(h \act (q_x \otimes \hat w_c)\big) =  (q_{hxh^{-1}} \otimes s_{h,x} \cdot \hat{w}_c) \cat h ,
\end{equation}
for every $h \in H$.

\bigskip \noindent
Specialising to $H=G$, the relative Drinfel'd centre reduces to the ordinary Drinfel'd centre so that $\VGG \simeq \mc Z(\Vect_G)$. Another formulation of the Drinfel'd centre $\mc Z(\Vect_G)$ is as the category $\Fun_{\Vect_G|\Vect_G}(\Vect_G,\Vect_G)$ of \emph{$\Vect_G$-bimodule endofunctors} of $\Vect_G$. Physically, this formulation should be interpreted as follows \cite{2002math......2130O}. As a $\Vect_G$-bimodule category, $\Vect_G$ encodes the trivial domain wall, i.e., the absence thereof. Objects in $\Fun_{\Vect_G|\Vect_G}(\Vect_G,\Vect_G)$ then encode localised excitations at the interface of two trivial domain walls, i.e., the deconfined excitations of the theory \cite{kongBdries}. 
Similarly, the relative Drinfel'd centre $\ZVHVG$ can be shown to be equivalent to the category $\Fun_{\Vect_H|\Vect_H}(\Vect_H,\Vect_G)$ of $\Vect_H$-bimodule functors from $\Vect_H$ to $\Vect_G$ \cite{Majid1991}. Objects in $\Fun_{\Vect_H|\Vect_H}(\Vect_H,\Vect_G)$ encodes localised excitations living at the interface of the trivial domain associated with $\Vect_H$ and a composite domain wall associated with $\Vect_G$. As a $\Vect_H$-bimodule, $\Vect_G$ decomposes over $H\setminus G/H$. Each element in $H \setminus G /H$ labels a type of domain wall, the trivial type being associated with $H$ itself. Bringing everything together, the relative Drinfel'd centre $\ZVHVG$ encodes localised excitations living at the endpoints of all the domain walls labelled by double cosets in $H \setminus G /H$, which includes both deconfined and confined excitations. This agree with the lattice picture of sec.~\ref{sec:condensation}.   

Interestingly, $\ZVHVG$ is not the only way to realise $\VGH$ as a relative Drinfel'd centre. Recall that the Morita equivalence between $\Vect_G$ and $\Rep(G)$---itself stemming from realising $\Rep(G)$ as the category $(\Vect_G)^*_{\Vect} := \Fun_{\Vect_G}(\Vect,\Vect)$ of $\Vect_G$-module  endofunctors of $\Vect$---implies $\mc Z(\Vect_G) \simeq \mc Z(\Rep(G)) \simeq \VGG$. Similarly, we show below that $\VGH \simeq \mc Z_{\Rep(G)}(\Rep(H))$. 

\bigskip \noindent
\paragraph{Relative centre over $\Rep(G)$:} 
Given a subgroup $H \leq G$, the fusion category $\Rep(H)$ is a $\Rep(G)$-bimodule category via the (monoidal) restriction functor $\Res_H^G$. 
Therefore, one can consider the relative Drinfel'd centre $\ZRGRH \simeq \Fun_{\Rep(G)|\Rep(G)}(\Rep(G),\Rep(H))$.
As advertised above, the relative Drinfel'd centre $\ZRGRH$ of $\Rep(G)$ over $\Rep(H)$ is also equivalent, as a fusion category, to the category $\VGH$ of $H$-equivariant $G$-graded vector spaces.\footnote{By analogy with $\mc Z_{\Vect_H}(\Vect_G)$, we may have expected the roles of $\Rep(H)$ and $\Rep(G)$ to be exchanged here. However, recall that $\Rep(H)$ is not a fusion subcategory of $\Rep(G)$ in general, and that the induction functor $\Ind_H^G$ is not a braided monoidal functor. Therefore, $\Rep(G)$ typically cannot be endowed with the structure of a $\Rep(H)$-bimodule category.}
The proof goes as follows.
Let $(W,R_{-,W})$ be an object in $\ZRGRH$.
Unpacking the definition, it consists of an $H$-representation $W$ and a collection of $H$-equivariant linear isomorphisms $R_{V,W} \colon V \act W \xrightarrow{\sim} W \cat V$ that are natural in $V$, for every $V \in \Rep(G)$. 
In particular, we have an isomorphism $R_{\CG,W} \colon \CG \act W \xrightarrow{\sim} W \cat \CG$, where $\CG = \mathbb C\{g\}_{g \in G}$ is treated here as the (left) regular $G$-representation. Recall that $\Res_H^G(\CG) \cong \bigoplus_{a = 1}^{(G:H)} \mathbb C[H]$. 
Pick any $w \in W$, write the expansion of $R_{\CG,W}({1_G} \act w) \in W \cat \CG$ in the basis of $\CG$ as $R_{\CG,W}({1_G} \act w) \equiv \sum_{x \in G} w_{x^{-1}} \cat x$, which defines $w_{x^{-1}} \in W$ uniquely. 
Now, let $\epsilon \colon \CG \to \mathbb C$ be the $G$-equivariant map $g \mapsto 1$, for every $g \in G$. It follows from the naturality of the mixed half-braiding that 
\begin{equation}
    ({\rm id}_W \cat \epsilon) \circ R_{\CG,W} = R_{\mathbb C,W} \circ (\epsilon \act {\rm id}_W) .
\end{equation}
Applying both sides of this equation to $1_G \act w$ yields
\begin{equation}
    \sum_{x \in G} w_{x^{-1}} \cat \epsilon(x) = \sum_{x \in G} w_x \cat 1_G = R_{\mathbb C,W}(1_G \act w) = w \cat 1_G ,
\end{equation}
Thus, the vector $w \in W$ decomposes as $w = \sum_{x \in G} w_x$.
Defining $W_x := \{w \in W \, | \, R_{\CG,W}({1_G} \act w) = w \cat x^{-1} \}$, it follows that $W = \bigoplus_{x \in G}W_x$ is $G$-graded. It remains to show that $W$ is endowed with a $H$-equivariant structure that is compatible with this $G$-grading. 
Let $w \in W_x$. By construction, $R_{\CG,W}({1_G}\act w) = w \cat {x^{-1}}$.
By $H$-equivariance of the linear isomorphism $R_{\CG,W}$,  
\begin{equation}
    \begin{split}
        R_{\CG,W}(h\act h\cdot w) = R_{\CG,W}(h\cdot({1_G}\act w)) = h\cdot R_{\CG,W}({1_G}\act w) = h \cdot w \cat {h x^{-1}} .
    \end{split}
\end{equation}
Next, define $R_h \colon \CG \to \CG, x \mapsto {xh^{-1}}$.
By naturality over $\Rep(H)$ of the mixed half-braiding,  
\begin{equation}
    ({\rm id}_W \cat R_h) \circ R_{\CG,W} = R_{\CG,W} \circ (R_h \act {\rm id}_W) . 
\end{equation}
Applying both sides of this equation to ${1_G} \act w$ results in $w \cat {x^{-1}h^{-1}} = R_{\CG,W}({h^{-1}} \act w)$ so that
\begin{equation}
    R_{\CG,Z}({1_G} \act h \cdot w)  = h \cdot R_{\CG,W}({h^{-1}} \act w) = h \cdot w \cat {hx^{-1}h^{-1}} .
\end{equation}
Therefore, for every $w \in W_x$, $h \cdot w \in W_{hxh^{-1}}$, as required. Conversely, given $W \in \VGH$, the corresponding object in $\ZRGRH$ has the same underlying $H$-representation and the mixed half-braiding natural isomorphism reads
\begin{equation}
    R_{V,W}(v \act w) = w \cat x^{-1} \cdot v ,
\end{equation}
for every $V \in \Rep(G)$, $v \in V$ and $w \in W_x$. 

\bigskip\noindent
The quantum double model with topological order $\VGG$ considered in sec.~\ref{sec:lattice} coincides with the string-net model whose input datum is the (spherical) fusion category $\Vect_G$. In this realisation, $\VGG$ is implicitly obtained as the Drinfel'd centre $\mc Z(\Vect_G)$. Since $\mc Z(\Rep(G)) \simeq \VGG$ as well, the same topological order is also produced by the string-net construction applied to $\Rep(G)$. This is the electromagnetic dual of the quantum double model \cite{Buerschaper:2010yf}.
Starting from the quantum double model, we obtained $\VGH \simeq (\VGG)_{A(H)}$ from $\VGG \simeq \mc Z(\Vect_G)$ by condensing the algebra of charges $A(H)$. The category-theoretic gymnastics performed above suggest an alternative construction. We could start instead from the topological order $\VHH \simeq \mc Z(\Rep(H))$, realised as a string-net model with input datum $\Rep(H)$, and then obtain $\VGH$ via some gauging operation in the spirit of ref.~\cite{Kawagoe:2024tgv}. That said, it is also possible to start from $\VGG \simeq \mc Z(\Rep(G))$, realised as a string net model with input datum $\Rep(G)$, and then condense the algebra $A(H)$ as before. This latter scenario would invoke yet another relative Drinfel'd centre (see sec.~\ref{sec:outlook}).

\subsection{Tube algebra\label{sec:Maths_tube}}

Using the equivalence $\VGH \simeq \ZVHVG$, we now connect with the tube algebra approach of sec.~\ref{sec:condensation}. From a conceptual standpoint, this relies on reinterpreting the $(G,H)$-coloured graphs introduced in sec.~\ref{sec:condensation} as morphisms in the category $\Vect_G$.

Let $W$ be the simple object 
$(\orb(f), \hat W)$, where $\orb(f) \in \orb(G)$ and 
$\hat W = \mathbb{C}\{\hat w_c\}_c$ is a simple object in $\Rep(\Stab_H(f))$. This object naturally carries the structure of a module over $\Tu_G^H$. The underlying vector space is $W$ itself, which is convenient to think of as $\bigoplus_{x \in G}\Hom_{\Vect_G}(\mathbb C_x,W)$, treating here $W$ as an object in $\Vect_G$. Recall that $W_x \cong \hat W$, for every $x \in \orb(f)$.  Graphically,
\begin{equation}
    \label{eq:modTube}
    W = \mathbb C\{q_x \otimes \hat w_c\}_{x \in \orb(f), \,  c= 1, \ldots, \dim_{\mathbb{C}} \hat W}
    \equiv 
    \mathbb{C}\left\{ 
    \modT{x}{W}{c}
    \right\}_{x \in \orb(f),\, c} .
\end{equation}
In other words, we reinterpret here the basis vector $q_x \otimes \hat w_c$ as a  morphism in $\Hom_{\Vect_G}(W_x,\mathbb C_x) \cong \hat W$. In particular, we have the following orthogonality condition:
\begin{equation}
    \innerModT{x}{c_1}{y}{c_2}{W} = \delta_{x,y} \, \delta_{c_1,c_2} \, {\rm id}_{\mathbb C_x},
\end{equation}
for every $x,y \in G$ and $c_1,c_2 \in \{1,\ldots,\dim_\mathbb C \hat W\}$.
As we established above, $W$ is also an object in $\ZVHVG$ and we depict the corresponding mixed half-braiding as
\begin{equation}
    \halfBraid{W}{h} \equiv R_{\mathbb C_h,W} \colon \mathbb C_h \act W \xrightarrow{\sim} W \cat \mathbb C_h.
\end{equation}
This graphical representation makes the naturality in $\mathbb C_h$ manifest.  In sec.~\ref{sec:condensation}, we constructed the vector space underlying $\Tu_G^H$ as the span of $(G,H)$-coloured graphs on the cylinder. In the current context, we reinterpret it as following hom-space in $\Vect_G$:
\begin{equation}
    \Tu_G^H = \mathbb C\{\mc T_y^h\}_{\substack{y \in G\\ h \in H}} = \mathbb C \left\{ \, \tubeT{h}{y}{{}^hy}{}  \right\}_{\substack{y \in G  \\ h \in H}} \! \cong \bigoplus_{\substack{y \in G \\ h \in H}} \Hom_{\Vect_G}(\mathbb C_{{}^hy} \otimes \mathbb C_h,\mathbb C_h \otimes \mathbb C_y).
\end{equation}
The $\Tu_G^H$-module structure of $W$ is then given by
\begin{equation}
    \tubeT{h}{y}{{}^hy}{} \cdot \modT{x}{W}{c}
    := \delta_{x,y} \, 
    \tubeActionT{h}{x}{{}^hx}{}{W}{c} \, .
\end{equation}
Here, the diagram on the right-hand side represents a morphism in $\Hom_{\Vect_G}(\mathbb C_{^hx}, W)$. Using the definition of the mixed half-braiding from eq.~\eqref{eq:ZVHVG_halfBraid}, this evaluates to
\begin{equation}
    \mc T^h_y \cdot (q_x \otimes \hat w_c) := \delta_{x,y} \, (q_{hxh^{-1}} \otimes s_{h,x} \cdot \hat w_c). 
\end{equation}
This captures the essence of the equivalences $\VGH \simeq \ZVHVG \simeq \Mod(\Tu_G^H)$.

Next, one introduces the so-called \emph{Kirby colour}, which can be thought of as a special kind of $H$-coloured graph:
\begin{equation}
    \lineT{kir}{} \equiv \frac{1}{|H|} \sum_{h \in H} \lineT{obj}{h} .
\end{equation}
We use it to formulate another orthogonality condition
\begin{equation}
    \orthoLemma{W_1}{c_1}{W_2}{c_2}{x} = \delta_{W_1,W_2} \, \delta_{c_1,c_2} \, \frac{{\rm id}_{W_1}}{\dim_\mathbb C {W_1}},
\end{equation}
which follows from \emph{Schur orthogonality} of the corresponding $G$-representations. 

Using this graphical calculus, the \emph{matrix units} of the tube algebra $\Tu_G^H$ can be concisely expressed as 
\begin{equation}
    \big(\mc E_W \big)^{x_1,c_1}_{x_2,c_2} := (\dim_\mathbb C W) \;  \matrixUnits{W}{x_2}{c_2}{x_1}{c_1}, 
\end{equation}
for every simple object $W$ in $\Mod(\Tu_G^H)$. By definition of the multiplication rule, these matrix units satisfy
\begin{equation}
    \big (\mc E_{W_1} \big)^{x_1,c_1}_{x_2,c_2} \cdot \big(\mc E_{W_2} \big)^{x_3,c_3}_{x_4,c_4}
    = \delta_{W_1,W_2} \, \delta_{x_2,x_3} \, \delta_{c_2,c_3} \, \big(\mc E_{W_1} \big)^{x_1,c_1}_{x_4,c_4},
\end{equation}
as well as the completeness relation
\begin{equation}
    \sum_{\orb(f),\hat W} \sum_{x \in \orb(f),c} \big( \mc E_{(
    \orb(f),\hat W)}\big)^{x,c}_{x,c} = {\rm id}_{\Tu_G^H}.
\end{equation}
In particular, the minimal central idempotents of $\Tu_G^H$ are given by 
\begin{equation}
    \mc E_{(\orb(f),\hat W)} := \sum_{x \in \orb(f),c} \big(\mc E_{(\orb(f),\hat W)} \big)^{x,c}_{x,c}. 
\end{equation}
Applying the definitions above, we recover
\begin{align}
    \mc E_{(\orb(f),\hat W)} 
    &= \frac{\dim_\mathbb C \hat W \cdot |\orb(f)|}{|H|} \sum_{x \in \orb(f)} \sum_{h \in \Stab_H(f)} \chi_{\hat W}(s_{h^{-1},hxh^{-1}}) \; \tub{x}{h}
    \\
    &= \frac{\dim_\mathbb C \hat W}{|\Stab_H(f)|} \sum_{x \in \orb(f)} \sum_{h \in \Stab_H(f)} \overline{\chi_{\hat W}(s_{h,x})} \; \tub{x}{h},
\end{align}
where we used the fact that $|H| = |\orb(f)| \cdot |\Stab_H(f)|$, for every $\orb(f) \in \orb(G)$, as well as $s_{h^{-1},hxh^{-1}} = s_{h,x}^{-1}$.

\subsection{Bimodule category decomposition}

So far we have been providing a category theoretic framework for the results of the previous section. We are now ready to analyse the category $\VGH$ in detail. In sec.~\ref{sec:condensation}, we alluded to the fact that $\VGH$ naturally decomposes over $H \setminus G /H$. Here, we refine this statement by showing that it provides the decomposition of $\VGH $ as a $\VHH$-bimodule category, the same way each component of a $G$-crossed braided fusion category is a bimodule over the trivial component. 

\bigskip \noindent
Recall that an object $W$ in $\VGH$ consists of a $G$-graded vector space $W = \bigoplus_{x \in G} W_x$ together with an $H$-action such that $h \cdot W_x \subseteq W_{hxh^{-1}}$, for every $h \in H$ and $x \in G$.
Moreover, the support $\supp(W)$ of any object $W \in \VGH$ is a union of $H$-conjugacy classes in $\orb(G)$. These $H$-conjugacy classes can be organised into double cosets in $\GH \equiv H \setminus G /H$, which partition the group. Because every double coset $[k] \equiv HkH \in 
\GH$ is closed under the action of $H$ by conjugation, we can decompose the category $\VGH$ as
\begin{equation}
    \label{eq:decomposition_VecGH}
    \VGH = \bigoplus_{[k] \in \GH} (\VGH)_{[k]} ,
\end{equation}
where $(\VGH)_{[k]} \equiv \Vect_{[k]}^H$ is the subcategory consisting of $H$-equivariant $G$-graded vector spaces with support contained within the double coset $[k]$. In particular, $[1] = H$ so that $\Vect_{[1]}^H = \VHH$, which we recall is equivalent to the category $(\VGG)_{A(H)}^{\rm loc}$ of local modules over $A(H)$ in $\VGG$.

Since $\VHH$ is a fusion subcategory of $\VGH$, $\VGH$ admits a $\VHH$-bimodule structure. Therefore, there is a decomposition of $\VGH$ into a direct sum of indecomposable $\VHH$-bimodule categories. We claim that eq.~\eqref{eq:decomposition_VecGH} provides this decomposition. For every $[k] \in \GH$, let us verify that $\Vect_{[k]}^H$ is equipped with the structure of an \emph{indecomposable} $\VHH$-bimodule category. 
By construction, 
\begin{equation}
    \mc Z(\VGH) \simeq \VHH \boxtimes \overline{\VGG} .
\end{equation}
Thus, there is a (fully faithful) braided tensor functor $\VHH \to \mc Z(\VGH)$ such that the inclusion $\VHH \hookrightarrow \VGH$ is obtained by postcomposing this braided tensor functor with the forgetful functor $\mc Z(\VGH) \to \VGH$. We say that $\VHH$ is \emph{central} in $\VGH$. Since $\VHH$ is central, any \emph{left} $\VHH$-module subcategory of $\VGH$ is also a \emph{right} $\VHH$-module category. Hence, it is enough to verify that $\Vect_{[k]}^H$ is an indecomposable left $\VHH$-module category.   

Indecomposable $\VHH$-module categories were classified by Ostrik in ref.~\cite{2002math......2130O}. These are labelled by conjugacy classes of pairs $(L,\beta)$ consisting of a subgroup $L \leq H \times H$ and a 2-cocycle $\beta$ in $H^2(L,\mathbb C^\times)$. In particular, the module category associated with the pair $(L,1)$ is equivalent to the category $\Vect_{(H \times H)/L}^{\Delta(H)}$ of $\Delta(H)$-equivariant $(H \times H)/L$-graded vector spaces, where $\Delta(H) < H \times H$ is the diagonal subgroup. For every $[k] \in \GH$, choose $L$ to be the subgroup $L_k := \{(h,k^{-1}hk) \, | \, h \in H \cap kHk^{-1}\} < H \times H$. We claim that $\Vect^H_{[k]}$ is equivalent to $\Vect_{(H \times H)/L_k}^{\Delta(H)}$, as a $\VHH$-module category. Here, the diagonal subgroup $\Delta(H)$ acts on $(H \times H)/L_k$ via $(h,h) \cdot (h_1,h_2)L_k = (hh_1,hh_2)L_k$, for every $h,h_1,h_2 \in H$. 

We are left to explain this equivalence. The group $H \times H$ acts on $G$ via $(h_1,h_2) \cdot k = h_1kh_2^{-1}$, for every $h_1,h_2 \in H$ and $k \in G$. For every $k \in G$, the orbit of $k$ under this action is the double coset $[k] \in \GH$. The stabiliser in $H \times H$ of $k$ is given by $\{(h_1,h_2) \in H \times H \, | \, h_1 k h_2^{-1}=k\}$, which is clearly isomorphic to the subgroup $L_k \leq H \times H$ introduced above. Thus, the \emph{orbit-stabiliser} theorem stipulates that $(H \times H) / L_k \cong [k]$. This isomorphism is realised by the assignment $(H \times H)/L_k \to [k]$, $(h_1,h_2)L_k \mapsto h_1 k h_2^{-1}$, which is well defined by virtue of $(h_1h,h_2k^{-1}h k )L_k \mapsto h_1 k h_2^{-1}$. Under this isomorphism, the $\Delta(H)$-action on $(H \times H)/L_k$ maps to $H$-conjugation action on $[k]$. Consequently, $\Vect_{(H \times H)/L_k}^{\Delta(H)} \simeq \Vect_{[k]}^H$ and eq.~\eqref{eq:decomposition_VecGH} does provide the decomposition of $\VGH$ into indecomposable $\VHH$-bimodule categories.

Importantly, note that $\Vect_{[k]}^H$ is typically a \emph{non-invertible} $\VHH$-bimodule category with respect to the Deligne product over the braided fusion category $\VHH$. Henceforth, we refer to $\Vect_{[k]}^H$ as the \emph{$[k]$-twisted sector}, and the corresponding $A(H)$-modules under $(\VGG)_{A(H)} \simeq \VGH$ as $[k]$-twisted $A(H)$-modules.

\subsection{Hypergroup-graded extension \label{sec:Maths_HyperGr}}

We have established that the indecomposable $\VHH$-bimodule categories in $\VGH$ are in one-to-one correspondence with the elements of $H \backslash G / H$. In the special case where $H$ is a normal subgroup $N$, the set of double cosets simplifies to $N \setminus G / N = G/N$, and the category $\Vect_G^N$ is naturally graded by the quotient group $G/N$. Our goal is to formulate an analogous result for the case where $H$ is not normal.

\bigskip \noindent
Let us first introduce some definitions. Given a finite set $\{k_i\}_{i \in I}$, a \emph{convex combination} of elements of $K$ is a (formal) linear combination $\sum_{i \in I}\lambda_ik_i$ such that $\lambda_i \geq 0$ and $\sum_{i \in I}\lambda_i=1$. We denote by $\text{Conv}(K)$ the \emph{convex hull} of $K$, defined as the set of all such convex combinations. For a convex combination $\sum_{i \in I}\lambda_i k_i \in \text{Conv}(K)$, we define its \emph{support} as $\supp \big(\sum_{i \in I} \lambda_i k_i \big) := \{k_{i} \in K \, | \, \lambda_i > 0\}$.

Following ref.~\cite{Bischoff:2016jmy}, a finite \emph{hypergroup} consists of a finite set $K = \{k_i\}_{i \in I}$ with a distinguished element $k_0 \in K$  together with a multiplication map $\star \colon K \times K \to \text{Conv}(K)$ and an involution $* \colon K \to K$ so that upon bilinear extension of $\star$ to $\mathbb C[K]$ and extension of $*$ to an anti-involution, the quadruple $(\mathbb C[K],\star,k_0,*)$ is an associative unital $*$-algebra such that $k_0 \in \supp(k_{1} \star k_{2})$ if and only if $k_{1} = k_{2}^*$, for every $k_{1},k_{2} \in K$.

Unpacking the definition, we find that the product of two hypergroup elements is of the  form
\begin{equation}
    k_{1} \star k_{2} = \sum_{k_3 \in K} C^{k_1 k_2}_{k_3}  \, k_{3}
\end{equation}
such that 
\begin{equation}
    \label{eq:hyperGr_StrCsts}
    C^{k_1k_2}_{k_3} \geq 0 \q \text{and} \q \sum_{k_3 \in K} C^{k_1k_2}_{k_3} = 1 ,     
\end{equation}
for every $k_{1},k_{2} \in K$. Moreover, for any $k_{1},k_{2} \in K$, $C^{k_1k_2}_{k_0} = C^{k_2k_1}_{k_0} > 0$ if and only if $k_{1} = k_{2}^*$. We refer to the collection $\{C^{k_1k_2}_{k_3}\}_{k_1,k_2,k_3}$ of non-integer numbers are the \emph{structure constants} of the hypergroup. 

\bigskip \noindent
\paragraph{Hypergroup of double cosets:}
Throughout this manuscript, we focus on the following type of hypergroup. Given $H \leq G$, the set $\GH$ of double cosets can be endowed the structure of a hypergroup with multiplication
\begin{equation}
    [k_1] \star [k_2] := \frac{1}{|H|} \sum_{h \in H}H(k_1hk_2)H \equiv \!\! \sum_{[k_3] \in \GH} \!\! C^{[k_1][k_2]}_{[k_3]} \; [k_3] ,
\end{equation}
unit $[1_G] = H$, and involution
\begin{equation}
    [k]^* := [k^{-1}] .
\end{equation}
By definition, the structure constants are given by
\begin{equation}
    \label{eq:hyperGr_StrCstsFormulaA}
    C^{[k_1][k_2]}_{[k_3]} = \frac{1}{|H|} \cdot \big| \big\{h \in H \, \big| \, k_1hk_2 \in [k_3]\big\} \big| ,
\end{equation}
which readily satisfy eq.~\eqref{eq:hyperGr_StrCsts}. Note that the multiplicity of $[k_3]$ inside $[k_1] \star [k_2]$ is provided by $|H| \cdot C_{[k_3]}^{[k_1][k_2]}$. In general, we denote by $[k_3] \prec [k_1] \star [k_2]$ the double cosets in $\supp([k_1] \star [k_2])$. Note that whenever $H = N$ is a normal subgroup, then the hypergroup $G/\!/N$ reduces to the quotient group $G/N$. 

Another useful formula for the structure constants is
\begin{equation}
    \label{eq:hyperGr_StrCstsFormulaB}
    C^{[k_1][k_2]}_{[k_3]} = \frac{1}{|[k_1]| \cdot |[k_2]|} \cdot \big| \big\{ (g_1,g_2) \in [k_1] \times [k_2] \, \big| \, g_1g_2 \in [k_3] \, \big\} \big| .
\end{equation}
This formula makes it explicit that the structure constants provide the \emph{probability} for the product of two elements in the double cosets in $[k_1]$ and $[k_2]$, respectively, to land in the double coset $[k_3]$. The proof of this formula goes as follows. 
On the one hand, suppose there exists $h \in H$ such that $k_1hk_2 \in [k_3]$, then $h_1k_1hk_2h_4 \in [k_3]$, for every $h_1,h_4 \in H$. Therefore, 
\begin{equation}
    C^{[k_1][k_2]}_{[k_3]} = \frac{1}{|H|^4} \cdot \big| \big\{ (h_1,\ldots,h_4) \in H^4 \, | \, h_1k_1h_2h_3k_2h_4 \in [k_3]\big\} \big| .
\end{equation}
On the other hand, for any $[k_1], [k_2] \in \GH$, let $g_1 \in [k_1]$ and $g_2 \in [k_2]$. 
By definition, there exist $h_1,\ldots,h_4 \in H$ such that $g_1 = h_1 k_1h_2$ and $g_2 = h_3k_2h_4$. Recall that, for every $[k] \in \GH$, $L_{k} = \{(h,k^{-1}hk) \, | \, h \in H \cap kHk^{-1}\}$ is the stabiliser of $k$ under the action of $H \times H$. Hence, there are $|L_{k_1}|$-many possible pairs of group elements $(h_1,h_2) \in H \times H$ such that $g_1 = h_1k_1h_2$, and similarly for $g_2$. Therefore,
\begin{equation*}
    \big|\big\{(h_1,\ldots,h_4) \in H^4 \, \big| \, h_1k_1h_2h_3k_2h_4 \in [k_3]\big\}\big|
    = |L_{k_1}| \cdot |L_{k_2}| \cdot \big| \big\{ (g_1,g_2) \in [k_1] \times [k_2] \, \big| \, g_1g_2 \in [k_3] \, \big\} \big| .
\end{equation*}
Invoking the orbit-stabiliser theorem, which stipulates that $|[k]|= |H|^2/|L_{k}|$, for every $[k] \in \GH$, concludes  the argument.

\bigskip \noindent
\paragraph{$\GH$-graded extension:} We are now ready to state that $\VGH$ is $\GH$-graded extension of $\VHH$, in the sense specified below. First of all, by definition of the monoidal structure of $\VGH$ and of the multiplication rule of $\GH$, if $W_1 \in \Vect_{[k_1]}^H$ and $W_2 \in \Vect_{[k_2]}^H$, then
\begin{equation}
    W_1 \otimes W_2 \in \!\!\! \bigoplus_{[k_3] \prec [k_1] \star [k_2]} \!\!\!\!\! \Vect^H_{[k_3]} .
\end{equation}
But there is a more stringent compatibility condition between the monoidal structure and the hypergroup multiplication, namely that for any \emph{simple} objects $W_1 \in \Irr(\Vect_{[k_1]}^H)$ and $W_2 \in \Irr(\Vect_{[k_2]}^H)$, we have \cite{hempel2023hypergroups}\footnote{We are grateful to Brandon Rayhaun for first pointing out this relation to us.}
\begin{equation}
    \label{eq:compMonHyperGr}
    \frac{\FPdim(\Pi_{[k_3]}(W_1 \otimes W_2))}{\FPdim(W_1 \otimes W_2)} = C^{[k_1][k_2]}_{[k_3]} ,
\end{equation}
where $\Pi_{[k_3]} \colon \VGH \to \Vect_{[k_3]}^H$ is the projection functor onto the $[k_3]$-twisted sector. The reasoning goes as follows. Let $W_1 \equiv (\orb(f_1),\hat W_1)$ and $W_2 \equiv (\orb(f_2),\hat W_2)$ such that $\orb(f_1) \subseteq [k_1]$ and $\orb(f_2) \subseteq [k_2]$. By definition, 
\begin{equation}
    \FPdim(W_1 \otimes W_2) = |\orb(f_1)| \cdot |\orb(f_2)| \cdot \dim_\mathbb C \hat W_1 \cdot \dim_\mathbb C \hat W_2 .    
\end{equation}
Moreover, since 
\begin{equation}
    (W_1 \otimes W_2)_x = \bigoplus_{\substack{x_1 \in \orb(f_1) \\ x_2 \in \orb(f_2) \\ x_1x_2=x}} (W_1)_{x_1} \otimes (W_2)_{x_2} ,
\end{equation}
for every $x \in G$, we have
\begin{equation*}
    \FPdim(\Pi_{[k_3]}(W_1 \otimes W_2)) = \big|\big\{ (x_1,x_2) \in \orb(f_1) \times \orb(f_2) \, \big| \, x_1x_2 \in [k_3]\big\}\big| \cdot \dim_\mathbb C \hat W_1 \cdot \dim_\mathbb C \hat W_2 .
\end{equation*}
Hence, 
\begin{equation}
    \label{eq:hyperGr_StrCstsFormulaC}
    \frac{\FPdim(\Pi_{[k_3]}(W_1 \otimes W_2))}{\FPdim(W_1 \otimes W_2)} = \frac{\big|\big\{ (x_1,x_2) \in \orb(f_1) \times \orb(f_2) \, \big| \, x_1x_2 \in [k_3]\big\}\big| }{|\orb(f_1)| \cdot |\orb(f_2)|} .
\end{equation}
Finally, the same way we equated eq.~\eqref{eq:hyperGr_StrCstsFormulaA} and eq.~\eqref{eq:hyperGr_StrCstsFormulaB}, we can equate eq.~\eqref{eq:hyperGr_StrCstsFormulaB} and eq.~\eqref{eq:hyperGr_StrCstsFormulaC}, which concludes the argument. It follows from the derivation that eq.~\eqref{eq:compMonHyperGr} does not depend on the choice of simple objects in the twisted sectors. 

\subsection{Non-invertible symmetry and monad\label{sec:Maths_monad}}

In sec.~\ref{sec:symmetry}, we identified the non-invertible symmetry of the condensed theory. In particular, for every element in $\GH$, we constructed a linear operator and computed its action on the localised excitations encoded in $\VGH \simeq \Mod(\Tu_G^H)$.  But this is not yet a \emph{categorical action}. We address this gap here from the perspective of the three equivalent formulations: $\VGH$, $\Mod(\Tu_G^H)$ and $(\VGG)_{A(H)}$.  

\paragraph{Monad:}
In the spirit of ref.~\cite{Cui:2018hxz}, let us formalise the non-invertible symmetry of $\VGH$ in terms of a \emph{monad} on $\VGH$. We employ the definition of ref.~\cite{BRUGUIERES2011745}. Given a monoidal category $\mc C \equiv (\mc C,\otimes ,1_\mc C,\alpha)$\footnote{The monoidal unit is always assumed to be strict.}, we define a \emph{monad} on $\mc C$ as an algebra object in the (strict) monoidal category $\End(\mc C)$ of endofunctors of $\mc C$. Concretely, a monad on $\mc C$ is a triple $(T,\mu,\eta)$ that consists of an endofunctor $T \colon \mc C \to \mc C$, and natural transformations $\mu \colon T \circ T \to T$ and $\eta \colon {\rm id}_\mc C \to T$ such that
\begin{equation}
    \mu_X \circ T(\mu_X) = \mu_X \circ \mu_{T(X)} 
    \q \text{and} \q
    \mu_X \circ \eta_{T(X)} = {\rm id}_{T(X)} = \mu_X \circ T(\eta_X) ,
\end{equation}
for every  $X \in \mc C$. 

\bigskip \noindent
Let us now apply this definition to our situation. Consider the endofunctor of $\VGH$ defined via the composition
\begin{equation}
    T \colon \VGH \xrightarrow{\Ind_H^G} \VGG \xrightarrow{\Res^G_H} \VGH .
\end{equation}
We have already studied the induction functor $\Ind_H^G \colon \VGH \to \VGG$. For every $W \in \VGH$, recall that $\Ind_H^G(W) = \CG \otimes_{\CH} W$, such that, for every $g \in G$ and homogeneous $w \in W$, $g \otimes w \in \Ind_H^G(W)$ is homogeneous with degree $|g \otimes w| = g|w|g^{-1} \in G$. Moreover, for every $g_1,g_2 \in G$ and $w \in W$, $g_1 \cdot (g_2 \otimes w) = g_1g_2 \otimes w$. The restriction functor $\Res^G_H \colon \VGG \to \VGH$ is defined as follows. For every $V \in \VGG$, $\Res^G_H(V) = \CG \otimes_{\CG} V$, where $\CG$ is treated here as $(\CH,\CG)$-bimodule. We define a $G$-grading on $\Res^G_H(V)$ by declaring that, for every homogeneous $v \in V$ and $g \in G$, $g \otimes v \in \Res_H^G(V)$ is homogeneous with degree $|g \otimes v| := g|v|g^{-1}$. This is compatible with the $G$-equivariant $G$-grading structure of $V$ because, for every $g \in G$ and homogeneous $v \in V$, $|g \otimes v| = |1_G \otimes g \cdot v| = |g \cdot v| = g|v|g^{-1}$. The $H$-equivariant structure is provided by the left $\CH$-module structure on the first factor, and reads $h \cdot (g \otimes v) = hg \otimes v = 1_G \otimes (hg)\cdot v$, for every $h \in H$, $g \in G$ and $v \in V$. Since $|hg \otimes v| = h|g \otimes v|h^{-1}$, we conclude that $\Res^G_H(V) \in \VGH$. Together, the functors $\Ind_H^G$ and $\Res_H^G$ form the \emph{adjunction} $\Ind_H^G \dashv \Res_H^G$. 

By composition, for every $W \in \VGH$, 
\begin{equation}
    T(W) := \CG \otimes_{\CG} \CG \otimes_{\CH} W \cong \CG \otimes_{\CH} W ,
\end{equation}
where $\CG$ in the last expression is treated as a $(\CH,\CH)$-bimodule. Applying the above definitions, the $G$-grading on $T(W)$ is such that, for every homogeneous $w \in W$ and $g \in G$, $g \otimes w$ is homogeneous with degree $|g \otimes w|=g|w|g^{-1}$. This is compatible with the $H$-equivariant $G$-graded structure of $W$ because, for every $h \in H$ and homogeneous $w \in W$, $|h \otimes w| = |1_G \otimes h \cdot w| = |h \cdot w| = h |w|h^{-1}$. The $H$-equivariant structure is provided by the left $\CH$-module structure on the first factor, and reads $h \cdot (g \otimes w) = hg \otimes w$. In particular, $h \cdot (1_G \otimes w) = h \otimes w = 1_G \otimes h \cdot w$, as expected. 

The \emph{unit} natural transformation $\eta \colon {\rm id}_{\VGH} \to \Res_H^G \circ \Ind_H^G$ of the adjunction $\Ind_H^G \dashv \Res_H^G$ is provided by 
\begin{equation}\label{eq:HopfMonadUnit}
    \eta_W \colon W \to \CG \otimes_{\CH} W, \q  w \mapsto 1_G \otimes w  , 
\end{equation} 
for every $W \in \VGH$. 
The \emph{counit} natural transformation $\epsilon \colon \Ind_H^G \circ \Res_H^G \to {\rm id}_{\VGG}$ of the adjunction $\Ind_H^G \dashv \Res_H^G$ is provided by 
\begin{equation}
    \epsilon_V \colon \CG \otimes_{\CH} V \to V, \q g \otimes v \mapsto g \cdot v,
\end{equation}
for every $V \in \VGG$.
These are used to endow $T$ with a monad structure. Specifically, the natural transformation $\eta \colon {\rm id}_{\VGH} \to T$ is directly provided by the unit of the adjunction, while the multiplication natural transformation $\mu \colon T \circ T \to T$ is defined via $\mu_W := \Res^G_H(\epsilon_{\Ind_H^G(W)}) \colon T^2(W) \to T(W)$. Explicitly, 
\begin{align}
    \label{eq:HopfMonadMult}
    \mu_W \colon \CG \otimes_{\CH} (\CG \otimes_{\CH} W) \to \CG \otimes_{\CH} W , \q g_1 \otimes ( g_2 \otimes w) \mapsto g_1g_2 \otimes w .
\end{align}
In particular, one verifies these maps are compatible with the $(\CH,\CH)$-bimodule structure of $\CG$.
Clearly, the triple $(T,\mu,\eta)$ defines a monad on $\VGH$.

The monad $(T,\mu,\eta)$ encodes the \emph{categorical action} of the hypergroup $\GH$. Indeed, as a $\CH$-bimodule, $\CG$ decomposes as $\CG \cong \bigoplus_{[k] \in \GH} \mathbb C[HkH]$. It implies that the endofunctor $T$ decomposes as
\begin{equation}
    T \cong \bigoplus_{[k] \in \GH} T_{[k]} \equiv \bigoplus_{[k] \in \GH} \mathbb C[HkH] \otimes_{\CH} - .
\end{equation}
The individual action of the hypergroup element $[k] \in \GH$ on the hypergroup $\GH$-graded fusion category $\VGH$ is provided by the endofunctor $T_{[k]}$. In sec.~\ref{sec:Maths_compat}, we make precise the sense in which this $\GH$-action is compatible with the $\GH$-grading. The monad structure on $T$ encodes a choice of \emph{symmetry fractionalisation}.\footnote{Whenever $H$ is chosen to be a normal subgroup $N \cat G$, then the multiplication $\mu$ induces isomorphisms $\mu_{[k_1],[k_2]}\colon T_{[k_1]} \circ T_{[k_2]} \xrightarrow{\sim} T_{[k_1k_2]}$, as expected.} As argued in ref.~\cite{KNBalasubramanian:2025vum}, the symmetry fractionalisation realised here is dictated by the monoidal structure of $\VGH$, which is one of several possible $\GH$-graded extensions of $\VHH$. In sec.~\ref{sec:Maths_frac} we explain how different  $\GH$-graded extensions of $\VHH$ give rise to $\GH$-actions whose distinct symmetry fractionalisations are encoded in the multiplication rule of the corresponding monad.

\bigskip \noindent
\paragraph{Tube algebra:} In order to confirm that the monad $(T,\mu,\eta)$ constructed above does encode the hypergroup action of the previous section, we need to reformulate it in terms of $\Mod(\TVHVG)$. 
First of all, recall that under the equivalence $\Mod(\TVHVG) \simeq \VGH$, the simple objects of $\VGH$ are in 
one-to-one correspondence with the minimal central idempotents of $\TVHVG$. 
Let $W \in \VGH$. As we explained in sec.~\ref{sec:Maths_tube}, the corresponding module over 
$\TVHVG = \mathbb{C}\{\mc T^h_y\}_{h \in H,\, y \in G}$---which we still denote by $W$---satisfies 
\begin{equation}
    \mc T^h_y \cdot w = \delta_{y,|w|} \, h \cdot w ,
\end{equation}
for every homogeneous $w \in W$.
By construction, for every $[k] \in \GH$, the $\TVHVG$-module associated with 
$T_{[k]}(W) = \mathbb{C}[HkH] \otimes_{\CH} W$ satisfies
\begin{equation}
    \mc T^h_y \cdot (g \otimes w) 
    = \delta_{y,\, g|w|g^{-1}} \, (hg \otimes w) ,
\end{equation}
for every $g \in [k]$ 
and homogeneous $w \in W$.
With this equivalence in hand, computing the action of the hypergroup on 
$\TVHVG$-modules  is straightforward. Equip $\mathbb{C}[HkH]$ with the structure 
of a $\TVHVG$-bimodule, which we denote by $M_{[k]}$. As a vector space, 
$M_{[k]}$ is spanned by elements $\mc T^g_x$ with $x \in G$ and 
$g \in [k]$. The $\TVHVG$-bimodule structure of $M_{[k]}$ is obtained by embedding it in $\Tu_G^G$ and using the multiplication rule, such that
\begin{equation}
    \mc T^{h_1}_{y_1} \cdot \tub{x}{g} \cdot \mc T^{h_2}_{y_2} = \delta_{y_1,gxg^{-1}} \, \delta_{g,h_2y_2h_2^{-1}} \, \tub{y_2}{h_1gh_2},
\end{equation}
for every $h_1,h_2 \in G$, $x,y_1,y_2 \in G$ and $g \in [k]$.
Then, we have
\begin{equation}
    T_{[k]}(W) = M_{[k]} \otimes_{\TVHVG} W \in \Mod(\TVHVG).
\end{equation}
Let us propose a graphical interpretation of this relative tensor product. For concreteness, suppose $W$ is a simple object 
$(\orb(f), \hat W)$, where $\orb(f) \in \orb(G)$ and 
$\hat W = \mathbb{C}\{\hat w_c\}_c \in \Irr(\Rep(\Stab_H(f)))$. The corresponding $\Tu_G^H$-module with underlying vector space $\mathbb C\{q_x \otimes \hat w_c\}$ is depicted in eq.~\eqref{eq:modTube}.
Recall that $|q_x \otimes \hat w_c| = x$ and $h \cdot (q_x \otimes \hat w_c)= q_{hxh^{-1}} \otimes s_{h,x} \cdot \hat w_c$, for every $h \in H$, where $hq_x = q_{hxh^{-1}}s_{h,x}$.
A general element of $T_{[k]}(W)$ then takes the form 
$g \otimes (q_x \otimes \hat w_c)$, with grading 
$|g \otimes (q_x \otimes \hat w_c)| = g\,|q_x \otimes \hat w_c|\,g^{-1} = gxg^{-1}$. Given a choice $\{o_d\}_{d=1,\ldots,([k]:H)}$ of representatives in $[k]$ of cosets in $[k]/H$, we have $T_{[k]}(W) = \mathbb C\{o_d \otimes (q_x \otimes \hat w_c)\}_{d,x \in \orb(f),c}$.
Stretching our graphical calculus, we depict the $\Tu_G^H$-module $T_{[k]}(W)$ in a way that is reminiscent of the action of a tube algebra element on $W$:
\begin{equation}
    T_{[k]}(W) = \mathbb C
    \left\{
    \tubeActionT{{o_d}}{x}{{o_dxo_d^{-1}}}{}{W}{c} 
    \right\}_{d, \, x \in \orb(f), \, c} .
\end{equation}
Using the tube action, one readily verifies that $T_{[k]}(W)$ transforms 
exactly as expected:
\begin{equation}
\begin{split}
    \mc T^h_y \cdot \big(o_d \otimes (q_x \otimes \hat w_c) \big) 
    &=
    \delta_{y,o_dxo_d^{-1}} \, \big(ho_d \otimes (q_x \otimes \hat w_c)\big),
\end{split}
\end{equation}
for every $h \in H$, $y \in G$, $x \in \orb(f)$, $d \in \{1,\ldots,([k]:H)\}$ and $c \in \{1,\ldots,\dim_\mathbb C \hat W\}$.
Graphically,
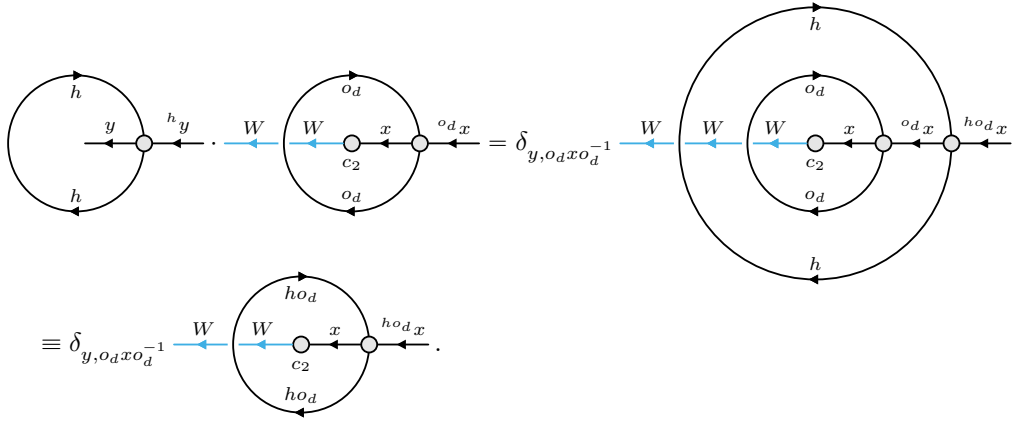
\begin{equation*}
\begin{split}
    &\tubeT{h}{y}{^hy}{} 
    \cdot
    \tubeActionT{o_d}{x}{^{o_d}x}{}{W}{c_2} 
    = \delta_{y,o_dxo_d^{-1}} \, 
    \tubeActionTMod{o_d}{h}{x}{^{o_d}x}{^{ho_d}x}{W}{c_2} \\[-2ex]
    & \q 
    \equiv \delta_{y,o_dxo_d^{-1}}  \, \tubeActionT{ho_d}{x}{^{ho_d}x}{}{W}{c_2} . 
\end{split}
\end{equation*}
We are now ready to recover the action of the hypergroup group element on the localised excitations, as computed in sec.~\ref{sec:symmetry}.
Given a simple module $W$ over the tube algebra $\TVHVG$, let $(\mc E_W)_1^1 \equiv (\mc E_W)^{1_G,1}_{1_G,1}$ be a corresponding primitive idempotent. By the \emph{Wedderburn--Artin theorem}, $W \cong \Tu_G^H \cdot (\mc E_W)_1^1$. Therefore, for every $[k]\in \GH$ and $W \in \Irr(\Mod(\TVHVG))$, we have $M_{[k]} \otimes_{\Tu_G^H} W \cong M_{[k]} \cdot (\mc E_W)_1^1$. It follows that
the dimension of the subspace $(\mc E_{W_2})_1^1 \cdot M_{[k]}\cdot (\mc E_{W_1})^1_1$ provides the multiplicity of the simple object $W_2$ in $T_{[k]}(W_1)$, i.e.
\begin{equation}
    \text{mult}_{W_2}(T_{[k]}(W_1)) = \dim_\mathbb C \big( (\mc E_{W_2})_1^1 \cdot M_{[k]}\cdot (\mc E_{W_1})^1_1 \big).
\end{equation}
In particular, let $\mc E_{\bm 1}$ be the minimal central idempotent associated with the trivial module over $\TVHVG$. Explicitly, it reads
\begin{equation}
    \mc E_{\bm 1} = \frac{1}{|H|} \sum_{h \in H} \mc T^h_{1_G}.
\end{equation}
By construction, the subspace $\mc E_{\bm 1} \cdot M_{[k]} \cdot \mc E_{\bm 1}$ is  one-dimensional, for every $[k]\in \GH$. It is spanned by the tube element $\mc T_{[k]} := \frac{1}{|[k]|}\sum_{g \in [k]}\mc T^g_{1_G} \in \TVHVG$, such that $\mc T_{[1_G]} = \mc E_{\bm 1}$. For every $[k_1],[k_2] \in \GH$, we find that
\begin{equation}
\begin{split}
    \mc T_{[k_1]} \cdot \mc T_{[k_2]} &= \frac{1}{|[k_1]| \cdot |[k_2]|} \sum_{\substack{g_1 \in [k_1]  \\ g_2 \in [k_2]}} \mc T^{g_1}_{1_G} \cdot \mc T^{g_2}_{1_G}
    \\ &=
    \frac{1}{|[k_1]|\cdot |[k_2]|} 
    \sum_{\substack{g_1 \in [k_1]  \\ g_2 \in [k_2]}}
    \sum_{\substack{[k_3] \in \GH \\ g_3 \in [k_3]}} \delta_{g_3,g_1g_2} \, \mc T^{g_3}_{1_G}
    = \sum_{[k_3] \in \GH} C^{[k_1][k_2]}_{[k_3]} \, \mc T_{[k_3]} ,
\end{split}
\end{equation}
where we made use of eq.~\eqref{eq:hyperGr_StrCstsFormulaB}. 

\bigskip \noindent
\paragraph{Equivalent definition:} 
Thus far, we have been relying on the equivalence $(\VGG)_{A(H)} \simeq \VGH$ in order to study the topological order enriched by a non-invertible symmetry. While this equivalence simplifies computations, it is very specific to our choices of parent topological order and pattern of anyon condensation. Therefore, it is desirable to perform the analysis directly at the level of $(\VGG)_{A(H)}$, paving the way towards generalising this approach to more intricate scenarios (see sec.~\ref{sec:outlook}). In that spirit, let us explain how to construct the hypergroup action at the level of $(\VGG)_{A(H)}$.

\bigskip \noindent
Overall, the strategy remains the same. Consider the endofunctor of $(\VGG)_{A(H)}$
\begin{equation}
    T \colon (\VGG)_{A(H)} \longrightarrow \VGG \longrightarrow (\VGG)_{A(H)} 
\end{equation}
obtained by composing the forgetful functor $(\VGG)_{A(H)} \to \VGG$, $(M,\rho) \mapsto M$ with the induction functor $\VGG \to (\VGG)_{A(H)}$, $M \mapsto (M \otimes A(H),{\rm id}_M \otimes \mu)$. When writing the induction functor as $- \otimes A(H)$, we implicitly treat $A(H)$ as an $(1,A(H))$-bimodule. Similarly, express the forgetful functor as $- \otimes_{A(H)} A(H)$, where $A(H)$ is here treated as an $(A(H),1)$-bimodule. Therefore, $T = - \otimes_{A(H)} (A(H) \otimes A(H))$, where $A(H) \otimes A(H)$ is an $(A(H),A(H))$-bimodule.

Following ref.~\cite{Bischoff:2016jmy,Riesen2025}, the  hypergroup action on $(\VGG)_{A(H)}$ should be obtainable by decomposing $A(H) \otimes A(H)$ into indecomposable $A(H)$-bimodules.
Recall that, as an object in $\Rep(G)$, $A(H)$ is isomorphic to the permutation representation $\mathbb C(G/H) \cong \Ind_H^G(\mathbb C)$. Therefore, let us consider $\mathbb C(G/H) \otimes \mathbb C(G/H) \cong \mathbb C(G/H \times G/H)$.
Orbits of $G/H \times G/H$ under the diagonal action of $G$ are in bijection with $\GH$. Given an orbit with representative $(r_{a_1}H,r_{a_2}H)$, the corresponding double coset in $\GH$ is given by $Hr_{a_1}^{-1}r_{a_2}H$. Conversely, the orbit in $G/H \times G/H$ associated with $[k] \in \GH$ is that with representative $(H,kH)$; the corresponding stabiliser in $G$ is provided by $H_k := H \cap kHk^{-1}$. It follows that
\begin{align}
    \label{eq:decompAA}
    \mathbb C(G/H) \otimes \mathbb C(G/H) &\cong \bigoplus_{[k] \in \GH} \Ind_{H_{k}}^G(\mathbb C) 
    \cong 
    \bigoplus_{[k] \in \GH} \mathbb C(G/H_{k}).
\end{align}
This provides the decomposition of $A(H) \otimes A(H)$ into $A(H)$-bimodules in $(\VGG)_{A(H)}$.
Indeed, introduce the \emph{projection maps}
\begin{equation}
    \begin{array}{rcccl}
        G/H &\stackrel{\Pi^{\rm l}}{\longleftarrow}& G/H_{k} &\stackrel{\Pi^{\rm r}}{\longrightarrow}& G/H
        \\
        gH &\longmapsfrom & gH_{k} &\longmapsto& gkH 
         .
    \end{array}
\end{equation}
The $A(H)$-bimodule structure of $\mathbb C(G/H_k)$ is given by
\begin{align}
    (\phi_1 \cdot \phi \cdot \phi_2)(gH_k) 
    &= \phi_1\big(\Pi^{\rm l}(gH_x)\big) \, \phi(gH_k) \, \phi_2\big(\Pi^{\rm r}(gH_k)\big)
    \\
    &= \phi_1(gH)\, \phi(gH_k)\, \phi_2(gkH) ,
\end{align}
for every $\phi_1,\phi_2 \in A(H)$ and $\phi \in \mathbb C(G/H_k)$. One can readily check that it is indecomposable. 
It implies that the endofunctor $T$ decomposes as $T \cong \bigoplus_{[k] \in \GH}T_{[k]}$, where $T_{[k]}((M,\rho)) := M \otimes_{A(H)} \mathbb C(G/H_k)$, for every $[k] \in \GH$. Note that the right $A(H)$-module structure of $T_{[k]}((M,\rho))$ is provided by the right $A(H)$-module structure of $\mathbb C(G/H_k)$ spelt out above. 

\bigskip \noindent
Suppose $H \leq G$ is isomorphic to a normal subgroup $N \lhd G$. Recall the hypergroup $G/\!/N$ boils down to the quotient subgroup $G/N$. In this case, the tensor product $\mathbb C(G/N) \otimes \mathbb C(G/N)$ decomposes into $|G/N|$ copies of $\mathbb C(G/N)$. Therefore, $T$ decomposes as $T \cong \bigoplus_{k \in G/N} T_k$, where $T_k$ sends $(M,\rho)$ to $(M,k(\rho))$, i.e., an $A(N)$-module in $\VGG$ with the same underlying object. The $A(N)$-module action $k(\rho) \colon M \otimes A(N) \to M$ of $T_k((M,\rho))$ is simply obtained by precomposing $\rho$ with the automorphism of $A(N)$ associated with the group element $k^{-1} \in G/N$ \cite{KirillovJr2002}.

\subsection{Compatibility conditions\label{sec:Maths_compat}}

One of the defining properties of a $G$-crossed braided fusion category $\mc B = \bigoplus_{x \in G} \mc B_x$ is that $g \cdot \mc B_x \subset \mc B_{gxg^{-1}}$. We identify analoguous---though not as constraining---properties in $\VGH$.

\bigskip\noindent
The $\GH$-hypergroup action on $\VGH$ is compatible with its $\GH$-grading structure in the following sense. For every $[k_1],[k_2] \in \GH$, write ${\rm Ad}_{[k_1]}([k_2]) := [k_1] \star [k_2] \star [k_1^{-1}]$. For every $W \in \Vect^H_{[k_2]}$, we have
\begin{equation}
    T_{[k_1]}(W) \in \!\!\! \bigoplus_{[k_3] \prec {\rm Ad}_{[k_1]}([k_2])} \!\!\!  \Vect_{[k_3]}^H.
\end{equation}
Essentially, this follows from the fact that $|g \otimes w| = g |w|g^{-1} \in \supp({\rm Ad}_{[k_1]}([k_2]))$, for every $g \otimes w \in T_{[k_1]}(W)$. As it stands, it is a rather weak statement. Indeed, it still allows for two simple objects in the same $[k_2]$-twisted sector to land in sectors twisted by distinct elements in $\supp({\rm Ad}_{[k_1]}([k_2]))$ under $T_{[k_1]}$.
But a slightly more stringent compatibility condition does hold. For every $[k_1],[k_2] \in \GH$, let us introduce a new set of structure constants via
\begin{equation}
    {\rm Ad}_{[k_1]}([k_2]) \equiv \sum_{[k_3] \in \GH}A^{[k_1][k_2]}_{[k_3]} \, [k_3],
\end{equation}
which are well defined thanks to the associativity of $\mathbb C[\GH]$. Besides, for every $[k] \in \GH$, let $R_{[k]}$ be some choice of \emph{$\VHH$-regular object} in the $[k]$-twisted sector, i.e., an object in $\Vect_{[k]}^H$ such that $V \act R_{[k]} = \FPdim(V) \cdot R_{[k]}$ for all $V \in \VHH$. We claim that 
\begin{equation}
    \label{eq:compAdHyperGr}
    \frac{\FPdim \big(\Pi_{[k_3]}(T_{[k_1]}(R_{[k_2]})) \big)}{\FPdim\big( T_{[k_1]}(R_{[k_2]}) \big)} = A^{[k_1][k_2]}_{[k_3]}.
\end{equation}
Let us verify this formula. First of all, the constants  $A^{[k_1][k_2]}_{[k_3]}$ are given by 
\begin{equation}
\begin{split}
    A^{[k_1][k_2]}_{[k_3]} 
    &= \frac{1}{|H|^2} \cdot \big| \big\{(h_1,h_2) 
    \in H \times H \, \big| \, k_1h_1k_2h_2k_1^{-1} \in [k_3] \big\} \big|
    \\
    &= \frac{1}{|H|^6} \cdot \big| \big\{(h_1,\ldots,h_6) 
    \in H^{6} \, \big| \, h_1k_1h_2h_3k_2h_4h_5k_1^{-1}h_6 \in [k_3] \big\} \big|.
\end{split}
\end{equation}
In the same vein as eq.~\eqref{eq:hyperGr_StrCstsFormulaB}, another formula for these constants is provided by
\begin{equation}
    \label{eq:hyperGrAdCstsAlt}
    A^{[k_1][k_2]}_{[k_3]} = \frac{1}{|[k_1]| \cdot |[k_2]|} \cdot \big| \big\{ (g_1,g_2) \in [k_1] \times [k_2] \, \big| \, g_1g_2g_1^{-1} \in [k_3] \, \big\} \big|.
\end{equation}
Indeed, for any $[k_1],[k_2] \in \GH$, let $g_1 \in [k_1]$ and $g_2 
\in [k_2]$. By definition, there exist $h_1,\ldots,h_4 \in H$ such that $g_1 = h_1k_1h_2$ and $g_2 = h_3k_2h_4$. As a matter of fact, there are $|L_{k_1}|$- and $|L_{k_2}|$-many possible pairs of group elements $(h_1,h_2), (h_3,h_4) \in H \times H$ such that $g_1 = h_1k_1h_2$ and $g_2=h_3k_2h_4$, respectively. Hence,
\begin{equation}
\begin{split}
    &\frac{1}{|H|^2} \cdot \big| \big\{(h_1,\ldots,h_6) 
    \in H^{6} \, \big| \, h_1k_1h_2h_3k_2h_4h_5k_1^{-1}h_6 \in [k_3] \big\} \big|
    \\[-.5ex]
    &\q =\big| \big\{(h_1,\ldots,h_4) 
    \in H^4 \, \big| \, h_1k_1h_2h_3k_2h_4(h_1k_1h_2)^{-1} \in [k_3] \big\} \big| 
    \\
    & \q = |L_{k_1}| \cdot |L_{k_2}| \cdot \big| \big\{(g_1,g_2) \in [k_1] \times [k_2] \, \big| \, g_1g_2g_1^{-1} \in [k_3] \big\} \big|,
\end{split}
\end{equation}
where the first step simply consists in reparametrisations within $[k_3]$. 
Applying the orbit-stabiliser theorem, which stipulates that $|L_k| = |H|^2/|[k]|$, produces eq.~\eqref{eq:hyperGrAdCstsAlt}.
Next, the $\VHH$-regularity of $R_{[k_2]}$ implies that $\FPdim(R_{[k_2]}) = |H| \cdot |[k_2]|$. Moreover, it requires the fibre dimension to be constant across $\supp(R_{[k_2]}) = [k_2]$, so $\dim_\mathbb C(R_{[k_2]})_x = \FPdim(R_{[k_2]})/|[k_2]|= |H|$, for every $x \in [k_2]$. It follows that when computing $T_{[k_1]}(R_{[k_2]}) = \mathbb C[Hk_1H] \otimes_{\mathbb C[H]} R_{[k_2]}$, 
each pair $(g,x) \in [k_1] \times [k_2]$ contributes a one-dimensional summand graded by $gxg^{-1}$. Summing over all such pairs yields
\begin{equation*}
    \FPdim\big(T_{[k_1]}(R_{[k_2]})\big) = |[k_1]| \cdot |[k_2]|.
\end{equation*}
Restricting to pairs landing in $[k_3]$ finally yields
\begin{equation*}
    \FPdim\big(\Pi_{[k_3]}(T_{[k_1]}(R_{[k_2]}))\big) = \big|\big\{(g,x) \in [k_1] \times [k_2] \, \big| \, gxg^{-1} \in [k_3] \big\}\big|.
\end{equation*}
Dividing this by $\FPdim\big( T_{[k_1]}(R_{[k_2]}) \big) = |[k_1]| \cdot |[k_2]|$ exactly recovers the formula for $A^{[k_1][k_2]}_{[k_3]}$ derived in eq.~\eqref{eq:hyperGrAdCstsAlt}, thereby verifying eq.~\eqref{eq:compAdHyperGr}.

\subsection{Gauging the non-invertible symmetry}

In sec.~\ref{sec:symmetry}, we explained how to recover the topological order $\VGG$ from the condensed theory $\VGH$ by gauging the non-invertible $\GH$-symmetry. Mathematically, this should amount to a hypergroup equivariantisation. This raises an important subtlety: the goal is not merely to recover $\VGG$ as a category, but as a fusion category. Reconstructing the monoidal structure of the $\GH$-equivariantisation requires some form of \emph{comultiplication rule}, which hypergroups typically do not possess. This is precisely where the monad formalism of sec.~\ref{sec:Maths_monad} comes into play. Indeed, one can show that the monad constructed there is further endowed with the structure of a \emph{Hopf monad}, which ultimately allows us to recover the fusion category $\VGG$ as the \emph{Eilenberg--Moore category} of modules over this Hopf monad.

\bigskip \noindent
\paragraph{Hopf monad:}  Following ref.~\cite{MR2355605,BRUGUIERES2011745}, a monad $(T,\mu,\eta)$ on $\mc C$ is said to be a \emph{bimonad} if $T$ is a \emph{comonoidal} functor and $(\mu,\eta)$ are comonoidal natural transformations. Here, comonoidality of $T \colon \mc C \to \mc C$ indicates that it is further equipped with natural transformations $\Delta_{-,-} \colon T(- \otimes -) \to T(-) \otimes T(-)$ and $\epsilon \colon T(1_\mc C) \to 1_\mc C$ such that
\begin{equation}
\begin{split}
    &({\rm id}_{T(X_1)} \otimes \Delta_{X_2,X_3}) \circ \Delta_{X_1,X_2 \otimes X_3} \circ T(\alpha_{X_1,X_2,X_3}) 
    \\
    & \q = \alpha_{T(X_1),T(X_2),T(X_3)} \circ (\Delta_{X_1,X_2} \otimes {\rm id}_{T(X_3)}) \circ \Delta_{X_1 \otimes X_2,X_3} ,
\end{split}
\end{equation}
and
\begin{equation}
     ({\rm id}_{T(X)} \otimes \epsilon) \circ \Delta_{X,1_\mc C} = {\rm id}_{T(X)} = (\epsilon \otimes {\rm id}_{T(X)}) \circ \Delta_{1_\mc C,X}  ,
\end{equation}
for every $X,X_1,X_2,X_3 \in \mc C$. Comonoidality of $\mu \colon T \circ T \to T$ and $\eta \colon {\rm id}_\mc C \to T$ is then the requirement that the following identities hold for every $X_1,X_2 \in \mc C$:
\begin{align}
    \Delta_{X_1,X_2} \circ \mu_{X_1 \otimes X_2} ,
    &= 
    (\mu_{X_1} \otimes \mu_{X_2}) \circ \Delta_{T(X_1),T(X_2)} \circ T(\Delta_{X_1,X_2}) ,
    \\
    \epsilon \circ \mu_{1_\mc C} &= \epsilon \circ T(\epsilon) ,
    \\
    \eta_{X_1} \otimes \eta_{X_2} &= \Delta_{X_1,X_2} \circ \eta_{X_1 \otimes X_2} ,
    \\
    \epsilon \circ \eta_{1_\mc C} &= {\rm id}_{1_\mc C} .
\end{align}
For any bimonad $(T,\mu,\eta)$, one can define natural transformations $\theta^{\rm l}_{-,-} \colon T(- \otimes T(-)) \to T(-) \otimes T(-)$ and $\theta^{\rm r}_{-,-} \colon T(T(-) \otimes -) \to T(-) \otimes T(-)$ according to
\begin{align}
    \theta^{\rm l}_{X_1,X_2} &:= ({\rm id}_{T(X_1)} \otimes \mu_{X_2}) \circ \Delta_{X_1 ,T(X_2)} \colon T(X_1 \otimes T(X_2)) \to T(X_1) \otimes T(X_2) ,
    \\
    \theta^{\rm r}_{X_1,X_2} &:= (\mu_{X_1} \otimes {\rm id}_{T(X_2)} ) \circ \Delta_{T(X_1) ,X_2} \colon T(T(X_1) \otimes X_2) \to T(X_1) \otimes T(X_2) ,
\end{align}
for every $X_1,X_2 \in \mc C$. These are subject to a number of coherence relations that we can infer from previous ones. Finally, we define a \emph{Hopf monad} as a bimonad $(T,\mu,\eta,\Delta,\epsilon)$ whose natural transformation $\theta^{\rm l}$ and $\theta^{\rm r}$ are natural \emph{isomorphisms}. Given a Hopf monad, the natural isomorphisms $\theta^{\rm l}$ and $\theta^{\rm r}$ can be employed to define a left and a right \emph{antipode} natural transformations $S^{\rm l}_- \colon T({}^*T(-)) \to {}^*-$ and $S^{\rm r}_- \colon T(T(-)^*) \to -^*$. 

\bigskip \noindent
\paragraph{Eilenberg--Moore category:} Given a monad $T \equiv (T,\mu,\eta)$ on a fusion category $\mc C$, a left module over $T$ in $\mc C$ is a pair $(M,\varrho)$ consisting of an object $M \in \mc C$ and a morphism $\varrho \colon T(M) \to M$---referred to as the \emph{$T$-action}---such that
\begin{equation}
    \label{eq:TModProp}
    \varrho \circ T(\varrho) = \varrho \circ \mu_M \q \text{and} \q \varrho \circ \eta_M = {\rm id}_M .
\end{equation}
We denote by $\mc C^T$ the (Eilenberg--Moore) category of (left) $T$-modules in $\mc C$ and morphisms of $T$-modules \cite{10.1215/ijm/1256068141}. Crucially, the category $\mc C^T$ of $T$-modules over a \emph{Hopf} monad is rigid monoidal \cite{MR2355605}. The tensor product of two $T$-modules $(M_1,\varrho_1)$ and $(M_2,\varrho_2)$ in $\mc C^T$ is provided by
\begin{equation}
    (M_1,\varrho_1) \otimes (M_2,\varrho_2) := \big(M_1 \otimes M_2, (\varrho_1 \otimes \varrho_2) \circ \Delta_{M_1,M_2} \big) . 
\end{equation}
The (right) dual of the $T$-module $(M,\varrho)$ is defined as $(M^*,S^r_M \circ T(\varrho^*))$.

\bigskip \noindent
\paragraph{Equivariantisation:}  
Above, we defined an action of the hypergroup $\GH$ on $\VGH$ by assigning to every $[k] \in \GH$ the endofunctor $T_{[k]} = \mathbb{C}[H k H] \otimes_{\mathbb{C}[H]} -$. Together, these endofunctors organise into a monad $(T, \mu, \eta)$, where $T = \bigoplus_{[k] \in \GH} T_{[k]}$ and $\mu$, $\eta$ are defined in eq.~\eqref{eq:HopfMonadMult} and eq.~\eqref{eq:HopfMonadUnit}, respectively. We took this monad structure to formalise the categorical action of the hypergroup $\GH$ on $\VGH$. Let us now endow $(T,\mu,\eta)$ with a bimonad structure. Define the natural transformations 
$\Delta_{-,-} \colon T(- \otimes -) \to T(-) \otimes T(-)$ and $\epsilon \colon T(\mathbb C) \to \mathbb C$ via
\begin{align}
    \Delta_{W_1,W_2} 
    &\colon 
    T(W_1 \otimes W_2) \to T(W_1) \otimes T(W_2) , \q g \otimes (w_1 \otimes w_2) \mapsto (g \otimes w_1) \otimes (g \otimes w_2) ,
    \\
    \epsilon &\colon \mathbb \Ind_H^G(\mathbb C) \to \mathbb C , \q g \otimes 1 \mapsto 1 .
\end{align}
It essentially follows from the bialgebra structure of the group algebra $\CG$ that $(T,\mu,\eta,\Delta,\epsilon)$ defines a bimonad. Furthermore, consider the natural transformation $\theta^{\rm l}_{-,-} \colon T(- \otimes T(-)) \to T(-) \otimes T(-)$ and $\theta^{\rm r}_{-,-} \colon T(T(-) \otimes -) \to T(-) \otimes T(-)$ defined according to
\begin{align*}
    \theta^{\rm l}_{W_1,W_2} &\colon T(W_1 \otimes T(W_2)) \to T(W_1) \otimes T(W_2) , \q g_1 \otimes (w_1 \otimes (g_2 \otimes w_2)) \mapsto (g_1 \otimes w_1) \otimes (g_1g_2 \otimes w_2) ,
    \\
    \theta^{\rm r}_{W_1,W_2} &\colon T(T(W_1) \otimes W_2) \to T(W_1) \otimes T(W_2) , \q g_2 \otimes ((g_1 \otimes w_1) \otimes w_2) \mapsto (g_2g_1 \otimes w_1) \otimes (g_2 \otimes w_2) .
\end{align*}
These are invertible with
\begin{align*}
    (\theta^{\rm l})_{W_1,W_2}^{-1} &\colon T(W_1) \otimes T(W_2)
     \to T(W_1 \otimes T(W_2)) , \q (g_1 \otimes w_1) \otimes (g_2 \otimes w_2) 
     \mapsto g_1 \otimes (w_1 \otimes (g_1^{-1}g_2 \otimes w_2)) ,
    \\
    (\theta^{\rm r})_{W_1,W_2}^{-1} &\colon T(W_1) \otimes T(W_2) \to T(T(W_1) \otimes W_2) , \q (g_1 \otimes w_1) \otimes (g_2 \otimes w_2) 
     \mapsto g_2 \otimes ((g_2^{-1}g_1 \otimes w_1) \otimes  w_2) .
\end{align*}
One needs to verify that these maps are indeed well defined. Take for instance $(g_1h \otimes w_1) \otimes (g_2 \otimes w_2) = (g_1 \otimes h \cdot w_1) \otimes (g_2 \otimes w_2)$. Applying $(\theta^{\rm l})^{-1}_{W_1,W_2}$ to both sides of this equation yields $g_1h \otimes (w_1 \otimes (h^{-1}g_1^{-1}g_2 \otimes w_2))$ and $g_1 \otimes (h \cdot w_1 \otimes (g_1^{-1}g_2 \otimes w_2))$, respectively, which are indeed equal since $H$ acts diagonally on $W_1 \otimes T(W_2)$. Bringing everything together, $(T,\mu,\eta,\Delta,\epsilon)$ defines a Hopf monad on $\VGH$.

\bigskip \noindent
By virtue of $T \equiv (T,\mu,\eta,\Delta,\epsilon)$ being a Hopf monad on $\VGH$, we know that $(\VGH)^T$ is a rigid monoidal category. Let us verify that $(\VGH)^T \simeq \VGG$. Let $(M,\varrho)$ be an object in $(\VGH)^T$. By definition, $M$ is $G$-graded. For every $x \in G$ and $m \in M_x$, define $g \cdot m := \varrho(g \otimes m)$. It follows from eq.~\eqref{eq:TModProp} that
\begin{equation}
\begin{split}
    g_1 \cdot (g_2 \cdot m) 
    = \varrho\big(g_1 \otimes \varrho(g_2 \otimes m)\big) 
    &= (\varrho \circ T(\varrho))\big(g_1 \otimes (g_2 \otimes m)\big)
    \\
    &=(\varrho \circ \mu_M)\big((g_1 \otimes g_2) \otimes m\big) = \varrho(g_1g_2 \otimes m) = g_1g_2 \cdot m ,
\end{split}
\end{equation}
for every $g_1,g_2 \in G$, so that $M$ is $G$-equivariant. It also follows from eq.~\eqref{eq:TModProp} that restricting this $G$-action to $H$ recovers the native $H$-equivariant structure of $M$. Indeed,
\begin{equation}
    h \cdot m = \varrho(h \otimes m) = \varrho (1_G \otimes h \cdot m) = (\varrho \circ \eta_M)(h \cdot m) = h \cdot m,
\end{equation}
for every $h \in H$. Finally, since $\varrho$ is a $G$-grading preserving linear map, $|g \cdot m| = |\varrho(g \otimes m)| = |g \otimes m|=gxg^{-1}$. Hence, $g \cdot M_x \subseteq M_{gxg^{-1}}$. Bringing everything together, one obtains that $M \in \VGG$. Conversely, let $V \in \VGG$. In particular, $V \in \VGH$. Define $\varrho \colon T(V) = \CG \otimes_{\CH} V \to V$, via $g \otimes v \mapsto g \cdot v$. Clearly, $(V,\varrho)$ defines a $T$-module. Moreover, one readily verifies that the two functors thus constructed are inverse to each other up to natural isomorphisms. This establishes  the equivalence $(\VGH)^T \simeq \VGG$ of categories. One promotes this equivalence to an equivalence of rigid monoidal categories---and ultimately fusion categories---by noticing that
\begin{equation}
\begin{split}
    g \cdot (m_1 \otimes m_2) &= \big((\varrho_1 \otimes \varrho_2) \circ \Delta_{M_1,M_2}\big)\big(g \otimes (m_1 \otimes m_2)\big)
    \\
    &= \varrho_1(g \otimes m_1) \otimes \varrho_2(g \otimes m_2) = (g \cdot m_1)\otimes (g \cdot m_2),
\end{split}
\end{equation}
for every $(M_1,\varrho_1),(M_2,\varrho_2) \in (\VGH)^T$, and 
\begin{equation}
\begin{split}
    (g \cdot \phi)(-)
    &= \big( S^{\rm r}_M \circ T(\varrho^*)\big)(g \otimes \phi)
    =
    S^{\rm r}_M\Big(g \otimes \sum_{x \in G}\big(\delta_x \otimes \phi(x \cdot -) \big)\Big)
    \\ &= \sum_{x \in G} \delta_x(g^{-1}) \, \phi(x \cdot -) =  \phi(g^{-1} \cdot -),
\end{split}
\end{equation}
for every $(M,\varrho) \in (\VGH)^T$ and $\phi \in M^*$.

\bigskip \noindent
Before concluding this part, let us provide a different perspective on this equivariantisation procedure. A closely related situation was considered in ref.~\cite{KNBalasubramanian:2025vum}. Starting from the topological order $\VGG$, which admits a $1$-form $\Rep(G)$ symmetry, the authors consider \emph{gauging} the algebra of lines provided by $A(H)$, producing a dual theory with topological order $\VHH$. The question they address is which gauging procedure takes one back to $\VGG$. More precisely, the symmetry structure of $\VHH$ is encoded in the fusion $2$-category $\Mod(\VHH)$ of (finite semisimple) module categories over $\VHH$, whose objects label \emph{condensation defects} generating a \emph{non-invertible} $0$-form symmetry. In this context, gauging operations are encoded in \emph{separable algebras} in $\Mod(\VHH)$, namely fusion $1$-categories $\mc C$ equipped with a central functor from $\VHH$, i.e., a (fully faithful) braided tensor functor $\VHH \to \mc Z(\mc C)$~\cite{DECOPPET2023108967}. The fusion category $\VGH$ provides such a separable algebra, and $\Mod(\VGH)$ defines a finite semisimple module $2$-category over $\Mod(\VHH)$. The result of the corresponding gauging operation is the \emph{Morita dual} of $\Mod(\VHH)$ with respect to $\Mod(\VGH)$, which is equivalent to the $2$-category of modules over the centraliser in $\mc Z(\VGH)$ of the image of $\VHH$ under the braided tensor functor $\VHH \to \mc Z(\VGH)$. Since $\mc Z(\VGH) \simeq \VHH \boxtimes \overline{\VGG}$, this centraliser is provided by $\VGG$, and the symmetry structure of the gauged theory is therefore $\Mod(\VGG)$. More generally, and as anticipated in ref.~\cite{Riesen2025}, we expect this mechanism to provide an alternative way of computing the result of a hypergroup equivariantisation. 

\subsection{Generalised fixed point theorem}

We conclude our study of the $\GH$-equivariant $\GH$-graded structure of $\VGH$ by proving a statement that further strengthens the compatibility between $\GH$-equivariance and $\GH$-grading. Let $\mc B = \bigoplus_{x \in G} \mc B_x$ be a $G$-crossed braided extension of a non-degenerate braided fusion category $\mc B_{1_G}$. Then, for every $g \in G$, the number $|\Irr(\mc B_{1_G})^g|$ of simple objects fixed by the action of the group element $g \in G$ equals the rank of $\mc B_g$ \cite{PhysRevB.100.115147,Bischoff2020,JONES2021}. In symbols, $\text{rank}(\mc B_g) = |\Irr(\mc B_{1_G})^g|$. 

A generalisation of the above formula to the case of a non-invertible symmetry was physically motivated in ref.~\cite{KNBalasubramanian:2025vum}. Within our context, it goes as follows. 
For every $[k] \in \GH$, define
\begin{equation}
    \Irr(\VHH)^{[k]} := \!\!\! \bigoplus_{V \in \Irr(\VHH)} \!\!\!  \Hom_{\VHH}\big(T_{[k]}(V),V\big) .
\end{equation}
The claim is that 
\begin{equation}
    \label{eq:fixedPtTh}
    \text{rank}(\Vect_{[k]}^H) = \dim_\mathbb C \Irr(\VHH)^{[k]}.
\end{equation}
Let us first compute the left-hand side of eq.~\eqref{eq:fixedPtTh}. By definition,
\begin{equation}
\begin{split}
    \text{rank}(\Vect_{[k]}^H) &= \! \sum_{\orb(f) \in [k]} \! \text{rank}\big(\Rep(\Stab_H(f)) \big) = \! \sum_{\orb(f) \in [k]} \! \big|\cl(\Stab_H(f))\big|
    \\
    &= \big| \big\{(g,h) \in [k] \times H \, | \, gh=hg\big\}/\sim\big|,
\end{split}
\end{equation}
where $(g_1,h_1) \sim (g_2,h_2)$ if and only if there exists $h \in H$ such that $(g_1,h_1) = (hg_1h^{-1},hh_1h^{-1})$. An application of the \emph{Burnside's lemma} finally yields
\begin{equation}
    \text{rank}(\Vect_{[k]}^H) \
    = \frac{1}{|H|} \cdot \big|\big\{(g,h_1,h_2) \in [k] \times H \times H \, | \, gh_1=h_1g, \; gh_2=h_2g, \; h_1h_2=h_2h_1 \big\} \big|.
\end{equation}
Let us now compare it to the right-hand side of eq.~\eqref{eq:fixedPtTh}. Let $V \equiv (\cl(f),\hat V) = \mathbb C\{p_x \otimes \hat v_b\}_{x \in \cl(f),b}$ be a simple object in $\VHH$. Let us treat it as an object in $\Mod(\Tu_H^H)$, whereby
\begin{equation}
    \tub{y}{h} \cdot (p_x \otimes \hat v_b) = \delta_{y,x} \, (p_{hxh^{-1}} \otimes z_{h,x} \cdot \hat v_b),
\end{equation}
for every $h,y \in H$, $x \in \cl(f)$ and $b \in \{1,\ldots,\dim_\mathbb C \hat V\}$, such that $hp_x = p_{hxh^{-1}}z_{h,x}$. Let $\chi_{V}$ be the corresponding character. Recall that
\begin{equation}
    \chi_V(\tub{y}{h}) = \delta_{hy,yh} \, \delta_{y \in \cl(f)}  \, \chi_{\hat V}(z_{h,y}),
\end{equation}
where $\chi_{\hat V}$ is the character of $\hat V \in \Irr(\Rep(Z_H(f)))$. Similarly, consider the $\Tu_H^H$-module associated with $T_{[k]}(V) = \mathbb C\{o_d \otimes (p_x \otimes \hat v_b)\}_{d,x \in \cl(f),c}$, where $\{o_d\}_{d=1,\ldots,([k]:H)}$ denotes a choice of representatives in $[k]$ of cosets in $[k]/H$. When acting with the tube element $\tub{y}{h}$, we are only interested in the components that map back to $\VHH$, which requires $h \cdot o_dH = o_d H$ and thus $o_d^{-1}ho_d \in H$.  The corresponding character reads
\begin{equation}
\begin{split}
    \chi_{T_{[k]}(V)}(\tub{y}{h}) &= \sum_{d=1}^{([k]:H)} \delta_{y \in o_dHo_d^{-1}} \, \delta_{h \in o_dHo_d^{-1}} \, \delta_{hy,yh} \, \delta_{o_d^{-1}yo_d \in \cl(f)} \, \chi_{\hat V}(z_{o_d^{-1}ho_d,o_d^{-1}yo_d})
    \\
    &= \sum_{d=1}^{([k]:H)} \delta_{y \in o_dHo_d^{-1}} \, \delta_{h \in o_dHo_d^{-1}} \,\chi_V \big(\tub{o_d^{-1}yo_d}{o_d^{-1}ho_d}\big).
\end{split}
\end{equation}
Replacing the sum over cosets in $[k]/H$ by a sum over group elements in $[k]$ yields
\begin{equation}
    \chi_{T_{[k]}(V)}(\tub{y}{h}) = \frac{1}{|H|}\sum_{g \in [k]} \delta_{y \in gHg^{-1}} \, \delta_{h \in gHg^{-1}} \, \chi_{V}\big(\tub{g^{-1}yg}{g^{-1}hg}\big).
\end{equation}
Therefore, for every $V \equiv (\cl(f),\hat V) \in \Irr(\VHH)$, 
\begin{align}
    \big\la \chi_{T_{[k]}(V)},\chi_V \big\ra_{\VHH} &= \frac{1}{|H|}\sum_{h,y \in H} {\chi_{T_{[k]}(V)}(\tub{y}{h})} \, \overline{\chi_V(\tub{y}{h})}
    \\ \nn
    &= \frac{1}{|H|^2} \sum_{h,y \in H} \sum_{g \in [k]} \delta_{y \in g H g^{-1}} \, \delta_{h \in g H g^{-1}}  \, {\chi_V\big(\tub{g^{-1}yg}{g^{-1}hg}\big)} \, \overline{\chi_V(\tub{y}{h})}.
\end{align}
It follows from \emph{Schur's orthogonality relations} that
\begin{equation}
\begin{split}
    &\delta_{y \in g H g^{-1}} \, \delta_{h \in g H g^{-1}} \!\!\! \sum_{V \in \Irr(\VHH)}\!\!\!  
    \chi_{V}\big(\tub{g^{-1}yg}{g^{-1}hg}\big) \, \overline{\chi_V(\tub{y}{h})} 
    \\ & \q = 
    \delta_{hy,yh}
    \begin{cases}
        |Z_H(h,y)| & \text{if $(h,y)$ and $(g^{-1}hg,g^{-1}yg)$ are $H$-conjugate} 
        \\
        0 & \text{otherwise},
    \end{cases}
\end{split}
\end{equation}
where $Z_H(h,y)$ is the centraliser in $H$ of the pair $(h,y) \in H \times H$. 
In order for $(h,y)$ and $(g^{-1}hg,g^{-1}yg)$ to be $H$-conjugate, there must exist a group element in $gH$ that commutes with $h$ and $y$. Provided that there exists at least one such group element, then the total number of group elements in $gH$ commuting with both $h$ and $y$ equals $|Z_H(h,y)|$. Therefore, for every $h,y \in H$ such that $h \in gHg^{-1}$ and $y \in gHg^{-1}$,  
\begin{equation}
    \sum_{V \in \Irr(\VHH)}\!\!\!  
    \chi_{V}\big(\tub{g^{-1}yg}{g^{-1}hg}\big) \, \overline{\chi_V(\tub{y}{h})}
    = \delta_{hy,yh}\, \big|\big\{x \in gH \, | \, xh=hx, \; xy=yx \big\} \big|.
\end{equation}
Performing the sum over the double coset $[k]$ yields
\begin{equation}
    \frac{1}{|H|} \sum_{g \in [k]}
    \big|\big\{x \in gH \, | \, xh=hx, \; xy=yx \big\} \big|
    =
    \big|\big\{g \in [k] \, | \, gh=hg, \; gy=yg \big\} \big|.
\end{equation}
Bringing everything together, one finally obtains
\begin{equation}
\begin{split}
    \dim_\mathbb C \Irr(\VHH)^{[k]} &= \!\!\!
    \sum_{V \in \Irr(\VHH)} \!\!\! 
    \big\la \chi_{T_{[k]}(V)},\chi_V \big\ra_{\VHH} 
    \\
    &=
    \frac{1}{|H|} \cdot \big|\big\{(g,h,y) \in [k] \times H \times H \, | \, gh=hg, \; gy=yg, \; hy=yh \big\} \big| .
\end{split}
\end{equation}
This establishes the equality \eqref{eq:fixedPtTh} and concludes our study of $\VGH$. 

\subsection{Twisted quantum double, symmetry fractionalisation and discrete torsion \label{sec:Maths_frac}}

The starting point of our construction is the quantum double model with input group $G$, whose topological order is encoded in the non-degenerate braided fusion category $\VGG \simeq \mc Z(\Vect_G)$. Instead, one could have started from one of its twisted variants, which further requires a choice of normalised representative $\alpha$ of a cohomology class in $H^3(G,{\rm U}(1))$. Such a lattice model has been investigated in great detail \cite{Hu:2012wx,Bullivant:2019fmk}. In particular, localised excitations are known to be encoded in the category $\Mod(\Tu_G^{G,\alpha})$ of modules over the \emph{twisted} tube algebra $\Tu_G^{G,\alpha}$. As a vector space, $\Tu_G^{G,\alpha} = \mathbb C\{\mc T_x^g\}_{g,x \in G}$. The multiplication rule is defined by 
\begin{equation}
    \mc T^{g_1}_{x_1} \cdot \mc T^{g_2}_{x_2} := \delta_{x_1,{}^{g_2}x_2} \, \alpha(g_1,g_2|x_2) \, \mc T_{x_2}^{g_1g_2},
\end{equation}
where one introduced
\begin{equation}
    \alpha(g_1,g_2|x) := \frac{\alpha(g_1,{}^{g_2}x,g_2)}{\alpha(g_1,g_2,x) \, \alpha(^{g_1g_2}x,g_1,g_2)}.
\end{equation}
It follows from the 3-cocycle condition satisfied by $\alpha$ that
\begin{equation}
    \alpha(g_1,g_2g_3|x) \, \alpha(g_2,g_3|x) = \alpha(g_1g_2,g_3|x) \, \alpha(g_1,g_2|{}^{g_3}x),
\end{equation}
for every $g_1,g_2,g_3,x \in G$, which guarantees that $\Tu_G^{G,\alpha}$ is associative. 

Given a $G$-graded vector space $V = \bigoplus_{x \in G} V_x$, one says that a $G$-action on $V$ such that $g \cdot V_x \subseteq V_{gxg^{-1}}$ is \emph{$\alpha$-projective} if $g_1 \cdot (g_2 \cdot v) = \alpha(g_1,g_2|x) \, (g_1g_2) \cdot v$, for every $g_1,g_2 \in G$ and $v \in V_x$. Let $\VGGt$ be the category of $G$-graded vector spaces with $\alpha$-projective $G$-action. The monoidal structure of $\VGGt$ is given by the monoidal structure of $\Vect_G^\alpha$---where $\Vect_G^\alpha$ is the category of $G$-graded vector spaces with monoidal associator $\alpha_{V_1,V_2,V_3} \colon (V_1 \otimes V_2) \otimes V_3 \xrightarrow{\sim} V_1 \otimes (V_2 \otimes V_3)$ defined by $\alpha_{V_1,V_2,V_3}((v_1 \otimes v_2) \otimes v_3) := \alpha(x_1,x_2,x_3) \, v_1 \otimes (v_2 \otimes v_3)$, for every $v_1 \in (V_1)_{x_1}$, $v_2 \in (V_2)_{x_2}$ and $v_3 \in (V_3)_{x_3}$---with the $\alpha$-projective $G$-action provided by 
\begin{equation}
    g \cdot (v_1 \otimes v_2) := \alpha(g|x_1,x_2) \, (g \cdot v_1) \otimes (g \cdot v_2),
\end{equation}
for every $v_1 \in (V_1)_{x_1}$ and $v_2 \in (V_2)_{x_2}$,
where one introduced
\begin{equation}
    \alpha(g|x_1,x_2) :=  \frac{\alpha({}^g x_1,g,x_2)}{\alpha(g,x_1,x_2) \, \alpha({}^g x_1, {}^g x_2,g)}.
\end{equation}
It follows from the 3-cocycle condition satisfied by $\alpha$ that
\begin{equation}
\begin{split}
    \alpha(g_1,g_2|x_1x_2) \, \alpha(g_1g_2 | x_1,x_2) = \alpha(g_1,g_2|x_1) \, \alpha(g_1,g_2|x_2) \, \alpha(g_2|x_1,x_2) \, \alpha(g_1|{}^{g_2} x_1, {}^{g_2} x_2),
\end{split}
\end{equation}
for every $g_1,g_2,x_1,x_2 \in G$, which guarantees that the monoidal structure is well defined. Finally, the braiding $R_{V_1,V_2} \colon V_1 \otimes V_2 \xrightarrow{\sim} V_2 \otimes V_1$ is given by $v_1 \otimes v_2 \mapsto x \cdot v_2 \otimes v_1$, for every $v_1 \in (V_1)_x$ and $v_2 \in V_2$.
One can verify that the non-degenerate braided fusion category $\Mod(\Tu_G^{G,\alpha})$ is equivalent to $\VGGt$, which is itself equivalent to $\mc Z(\Vect_G^\alpha)$ (see e.g. \cite{Majid1991,DAVYDOV2017149,Bullivant:2021pkd}). 

The analysis of $\VGGt$ follows the same steps as for its untwisted counterpart. In particular, it decomposes as 
\begin{equation}
    \VGGt = \bigoplus_{\cl(f) \in \cl(G)} (\VGGt)_{\cl(f)},
\end{equation}
where $(\VGGt)_{\cl(f)}$ is the subcategory of $\VGGt$ consisting of objects $V \in \VGGt$ such that $\supp(V) = \cl(f)$. Given $\cl(f) \in \cl(G)$, there is an equivalence \cite{DIJKGRAAF199160}
\begin{equation}
    (\VGGt)_{\cl(f)} \simeq \Rep^{\alpha(-,-|f)}(Z_G(f)) \simeq \Mod(\mathbb C[Z_G(f)]^{\alpha(-,-|f)}),
\end{equation}
where $\mathbb C[Z_G(f)]^{\alpha(-,-|f)}$ is the group algebra of the centraliser $Z_G(f)$, twisted by the 2-cocycle $\alpha(-,-|f)$. It follows from this equivalence that simple objects in $\VGGt$ are labelled by pairs $(\cl(f),\hat V)$ consisting of $\cl(f) \in \cl(G)$ and $\hat V \in \Irr(\Rep^{\alpha(-,-|f)}(Z_G(f)))$.

\bigskip \noindent
Given a subgroup $H \leq G$, consider as before the algebra $\mathbb C(G/H)$ of functions $\phi \colon G/H \to \mathbb C$ with pointwise multiplication. As an object in $\Vect_G$, it is homogeneous of degree $1_G$, so that $(g \cdot \phi)(-) = \phi(g^{-1}-)$ equips $\mathbb C(G/H)$ with an $\alpha$-projective $G$-action. It follows that $\mathbb C(G/H)$ also provides a condensable algebra in $\VGGt$, which we still denote by $A(H)$. 

Given a $G$-graded vector space $W = \bigoplus_{x \in G}W_x$, one says that an $H$-action on $W$ such that $h \cdot W_x \subseteq W_{hxh^{-1}}$ is $\alpha$-projective if $h_1 \cdot (h_2 \cdot w) = \alpha(h_1,h_2|x) \, (h_1h_2) \cdot w$, for every $h_1,h_2 \in H$ and $w \in W_x$.\footnote{Notice that we are conflating $\alpha(-,-|-) \colon G \times G \times G \to {\rm U(1)}$ and its restriction $\alpha(-,-|-) {\sss |}_{H \times H \times G}$.}. Let $\VGHt$ be the category of $G$-graded vector spaces with $\alpha$-projective $H$-action. The monoidal structure of $\VGHt$ is given by the monoidal structure of $\Vect^\alpha_G$ with the $\alpha$-projective $H$-action provided by $h \cdot (w_1 \otimes w_2) := \alpha(h|x_1,x_2) \, (h \cdot w_1) \otimes (h \cdot w_2)$, for every $w_1 \in (W_1)_{x_1}$ and $w_2 \in (W_2)_{x_2}$.\footnote{Notice that we are conflating $\alpha(-|-,-) \colon G \times G \times G \to {\rm U(1)}$ and its restriction $\alpha(-|-,-) {\sss |}_{H \times G \times G}$.} It is equivalent to the category $\Mod(\Tu_G^{H,\alpha})$ of modules over a twisted tube algebra $\Tu_G^{H,\alpha}$, defined in an obvious way. Proceeding as before, one can then show that $(\VGGt)_{A(H)} \simeq \VGHt$ and $(\VGGt)_{A(H)}^{\rm loc} \simeq \VHHt$, with $\VHHt$ central in $\VGHt$. 

The same way $\VGH$ defines a $\GH$-graded extension of $\VHH$, $\VGHt$ defines a $\GH$-graded extension of $\VHHt$. In particular, one can readily verify that the compatibility condition \eqref{eq:compMonHyperGr} still holds. Let us now derive the categorical $\GH$-action. As before, we consider the endofunctor of $\VGHt$ defined via the composition 
\begin{equation}
    T \colon \VGHt \xrightarrow{\Ind_H^G} \VGGt \xrightarrow{\Res^G_H} \VGHt .
\end{equation}
Here, $\Ind_H^G$ and $\Res^G_H$ are defined in such a way that, for every $W \in \VGHt$,  
\begin{equation}
    T(W) = \CG \otimes^\alpha_{\CH} W,
\end{equation}
where the relative tensor product $\otimes^\alpha_{\CH}$ satisfies 
\begin{equation}
    g \otimes h \cdot w = \alpha(g,h|x) \, gh \otimes w,
\end{equation}
for every $g \in G$, $h \in H$ and $w \in W_x$.
We define a $G$-grading on $T(W)$ by declaring that, for every homogeneous $w \in W$ and $g \in G$, $g \otimes w$ is homogeneous with degree $|g \otimes w| = g |w| g^{-1}$. The $\alpha$-projective $H$-action is provided by $h \cdot (g \otimes w) := \alpha(h,g|x) \, (hg \otimes w)$. 

As  in the untwisted case, the endofunctor $T$ possesses the structure of a monad with unit $\eta \colon {\rm id}_{\VGHt} \to T$ and multiplication $\mu \colon T \circ T \to T$ natural transformations defined in terms of  the unit $\eta \colon {\rm id}_{\VGHt} \to \Res^G_H \circ \Ind_H^G$ and the counit $\epsilon \colon \Ind_H^G \circ \Res_H^G \to {\rm id}_{\VGGt}$ of the adjunction $\Ind_H^G \dashv \Res_H^G$. More precisely, we have 
\begin{align}
    \eta_W &\colon W \to T(W), \q w \mapsto 1_G \otimes w,
    \\
    \mu_W &\colon T(T(W)) \to T(W), \q g_1 \otimes (g_2 \otimes w)\mapsto \alpha(g_1,g_2|x) \, g_1g_2 \otimes w, 
\end{align}
for every $g_1,g_2 \in G$, $W \in \VGHt$ and $w \in W_x$. As in the untwisted case, the endofunctor $T$ decomposes over the hypergroup as
\begin{equation}
    T \cong \bigoplus_{[k] \in \GH} T_{[k]} \equiv \bigoplus_{[k] \in \GH} \mathbb C[HkH] \otimes_{\CH}^\alpha - .
\end{equation}
The summand $T_{[k]}$ implements the action of the hypergroup element $[k]$ and the monad $(T,\mu,\eta)$ encodes the categorical $\GH$-action on $\VGHt$. Notice in particular how the multiplication of the monad depends on $\alpha(-,-|-)$. Consider for instance the following scenario. Suppose the group $G$ admits a proper subgroup $H < G$ and a 3-cocycle $\alpha$ representing a non-trivial class in $H^3(G,{\rm U}(1))$ such that $\alpha(-,-|-)$ is non-trivial and $(\VGGt)_{A(H)}^{\rm loc} \simeq \VHH$. In this case, $\VGHt$ and $\VGH$ encode the localised excitations of two distinct $\GH$-enrichments of the same topological order $\VHH$, which are distinguished by their symmetry fractionalisation, as captured by their respective monads.\footnote{Note that different symmetry fractionalisations imply distinct fusion rings.}
This corroborates the notion that the multiplication of the monad encodes a choice of symmetry fractionalisation. If $\alpha(-,-|-)$ happens to be trivial, the categories $\VGHt$ and $\VGH$ are still distinguished by their monoidal associator, which is interpreted as a choice of \emph{discrete torsion} \cite{KNBalasubramanian:2025vum}.

\subsection{Example\label{sec:Maths_example}}

To conclude this section, we illustrate the general structure outlined above by returning to the example introduced in sec.~\ref{sec:example}. This  amounts to choosing $G = \mathbb D_6$ and $H = \mathbb Z_2$. After providing additional  definitions, we derive the monoidal structure of $\VDZ$, check its compatibility with the hypergroup $\DZ$-grading, compute the hypergroup $\DZ$-action both from the viewpoint of $\VDZ$ and $(\VDD)_{A(\mathbb Z_2)}$, verify the various compatibility conditions, and finally perform explicitly the equivariantisation.

\paragraph{$\VDD$ definitions:} Recall from sec.~\ref{sec:example} that we count eight simple objects in $\VDD$. We introduce the following shorthand notation: 
\begin{gather}
    \label{eq:anyons_short}
    (\cl(1),1) \equiv (1)_1 ,\q  
    (\cl(1),e) \equiv  (1)_e , \q
    (\cl(1),\pi) \equiv  (1)_\pi, 
    \\
    (\cl(s),\pm) \equiv  (s)_\pm, \q (\cl(r),1/\omega/\bar \omega) \equiv  (r)_{1/\omega/\bar \omega}.
\end{gather}
We reproduce below the fusion rules:
\begin{equation}
\begin{split}
    (1)_e \otimes (1)_e &\cong (1)_1 , \q 
    (1)_e \otimes (1)_\pi \cong (1)_\pi
    , \q
    (1)_\pi \otimes (1)_\pi \cong (1)_1 \oplus (1)_e \oplus (1)_\pi ,
    \\ 
    (r)_{1/\omega/\bar \omega} \otimes (1)_e &\cong (r)_{1/\omega/\bar \omega} , \q
    (r)_{1/\omega/\bar \omega} \otimes (r)_{1/\omega/\bar \omega} \cong (1)_1 \oplus (1)_e \oplus (r)_{1/\omega/\bar \omega} ,
    \\ 
    (r)_{1/\omega/\bar \omega} \otimes (1)_\pi &\cong (r)_{\omega/\bar \omega/1}  \oplus (r)_{\bar \omega/1/\omega} , \q
    (r)_{1/\omega/\bar \omega} \otimes (r)_{\omega/\bar \omega/1}  \cong (1)_\pi \oplus (r)_{\bar \omega/1/\omega}  ,
    \\ 
    (1)_e \otimes (s)_{\pm} &\cong (s)_{\mp} ,
    \\ 
    (s)_{\pm} \otimes (s)_{\pm} &\cong (1)_1 \oplus (1)_\pi \oplus (r)_1 \oplus (r)_\omega \oplus (r)_{\bar \omega} ,
    \\ 
    (s)_+ \otimes (s)_- &\cong (1)_e \oplus (1)_\pi \oplus (r)_1 \oplus (r)_\omega \oplus (r)_{\bar \omega} ,
    \\ 
    (r)_{1/\omega/\bar \omega} \otimes (s)_{\pm} &\cong (1)_\pi \otimes (s)_{\pm} \cong (s)_+ \oplus (s)_- .
\end{split}
\end{equation}
The topological spins are given by
\begin{equation}
\begin{split}
\theta_{(1)_1} &= \theta_{(1)_e} = \theta_{(1)_\pi} = \theta_{(s)_+} = \theta_{(r)_1} = 1 ,
\\
\theta_{(s)_-} &= -1 , \q \theta_{(r)_\omega} = e^\frac{2 \pi i}{3} , \q \theta_{(r)_{\bar \omega}} = e^{-\frac{2 \pi i}{3}} .
\end{split}
\end{equation}
Finally, we need to list some non-trivial $F$-symbols that are used in the following (see ref.~\cite{Cui:2014sfa}). Note that the fusion category $\VDD$ is multiplicity free. The matrices below are expressed in the same bases as those used in sec.~\ref{sec:example}. The non-trivial $F$-symbols involving the $\Rep(\mathbb{D_6})$ subcategory are 
\begin{equation}
    F^{(1)_e(1)_\pi (1)_\pi}_{(1)_\pi} = F^{(1)_\pi(1)_\pi (1)_\pi}_{(1)_e} = -1 ,\q F^{(1)_\pi(1)_\pi (1)_\pi}_{(1)_\pi} = \frac{1}{2}\begin{pmatrix}
       1 & 1 & \sqrt{2} \\   1 & 1 & -\sqrt{2} \\  \sqrt{2} & -\sqrt{2} & 0
    \end{pmatrix} .
\end{equation}
The $F$-symbols involving $(s)_\pm$ and $(1)_\pi$ are 
\begin{equation}
    F^{(s)_+(1)_\pi (1)_\pi}_{(s)_+} = F^{(s)_-(1)_\pi (1)_\pi}_{(s)_-} = \frac{1}{\sqrt{2}}\begin{pmatrix}
       1 & 1  \\   1 & -1 
    \end{pmatrix} , \q F^{(s)_+(1)_\pi (1)_\pi}_{(s)_-} = F^{(s)_-(1)_\pi (1)_\pi}_{(s)_+} = \frac{1}{\sqrt{2}}\begin{pmatrix}
       1 & -1  \\   1 & 1 
    \end{pmatrix} .
\end{equation}
The $F$-symbols involving $(r)_{1/\omega/\bar \omega}$ and $(1)_\pi$ are 
\begin{equation}
     F^{(r)_1(1)_\pi (1)_\pi}_{(r)_1}  = F^{(r)_{\bar \omega}(1)_\pi (1)_\pi}_{(r)_{\bar \omega}}= \frac{1}{\sqrt{2}}\begin{pmatrix}
       1 & -1  \\   1 & 1 
    \end{pmatrix} , \q F^{(r)_\omega(1)_\pi (1)_\pi}_{(r)_\omega}= \frac{1}{\sqrt{2}}\begin{pmatrix}
       1 & 1  \\   1 & -1 
    \end{pmatrix} .
\end{equation} 

\bigskip\noindent
\paragraph{$\VDZ$ definitions:}  Recall that simple objects in $\VDZ$ are labelled by pairs $(\text{cl}_{\Z_2}(f),\hat W)$ consisting of $\text{cl}_{\Z_2}(f) \in \text{cl}_{\Z_2}(\mathbb D_6)$ and $\hat W \in \Irr(\Rep(\Stab_{\mathbb Z_2}(f)))$. There are four $\mathbb Z_2$-conjugacy classes: $\cl_{\Z_2}(1)$, $\cl_{\Z_2}(s)$, $\cl_{\Z_2}(r)$ and $\cl_{\Z_2}(rs)$ with fixed representatives. Moreover $\Stab_{\Z_2}(f) = \mathbb Z_2$ for $f=1,s$ and $\Stab_{\Z_2}(f) = \mathbb Z_1$ for $f=r,rs$. Therefore, $\Vect_{\mathbb D_6}^{\mathbb Z_2}$ has six simple objects. Writing $\Irr(\Rep(\Z_2)) = \{\mathbb C_+,\mathbb C_-\}$ with $\mathbb C_\pm=\mathbb C\{\hat{w}_\pm\}$, the six simple objects are given by $(\cl_{\Z_2}(1),\mathbb C_+) = \mathbb C[\mathbb Z_2]\otimes_{\mathbb C[\mathbb Z_2]} \mathbb C_+ =\mathbb C \{q_1 \otimes \hat{w}_+ \}\equiv\mathbb C\{w_1^+\}$, $(\cl_{\Z_2}(1),\mathbb C_-)=\mathbb C \{q_1 \otimes \hat{w}_- \}\equiv\mathbb C\{w_1^-\}$, $(\cl_{\Z_2}(s),\mathbb C_+)=\mathbb C \{q_s \otimes \hat{w}_+ \}\equiv\mathbb C\{w_s^+\}$, $(\cl_{\Z_2}(s),\mathbb C_-) =\mathbb C \{q_s \otimes \hat{w}_- \}\equiv\mathbb C\{w_s^-\}$, $c_2 =(\cl_{\Z_2}(r),\mathbb C_+) = \mathbb C[\mathbb Z_2]\otimes_{\mathbb C[\mathbb Z_1]}\mathbb C_+ = \mathbb C\{q_r\otimes \hat{w}_+,q_{r^2}\otimes \hat{w}_+\}\equiv\mathbb C\{w_r^+,w_{r^2}^+\}$ and $c_1 = (\cl_{\Z_2}(rs),\mathbb C_+) = \mathbb C\{q_{rs}\otimes \hat{w}_+,q_{r^2s}\otimes \hat{w}_+\}\equiv\mathbb C\{w_{rs}^+,w_{r^2s}^+\}$. In terms of the notation of sec.~\ref{sec:example}, we identify
\begin{gather}
\label{eq:VDZ_short}
    \bm{1} \equiv\mathbb C\{w_1^+\},\q \bm{e} \equiv\mathbb C\{w_1^-\},\q \bm{m}\equiv\mathbb C\{w_s^+\} ,\q \bm{f}\equiv\mathbb C\{w_s^-\} , 
    \\
    \bm{c_1}\equiv\mathbb C\{w_{rs}^+,w_{r^2s}^+\},\q \bm{c_2}\equiv\mathbb C\{w_r^+,w_{r^2}^+\}.
\end{gather}
The basis vector appearing above are homogeneous of degree $|w_1^\pm| = 1$, $|w_s^\pm| = s$, $|w_r^+| = r$, $|w_{r^2}^+| = r^2$, $|w_{rs}^+| = rs$, $|w_{r^2s}^+| = r^2s$. Moreover, the $\Z_2$-action is $s \cdot w_{1}^{\pm} = \pm w_1^\pm$, $s \cdot w_s^\pm = \pm w_s^\pm$, $s \cdot \{w_r^+,w_{r^2}^+\} = \{w_{r^2}^+,w_r^+\}$ and $s \cdot \{w_{rs}^+,w_{r^2s}^+\} = \{w_{r^2s}^+,w_{rs}^+\}$. 

Before proceeding further, we rederive below this classification of simple objects by explicitly computing modules over the relevant condensable algebra in $\VDD$.The methods that we use here can be applied to more general parent topological orders than $\VGG$.

\bigskip \noindent
\paragraph{$(\VDD)_{A(\mathbb Z_2)}$ formulation:}
Recall that the condensable algebra is $A(\mathbb Z_2)$ is provided by the algebra $\mathbb C(\mathbb D_6 / \mathbb Z_2)$ of functions on $\mathbb D_6 / \mathbb Z_2$  with pointwise multiplication. As an object in $\VDD$, $A(\mathbb Z_2) \cong (1)_1 \oplus (1)_\pi$. 
Let us fix orthonormal bases 
\begin{equation}
    \Hom_{\VDD}(X_1 \otimes X_2,X_3) = \mathbb C\{ \varphi_{X_1 X_2}^{X_3} \}.
\end{equation}
In terms of these bases, we write explicitly the multiplication map $\mu \colon A(\mathbb Z_2) \otimes A(\mathbb Z_2) \to A(\mathbb Z_2) $  as 
\begin{equation}\label{eq:A_mult}
    \mu = \varphi_{\ay{1}{1}\ay{1}{1}}^{\ay{1}{1}} \oplus \varphi_{\ay{1}{\pi}\ay{1}{1}}^{\ay{1}{\pi}} \oplus \varphi_{\ay{1}{1}\ay{1}{\pi}}^{\ay{1}{\pi}} \oplus \sqrt{2} \,  \varphi_{\ay{1}{\pi} \ay{1}{\pi}}^{\ay{1}{1}} \oplus \varphi_{\ay{1}{\pi} \ay{1}{\pi}}^{\ay{1}{\pi}} .
\end{equation}
Let us justify this formula, working in fusion subcategory $\Rep(\mathbb D_6)$ for simplicity.  We  write $1 = \mathbb C \{v_1\},$ $\pi = \mathbb C\{v_{\pi,1},v_{\pi,2}\}$. We have $r \cdot v_{\pi,1} = \omega \,v_{\pi,1}$, $r \cdot v_{\pi,2} = \bar \omega \,v_{\pi,2}$ and $s \cdot v_{\pi,1/2}  = \, v_{\pi,2/1}$. Moreover, $\mathbb C(\mathbb D_6/\mathbb Z_2) = \mathbb C\{\delta_{\Z_2},\delta_{r\Z_2},\delta_{r^2\Z_2}\}$ with $r \cdot \{\delta_{\Z_2},\delta_{r\Z_2},\delta_{r^2\Z_2}\} = \{\delta_{r\Z_2},\delta_{r^2 \Z_2},\delta_{\Z_2}\}$ and $s \cdot \{\delta_{\Z_2},\delta_{r\Z_2},\delta_{r^2\Z_2}\} = \{\delta_{\Z_2},\delta_{r^2 \Z_2},\delta_{r\Z_2}\}$. It follows that 
\begin{equation}
\begin{split}
    \mathbb C(\mathbb D_6/\mathbb Z_2) &\cong \Ind_{\mathbb Z_2}^{\mathbb D_6}(1) \cong 1 \oplus \pi \\
    &\cong \mathbb C\{\delta_{\Z_2} + \delta_{r\Z_2} + \delta_{r^2 \Z_2}\} \oplus \mathbb C\{\delta_{\Z_2} + \bar \omega \delta_{r\Z_2} + \omega \delta_{r^2 \Z_2}, \delta_{\Z_2}+\omega \delta_{r\Z_2} + \bar \omega \delta_{r^2 \Z_2}\}
\end{split}
\end{equation}
and
\begin{equation}
\begin{split}
    \pi \otimes \pi &\cong 1 \oplus e \oplus \pi
    \\
    & \cong \mathbb C\{\frac{1}{\sqrt 2}v_{\pi,1}\otimes v_{\pi,2} + \frac{1}{\sqrt 2}v_{\pi,2} \otimes v_{\pi,1}\}
    \oplus
    \mathbb C\{-\frac{1}{\sqrt 2}v_{\pi,1} \otimes v_{\pi,2} + \frac{1}{\sqrt 2}v_{\pi,2} \otimes v_{\pi,1}\}
    \\
    & \q\, \oplus \mathbb C\{v_{\pi,2} \otimes v_{\pi,2} ,v_{\pi,1} \otimes v_{\pi,1}\}.
\end{split}
\end{equation}
We choose this decomposition to define the basis maps appearing in eq.~\eqref{eq:A_mult}. In this basis, the multiplication rule of $\mathbb C(\mathbb D_6/\mathbb Z_2)$ is simply given by $\delta_{r_1 \Z_2} \otimes \delta_{r_2\Z_2} \mapsto \delta_{r_1,r_2} \delta_{r_1\Z_2}$. Using the decomposition above, under this multiplication, we have
\begin{align}
    v_{\pi,1}  \otimes v_{\pi,1}  \mapsto v_{\pi,2}
    ,\q
    v_{\pi,1} \otimes v_{\pi,2} \mapsto v_1
    ,\q
    v_{\pi,2} \otimes v_{\pi,1}  \mapsto v_1
    ,\q
    v_{\pi,2} \otimes v_{\pi,2}  \mapsto v_{\pi,1} ,
\end{align}
which when expressed in the intertwiners bases above produce eq.~\eqref{eq:A_mult}.

Let us now  compute the simple right $A(\Z_2)$-modules $(M,\rho)$ with $\rho: M \otimes A(\Z_2) \rightarrow M$. We list below the four simple local modules and the two (non-local) simple modules: 
\begin{itemize}
    \item[\bul] $(1)_1 \oplus (1)_\pi$ gives one local module $\bm 1$ in $(\VDD)_{A(\mathbb Z_2)}$ with module action $\rho = \mu$.
    \item[\bul] $(1)_e \oplus(1)_\pi$ gives one module $\bm e$ with module action 
    \begin{equation}
        \rho = \varphi_{\ay{1}{e}\ay{1}{1}}^{\ay{1}{e}} \oplus \varphi_{\ay{1}{\pi}\ay{1}{1}}^{\ay{1}{\pi}} \oplus \varphi_{\ay{1}{e} \ay{1}{\pi}}^{\ay{1}{\pi}} \oplus \sqrt 2 \varphi_{\ay{1}{\pi} \ay{1}{\pi}}^{\ay{1}{e}} \oplus \big( - \varphi_{\ay{1}{\pi} \ay{1}{\pi}}^{\ay{1}{\pi}} \big).
    \end{equation}
    It follows from $R_{\ay{1}{\pi},\ay{1}{\pi}}^{\ay{1}{e}} = -1$ and $R_{\ay{1}{e},\ay{1}{\pi}}^{\ay{1}{\pi}} = R_{\ay{1}{\pi},\ay{1}{e}}^{\ay{1}{\pi}} =-1$ that it is a local module.
    \item[\bul] $(s)_+$ gives one module $\bm m$ with module action 
    \begin{equation}
        \rho = \varphi_{\ay{s}{+}\ay{1}{1}}^{\ay{s}{+}} \oplus \big(-\frac{1}{\sqrt{2}} \varphi_{\ay{s}{+}\ay{1}\pi}^{\ay{s}{+}} \big),
    \end{equation}
    which is also local thanks to $R_{\ay{s}{+},\ay{1}{\pi}}^{\ay{s}{+}} = R_{\ay{1}{\pi},\ay{s}{+}}^{\ay{s}{+}} = 1$.
    \item[\bul] $(s)_-$ gives one module $\bm f$ with module action 
    \begin{equation}
        \rho
        = \varphi_{\ay{s}{-}\ay{1}{1}}^{\ay{s}{-}} \oplus \frac{1}{\sqrt{2}}\varphi_{\ay{s}{-}\ay{1}\pi}^{\ay{s}{-}},
    \end{equation}
    which can also be checked to be local.
    \item[\bul] $(s)_+ \oplus (s)_-$ gives one module $\bm{c_1}$ with module action 
    \begin{equation}
        \rho = \varphi_{\ay{s}{+}\ay{1}{1}}^{\ay{s}{+}} \oplus \varphi_{\ay{s}{-}\ay{1}{1}}^{\ay{s}{-}} \oplus \big( -\frac{1}{\sqrt{2}} \varphi_{\ay{s}{+}\ay{1}{\pi}}^{\ay{s}{+}}\big) \oplus \frac{1}{\sqrt{2}} \varphi_{\ay{s}{-}\ay{1}{\pi}}^{\ay{s}{-}} \oplus {\frac{\sqrt{3}}{\sqrt{2}}}\varphi_{\ay{s}{+}\ay{1}{\pi}}^{\ay{s}{-}} \oplus {\frac{\sqrt{3}}{\sqrt{2}}}\varphi_{\ay{s}{-}\ay{1}{\pi}}^{\ay{s}{+}}.
    \end{equation}
    This is the first module that is not local since 
    \begin{equation}
        R_{\ay{s}{-},\ay{1}{\pi}}^{\ay{s}{+}} = R_{\ay{1}{\pi},\ay{s}{-}}^{\ay{s}{+}} = i \q \text{and} \q R_{\ay{s}{+},\ay{1}{\pi}}^{\ay{s}{-}} = R_{\ay{1}{\pi},\ay{s}{+}}^{\ay{s}{-}} = -i.
    \end{equation}
        Notice that as an object in $\VDD$, $\bm{c_1}$ is isomorphic to $\bm m \oplus \bm f$,  but not as an object in $(\VDD)_{A(\mathbb Z_2)}$ as their module structure differs.
    \item[\bul] $(r)_1\oplus (r)_\omega \oplus (r)_{\bar \omega}$ gives one module $\bm{c_2}$ with action
    \begin{equation}
    \begin{split}
        \rho = &\varphi_{\ay{r}{1}\ay{1}{1}}^{\ay{r}{1}} \oplus \varphi_{\ay{r}{\omega}\ay{1}{1}}^{\ay{r}{\omega}} \oplus \varphi_{\ay{r}{\bar \omega}\ay{1}{1}}^{\ay{r}{\bar \omega}} \oplus \varphi_{\ay{r}{1}\ay{1}{\pi}}^{\ay{r}{\omega}} \oplus \varphi_{\ay{r}{1}\ay{1}{\pi}}^{\ay{r}{\bar \omega}}  \oplus \varphi_{\ay{r}{\omega}\ay{1}{\pi}}^{\ay{r}{1}} \oplus \varphi_{\ay{r}{\omega}\ay{1}{\pi}}^{\ay{r}{\bar \omega}} 
        \\
        &\oplus \varphi_{\ay{r}{\bar \omega}\ay{1}{\pi}}^{\ay{r}{1}} 
        \oplus \varphi_{\ay{r}{\bar \omega}\ay{1}{\pi}}^{\ay{r}{\omega}} ,
    \end{split}
    \end{equation}
     which is not local because e.g., $R_{\ay{r}{1},\ay{1}{\pi}}^{\ay{r}{\omega}} = R_{\ay{1}{\pi},\ay{r}{1}}^{\ay{r}{\omega}} = \bar \omega$.
\end{itemize}
In summary, we get four local modules $\{\bm1,\bm e,\bm m,\bm f \}$ and two non-local modules $\{ \bm{c_1},\bm{c_2}\}$.
The module action should be associative, i.e., for a module $(M,\rho)$ we should verify that 
\begin{equation}
    \rho \circ (\rho \otimes 1_A) = \rho \circ (1_M \otimes \mu) \circ (\alpha_{M,A,A}) , 
\end{equation}
where $\alpha$ denotes the module associator $\VDD$.
One can check this is indeed the case by working component-wise in terms of the $F$-symbols, the structure constants of the algebra, and the constants defining the module action.
It is useful to remember that all the simple objects in $(\VDD)_{A(\mathbb Z_2)}$ can be realised  as a direct summand of $X \otimes A$, with $X$ a simple object in $\VDD$. The right action is defined in this case as $  (1_X \otimes \mu) \circ \alpha_{X,A,A}$. In particular, we have 
\begin{equation}
\begin{aligned}
    (1)_1 &\otimes A(\Z_2) \cong \bm 1 ,\q  (1)_e \otimes A(\Z_2) \cong \bm e  ,\q (1)_\pi \otimes A(\Z_2) \cong \bm 1 \oplus \bm e ,\q (s)_+ \otimes A(\Z_2) \cong\bm m \oplus \bm{c_1} 
    \\ 
    (s)_- &\otimes A(\Z_2) \cong \bm f \oplus \bm{c_1} ,\q  (r)_1 \otimes A(\Z_2) \cong (r)_\omega \otimes A(\Z_2) \cong (r)_{\bar \omega} \otimes A(\Z_2)  \cong \bm {c_2} 
\end{aligned} 
\end{equation}
as right $A(\Z_2)$-modules. 

Fusion rules in $(\VDD)_{A(\Z_2)}$ are given by the relative tensor product $\otimes_{A(\Z_2)}$, which we can compute using some simplifying tricks. For example, let us consider 
\begin{equation}
   \big((1)_e \otimes A(\Z_2)\big) \otimes_{A(\Z_2)} \big((1)_e \otimes A(\Z_2)\big) \cong (1)_1 \otimes A(\Z_2)
\end{equation}
immediately gives $\bm e \otimes\bm  e \cong \bm 1$. Now consider 
\begin{equation}
   \big((1)_e \otimes A(\Z_2) \big) \otimes_{A(\Z_2)} \big( (s)_+ \otimes A(\Z_2) \big) \cong (s)_- \otimes A(\Z_2) ,
\end{equation}
which implies  $\bm e \otimes (\bm m \oplus \bm{c_1}) \cong \bm f \oplus \bm{c_1}$. Quantum dimension arguments imply $\bm e \otimes \bm m \cong \bm f$ and $\bm e \otimes \bm{c_1} \cong \bm{c_1}$. Similarly, one can compute $\bm e \otimes \bm{c_2} = \bm{c_2}$. Now we can consider 
\begin{equation}
\begin{aligned}
    & \big((s)_+ \otimes A(\Z_2)\big) \otimes_{A(\Z_2)} \big( (s)_+ \otimes A(\Z_2) \big) \cong (s)_+ \otimes (s)_+ \otimes A(\Z_2) \cong \\
    &\big((1)_1 \oplus (1)_\pi \oplus (r)_1 \oplus (r)_\omega \oplus (r)_{\bar \omega} \big)\otimes A(\Z_2) .
\end{aligned}
\end{equation}
This implies $(\bm m \oplus \bm{c_1}) \otimes (\bm m\oplus \bm{c_1}) \cong \bm 1\oplus\bm 1\oplus\bm e\oplus 3 \,\bm{c_2}$. 
Associativity with the fusion rules we determined earlier fixes $\bm m \otimes \bm m \cong \bm 1$, $\bm m \otimes \bm{c_1}  \cong \bm{c_2}$, $\bm{c_1} \otimes \bm{c_1} \cong \bm 1 \oplus \bm e \oplus \bm{c_2}$. Now consider
\begin{equation}
     \big( (r)_1 \otimes A(\Z_2) \big) \otimes_{A(\Z_2)} \big( (r)_1 \otimes A(\Z_2) \big) \cong \big( (1)_1 \oplus(1)_e \oplus (r)_1 \big) \otimes A(\Z_2) ,
\end{equation}
which implies $\bm{c_2} \otimes \bm{c_2} \cong \bm 1 \oplus \bm e \oplus \bm{c_2}$. The remaining fusion rules $\bm m \otimes \bm{c_2} \cong \bm{c_1}$ and $\bm{c_1} \otimes \bm{c_2} \cong \bm m \oplus \bm f \oplus \bm{c_1}$ follow from 
\begin{equation}
    \big( (s)_+\otimes A(\Z_2) \big) \otimes_{A(\Z_2)}\big( (r)_1 \otimes A(\Z_2) \big) \cong \big( (s)_+ \oplus (s)_- \big)\otimes A(\Z_2).
\end{equation}
We summarise these (non-trivial) fusion rules below:
\begin{equation}\label{eq:VDZ_fusions}
\begin{aligned}
    &\bm{e} \otimes \bm{e} \cong \bm{m} \otimes \bm{m} \cong \bm{1},\q \bm{e} \otimes \bm{m} \cong \bm{f} ,\q \bm{e} \otimes \bm{c_1} \cong \bm{c_1} ,\q \bm{e} \otimes \bm{c_2} \cong \bm{c_2} ,\q \bm{m} \otimes \bm{c_1} \cong \bm{c_2}  \\
    &\bm{m} \otimes \bm{c_2} \cong \bm{c_1} ,\q \bm{c_1} \otimes \bm{c_1} \cong \bm{c_2} \otimes \bm{c_2} \cong \bm{1} \oplus \bm{e} \oplus \bm{c_2} ,\q \bm{c_1} \otimes \bm{c_2} \cong \bm{m} \oplus \bm{f} \oplus \bm{c_1}.
\end{aligned}    
\end{equation}
Going back to the $\VDZ$ formulation, let us now study the hypergroup grading and hypergroup equivariant structure of the condensed theory.

\bigskip \noindent
\paragraph{Hypergroup grading:} The hypergroup grading of $\VDZ$ is given by the double coset $\mathbb{Z}_2 \setminus \mathbb{D}_6 / \mathbb{Z}_2 \equiv \DZ$. Explicitly, there are two double coset elements $[1]  = \mathbb{Z}_2 1 \mathbb{Z}_2 = \{ 1,s\}$ and $[r] = \mathbb{Z}_2 r \mathbb{Z}_2 =\{ r,r^2,sr,sr^2\}$.  This can be given the structure of an hypergroup with the usual multiplication
\begin{equation}
    [k_1] \star [k_2] = \frac{1}{2} \sum_{h \in \{1,s\}} \{1,s\} (k_1 h k_2) \{1,s\}.
\end{equation}
One can check $[1]$ is the identity and the non-trivial fusion rule reads $[r] \star [r] = \frac{1}{2} [1] + \frac{1}{2} [r]$. Then the hypergroup grading of $\VDZ$ is as follows: $\{\bm{1},\bm e,\bm m,\bm f \} \in \Irr (\Vect^{\mathbb Z_2}_{[1]}) = \Irr (\Vect^{\mathbb Z_2}_{\mathbb Z_2})$, while $\{ \bm {c_1},\bm {c_2}\} \in \Irr (\Vect^{\mathbb Z_2}_{[r]})$. The compatibility between monoidal structure of $\VDZ$ and hypergroup grading \eqref{eq:compMonHyperGr} is immediate to check for the two non-trivial structure constants $C^{[r][r]}_{[1]} =C^{[r][r]}_{[r]} =1/2 $  given the fusion rules listed in eq.~\eqref{eq:VDZ_fusions}

\bigskip \noindent
\paragraph{Hypergroup action on $\VDZ$:} Let us work out explicitly the hypergroup $\DZ$ symmetry action on $\VDZ$. 
Following the general discussion in sec.\ \ref{sec:Maths_monad}, the action $T_{[k]}$ of the hypergroup element $[k]$ on an element $W \in \VDZ$ is given by 
\begin{equation}
    T_{[k]}(W) \equiv \mathbb C [\Z_2 k \Z_2] \otimes_{\mathbb{C}[\Z_2]} \otimes W .
\end{equation}
Notice that by definition 
\begin{equation}
    T_{[1]} (W) := \mathbb C[\Z_2] \otimes_{\mathbb C[\Z_2]} W \cong W .
\end{equation}
Let us now compute the action $T_{[r]}$ of the non-trivial hypergroup element in $\DZ$. To do so, recall the defining property of the relative tensor product over $\mathbb C[\Z_2]$:
\begin{equation}
    h \otimes w = 1_{\mathbb D_6} \otimes (h \cdot w) ,
\end{equation}
for every $h \in \Z_2$, 
and that the grading of $g \otimes w$ is $|g \otimes w|=g |w| g^{-1}$ for every homogenous $w \in W$. Moreover, the $\Z_2$-action reads $h \cdot (g \otimes w)=hg \otimes w$. 
 
We start by the action on $\bm{1}$:
\begin{equation}
    T_{[r]} (\bm 1) \cong \mathbb C \{ r, r^2, rs , r^2s\} \otimes_{\mathbb C[\Z_2]} \mathbb C\{w_1^+\} .
\end{equation}
Notice that since $rs \otimes w_1^+ = r \otimes (s \cdot w_1^+ )= r \otimes w_1^+$ and similarly $r^2s \otimes w_1^+ = r^2 \otimes w_1^+$, we have only two basis vectors, namely $r \otimes w_1^+$ and $r^2 \otimes w_1^+$, with grading $|r \otimes w_1^+|=1$, $|r^2 \otimes w_1^+|=1$. Moreover, the two are swapped by $s$, as $s \cdot (r \otimes w_1^+) = sr \otimes w_1^+ = r^2s \otimes w_1^+ = r^2 \otimes w_1^+$. Thus, by forming even and odd combinations, we identify 
\begin{equation}
    T_{[r]} (\bm 1) \cong \mathbb C \{r \otimes w_1^+ + r^2 \otimes w_1^+ \} \oplus \mathbb C \{r \otimes w_1^+ - r^2 \otimes w_1^+ \} \cong \bm 1 \oplus \bm e.
\end{equation}
The action on $\bm e$
\begin{equation}
    T_{[r]} (\bm e) \cong \mathbb C \{ r, r^2, rs , r^2s\} \otimes_{\mathbb C[\Z_2]} \mathbb C\{w_1^-\}
\end{equation}
can be computed very similarly. Here $s \cdot (r \otimes w_1^-) = - r^2 \otimes w_1^-$ and $s \cdot (r^2 \otimes w_1^-) = - r \otimes w_1^-$. Thus we find again 
\begin{equation}
    T_{[r]} (\bm e) \cong \mathbb C \{r \otimes w_1^- - r^2 \otimes w_1^- \} \oplus \mathbb C \{r \otimes w_1^- + r^2 \otimes w_1^- \} \cong \bm1 \oplus \bm e .
\end{equation}
Now, we compute the action on $\bm m$
\begin{equation}
    T_{[r]} (\bm m) \cong \mathbb C \{ r, r^2, rs , r^2s\} \otimes_{\mathbb C[\Z_2]} \mathbb C\{w_s^+\}.
\end{equation}
Again we have two basis vectors $r \otimes w_s^+$ and $r^2 \otimes w_s^+$ with grading $|r \otimes w_s^+|=r^2 s$ and $|r^2 \otimes w_s^+| = rs$. Therefore, we identify 
\begin{equation}
    T_{[r]}(\bm m) \cong \mathbb C \{r \otimes w_s^+ , r^2 \otimes w_s^+\} \cong \bm{c_1} .
\end{equation}
Similarly, for $\bm f$ we have 
\begin{equation}
    T_{[r]}(\bm f) \cong  \mathbb C \{ r, r^2, rs , r^2s\} \otimes_{\mathbb C[\Z_2]} \mathbb C\{w_s^-\} \cong \mathbb C \{r \otimes w_s^- , -r^2 \otimes w_s^-\} \cong \bm{c_1} .
\end{equation}
Finally, for the two $[r]$-twisted sectors $\bm{c_1}$ and $\bm{c_2}$, we have 
\begin{equation}
\begin{aligned}
    T_{[r]} (\bm{c_1}) &\cong  \mathbb C \{ r, r^2, rs , r^2s\} \otimes_{\mathbb C[\Z_2]} \mathbb C\{w_{rs}^+,w_{r^2s}^+ \} \\
    &\cong \mathbb C \{ r \otimes w_{rs}^+ + r^2 \otimes w_{r^2s}^+ \} \oplus \mathbb C \{ r \otimes w_{rs}^+ - r^2 \otimes w_{r^2s}^+ \} \oplus \mathbb C \{ r^2 \otimes w_{rs}^+ , r \otimes w_{r^2s}^+ \} \\ 
    &\cong \bm m \oplus \bm f \oplus \bm{c_1} 
\end{aligned}    
\end{equation}
and 
\begin{equation}
\begin{aligned}
    T_{[r]} (\bm{c_2}) &\cong  \mathbb C \{ r, r^2, rs , r^2 s\} \otimes_{\mathbb C[\Z_2]} \mathbb C\{w_{r}^+,w_{r^2}^+ \} \\
    &\cong \mathbb C \{ r \otimes w_{r}^+, r^2 \otimes w_{r^2}^+\} \oplus C \{ r^2 \otimes w_{r}^+, r \otimes w_{r^2}^+\} \cong \bm{c_2} \oplus \bm{c_2} .
\end{aligned}    
\end{equation}
To check the compatibility between hypergroup grading and hypergroup action, note that in this case the non-trivial adjoint structure constants are $A^{[r][r]}_{[1]}=1/4$ and $A^{[r][r]}_{[r]}=3/4$. Then eq.~\eqref{eq:compAdHyperGr} follows immediately by considering $T_{[r]} (R_{[r]})$, where the regular object in the $[r]$-sector is given by $R_{[r]} =\bm{c_1}\oplus \bm{c_2}$.

\bigskip \noindent
\paragraph{Hypergroup action on $(\VDD)_{A(\mathbb Z_2)}$:} Let us reproduce the hypergroup $\VDZ$-equivariant structure from the viewpoint of $(\VDD)_{A(\mathbb Z_2)}$. We do so in two equivalent ways, the latter being more generic. We have the following $A(\mathbb Z_2)$-bimodule decomposition:
\begin{equation}
    \mathbb{C}(\mathbb{D}_6 / \mathbb{Z}_2) \otimes \mathbb{C}(\mathbb{D}_6 / \mathbb{Z}_2) \cong \bigoplus_{[k] \in \DZ} \mathbb C(\mathbb D_6/(\Z_2)_{k})\cong \mathbb{C}(\mathbb{D}_6 / \mathbb{Z}_2) \oplus \mathbb{C}(\mathbb{D}_6 ) ,
\end{equation}
where $\mathbb{C}(\mathbb{D}_6 / \mathbb{Z}_2)$ corresponds to the trivial hypergroup element $[1]$ and $\mathbb{C}(\mathbb{D}_6 )$ corresponds to the non-trivial element $[r]$. We then define the action of $T_{[k]}$ on a right $A(\Z_2)$-module $(M,\rho)$  as
\begin{equation}
    T_{[k]} \big((M,\rho)\big) = M \otimes_{A(\mathbb Z_2)} \mathbb C(\mathbb D_6/(\Z_2)_{k}) .
\end{equation}
We now compute  the action of the non-trivial component $T_{[r]}$. To do this, we can use the fact that the $A(\mathbb Z_2)$-bimodule $\mathbb{C}(\mathbb{D}_6 )$ decomposes as $ (1)_\pi\otimes A(\Z_2)$ as a right $A(\mathbb Z_2)$-module. To verify this claim, let us work out the $A(\mathbb Z_2)$-bimodule structure of $\mathbb C(\mathbb D_6)$ as explicitly as possible. The decomposition of the regular representation $\mathbb C(\mathbb D_6) = \mathbb C\{\delta_1,\delta_r,\delta_{r^2},\delta_s,\delta_{rs},\delta_{r^2s}\}$ into irreducible $\mathbb D_6$-representations reads  
\begin{equation}
\begin{split}
    \mathbb C(\mathbb D_6) 
    &\cong \mathbb C\{\delta_1 + \delta_s, \delta_r + \delta_{rs}, \delta_{r^2} + \delta_{r^2s}\} \oplus \mathbb C\{\delta_1 - \delta_s, \delta_r - \delta_{rs}, \delta_{r^2} - \delta_{r^2s}\}
    \\
    &\cong \Ind_{\mathbb Z_2}^{\mathbb D_6}(1) \oplus \Ind_{\mathbb Z_2}^{\mathbb D_6}(e) \cong (1 \oplus \pi) \oplus (e \oplus \pi).
\end{split}
\end{equation}
Next, recall that the $A(\mathbb Z_2)$-bimodule structure of $\mathbb C(\mathbb D_6/(\mathbb Z_2)_k)$ is given by $(\phi_1 \cdot \phi \cdot \phi_2)(g(\mathbb Z_2)_k) = \phi_1(g\Z_2) \phi(g(\Z_2)_k) \phi_2(gk\Z_2)$. Here $\mathbb C(\mathbb D_6) = \mathbb C(\mathbb D_6 / (\mathbb Z_2 \cap r \mathbb Z_2 r^{-1}))$, so $k=r$ and $(\mathbb Z_2)_r = \mathbb Z_1$. 
As a left $A(\mathbb Z_2)$-module in $\VDZ$, $\mathbb C(\mathbb D_6$) decomposes as $\big((1)_1 \oplus (1)_\pi\big) \oplus \big((1)_e \oplus (1)_\pi\big) \cong (1)_\pi \otimes A(\Z_2)$ with respect to the above decomposition. As a right $A(\mathbb Z_2)$-module, it also decomposes as $(1)_\pi \otimes A(\Z_2)$, but with respect to a distinct decomposition, namely
\begin{equation}
\begin{split}
    \mathbb C(\mathbb D_6) 
    &\cong \mathbb C\{\delta_{r^2} + \delta_{rs}, \delta_1 + \delta_{r^2s}, \delta_{r} + \delta_{s}\} \oplus \mathbb C\{\delta_{r^2} - \delta_{rs}, \delta_1 - \delta_{r^2s}, \delta_{r} - \delta_{s}\} ,
\end{split}
\end{equation}
so that it is indeed indecomposable as a bimodule.

Going back to the hypergroup action, we start by computing 
\begin{equation}
    \big( (1)_1 \otimes A(\Z_2)\big) \otimes_{A(\Z_2)} \mathbb C(\mathbb D_6) \cong (1)_1 \otimes \big((1)_\pi \otimes A(\Z_2)\big) \cong (1)_\pi \otimes A(\Z_2),
\end{equation}
which implies $T_{[r]}(\bm 1) = \bm 1 \oplus \bm e$. Similarly, one can show $T_{[r]}(\bm e) = \bm 1 \oplus \bm e$ and  $T_{[r]}(\bm{c_2}) =  \bm{c_2} \oplus \bm{c_2}$.
Then consider 
\begin{equation}
     \big( (s)_+ \otimes A(\Z_2) \big) \otimes_{A(\Z_2)} \mathbb C(\mathbb D_6)\cong (s)_+ \otimes \big( (1)_\pi\otimes A(\Z_2) \big) \cong \big( (s)_+ \oplus (s)_- \big)  \otimes A(\Z_2),
\end{equation}
which implies $T_{[r]} (\bm m \oplus \bm{c_1}) = \bm m \oplus \bm{c_1}  \oplus \bm f \oplus \bm{c_1}$. 
Similarly, we have $T_{[r]} (\bm f \oplus \bm{c_1}) = \bm m \oplus \bm{c_1}  \oplus \bm f \oplus \bm{c_1}$.
Notice this does not fix uniquely the action of $T_{[r]}$ on $\bm f,\bm{c_1},\bm m$ individually. To do this, we can compute the full hypergroup action  $A(\mathbb Z_2) \otimes A(\mathbb Z_2) = \mathbb{C}(\mathbb{D}_6 / \mathbb{Z}_2) \otimes \mathbb{C}(\mathbb{D}_6 / \mathbb{Z}_2)$ rather than the action of the single  component, that is 
\begin{equation}
    \left( T_{[1]} \oplus T_{[r]} \right)(M,\rho) = M \otimes_{A(\Z_2)} (A(\Z_2) \otimes A(\Z_2)) .
\end{equation}
For example, we can compute 
\begin{equation}
    (s)_+ \otimes_{A(\Z_2)} A(\Z_2)\otimes A(\Z_2) \cong (s)_+ \otimes A(\Z_2) ,
\end{equation}
which immediately implies $(T_{[1]} \oplus T_{[r]}) (\bm m) = \bm m \oplus \bm{c_1}$. Together with the previous observation of $T_{[r]} (\bm m \oplus \bm{c_1})$, this fixes $T_{[r]} (\bm m) = \bm{c_1}$, $T_{[r]} (\bm f) = \bm{c_1}$ and $T_{[r]} (\bm{c_1}) = \bm m \oplus \bm f \oplus \bm{c_1}$. 

Since this last trick relies on the fact we have only two double cosets, it is enlightening to compute the action more explicitly.
To do so, recall again the equivalence $\Vect_{\mathbb D_6}^{\mathbb Z_2} \simeq (\VDD)_{A(\mathbb Z_2)}$ and that the functor $\Vect_{\mathbb D_6}^{\mathbb Z_2} \to (\VDD)_{A(\mathbb Z_2)}$ is given by $\Ind_{\mathbb Z_2}^{\mathbb D_6}$. In general, $\Ind_{\mathbb Z_2}^{\mathbb D_6}(W) = \mathbb C\{r_a \otimes (q_x\otimes\hat{w}_c)\}_{r_a\mathbb Z_2 \in \mathbb D_6 / \mathbb Z_2,x \in \cl_{\Z_2}(f),c}$ such that $|r_a \otimes (q_x\otimes\hat{w}_c)| = r_a xr_a^{-1}$ and $g \cdot \big( r_a \otimes (q_x\otimes\hat{w}_c)\big) = r_{g(a)}\otimes h_{g,a}\cdot(q_x \otimes \hat{w}_c)$ with $gr_a = r_{g(a)} h_{g,a}$. For future reference, we have $s \cdot \mathbb Z_2 = \mathbb Z_2$, $s \cdot r \mathbb Z_2 = r^2 \mathbb Z_2$ and $s \cdot r^2\mathbb Z_2 = r \mathbb Z_2$. Besides $h_{r,r} = h_{r,r^2} = 1 \in \mathbb Z_2$ and $h_{s,r} = h_{s,r^2} = s \in \mathbb Z_2$. Finally, we write $A(\mathbb Z_2) = \mathbb C(\mathbb D_6/\mathbb Z_2) = \mathbb C\{\delta_{\Z_2},\delta_{r\Z_2},\delta_{r^2\Z_2}\}$. Using the same notation as introduced around eq.~\eqref{eq:VDZ_short} for the simple objects in $\VDZ$, we then have 
\begin{itemize}
    \item[\bul] $\bm 1 := \Ind_{\mathbb Z_2}^{\mathbb D_6}(\mathbb C\{w_1^+\}) = \mathbb C\{1\otimes w_1^+, r \otimes w_1^+, r^2\otimes w_1^+\}$.

    \item[\bul] $\bm e := \Ind_{\mathbb Z_2}^{\mathbb D_6}(\mathbb C\{w_1^-\}) = \mathbb C\{1\otimes w_1^-, r \otimes w_1^-, r^2\otimes w_1^-\}$. 
    
    \item[\bul] $\bm m := \Ind_{\mathbb Z_2}^{\mathbb D_6}(\mathbb C\{w_s^+\}) = \mathbb C\{1 \otimes w_s^+, r \otimes w_s^+, r^2 \otimes w_s^+\}$. 

    \item[\bul] $\bm f := \Ind_{\mathbb Z_2}^{\mathbb D_6}(\mathbb C\{w_s^-\}) = \mathbb C\{1 \otimes w_s^-, r \otimes w_s^-, r^2 \otimes w_s^-\}$. 

    \item[\bul] $\bm{c_1} := \Ind_{\mathbb Z_2}^{\mathbb D_6}(\mathbb C\{w_{rs}^+,w_{r^2s}^+\})= \mathbb C\{ 1\otimes w_{rs}^+, r\otimes w_{rs}^+, r^2 \otimes w_{rs}^+, 1 \otimes w_{r^2s}^+, r \otimes w_{r^2s}^+, r^2 \otimes w_{r^2s}^+\}$. 
    
    \item[\bul] $\bm{c_2} := \Ind_{\mathbb Z_2}^{\mathbb D_6}(\mathbb C\{w_{r}^+,w_{r^2}^+\})= \mathbb C\{ 1\otimes w_{r}^+ , r\otimes w_{r}^+, r^2\otimes w_{r}^+, 1\otimes w_{r^2}^+ , r\otimes w_{r^2}^+, r^2\otimes w_{r^2}^+\}$. 

\end{itemize}
As an example of computation in this framework, we verify the hypergroup action $T_{[r]}(\bm m) = \bm{c_1}$. By definition, we have
\begin{equation}
    T_{[r]}(\bm m) \cong \Ind_{\mathbb Z_2}^{\mathbb D_6}(\mathbb C\{w_s^+\}) \otimes_{A(\mathbb Z_2)} \mathbb C(\mathbb D_6) .
\end{equation}
By inspection, we find that as a vector space 
\begin{equation}   
\begin{split}
    T_{[r]}(\bm m) \cong \mathbb C\{&(1\otimes w_s^+) \otimes \delta_1, (r\otimes w_s^+) \otimes \delta_{r}, (r^2\otimes w_s^+) \otimes \delta_{r^2},
    \\
    &(1\otimes w_s^+) \otimes \delta_s, (r\otimes w_s^+) \otimes \delta_{rs}, (r^2\otimes w_s^+) \otimes \delta_{r^2s}\}.
\end{split}
\end{equation}
By definition, $|(1\otimes w_s^+) \otimes \delta_1| = s$, $|(r\otimes w_s^+) \otimes \delta_{r}| = r^2s$, $|(r^2\otimes w_s^+)\otimes \delta_{r^2}| = rs$, $|(1\otimes w_s^+) \otimes \delta_s| = s$, $|(r\otimes w_s^+) \otimes \delta_{rs}| = r^2s$ and $|(r^2\otimes w_s^+) \otimes \delta_{r^2s}| = rs$.
The right $A(\mathbb Z_2)$-module  structure is given by the right $A(\mathbb Z_2)$-module structure of $\mathbb C(\mathbb D_6)$, which explicitly reads  
\begin{equation*}
 \begin{split}
    T_{[r]}(\bm m) \cdot \delta_{\Z_2}
    &\cong \{0,0,(r^2\otimes w_s^+) \otimes \delta_{r^2},0,(r\otimes w_s^+) \otimes \delta_{rs},0\}
    \\
    T_{[r]}(\bm m) \cdot \delta_{r \Z_2}
    &\cong \{(1 \otimes w_s^+) \otimes \delta_{1},0,0,0,0,(r^2 \otimes w_s^+) \otimes \delta_{r^2s}\}
    \\
    T_{[r]}(\bm m) \cdot \delta_{r^2 \Z_2}
    &\cong \{0,(r \otimes w_s^+) \otimes \delta_{r},0,(1\otimes w_s^+) \otimes \delta_{s},0,0\} . 
\end{split}
\end{equation*}
On the other hand, $\bm{c_1}=\mathbb C\{ 1\otimes w_{rs}^+, r\otimes w_{rs}^+, r^2 \otimes w_{rs}^+, 1 \otimes w_{r^2s}^+, r \otimes w_{r^2s}^+, r^2 \otimes w_{r^2s}^+\}$ is defined such that $|1\otimes w_{rs}^+| = rs$, $|r\otimes w_{rs}^+| = s$, $|r^2 \otimes w_{rs}^+|=r^2s$, $|1 \otimes w_{r^2s}^+| =r^2s$, $|r \otimes w_{r^2s}^+| =rs$ and $|r^2 \otimes w_{r^2s}^+| =s$, while the right $A(\mathbb Z_2)$-module structure is given by
\begin{equation*}
    \begin{split}
        \{1\otimes w_{rs}^+, r\otimes w_{rs}^+, r^2 \otimes w_{rs}^+, 1 \otimes w_{r^2s}^+, r \otimes w_{r^2s}^+, r^2 \otimes w_{r^2s}^+\} \cdot \delta_{\Z_2}
        &= \{1\otimes w_{rs}^+, 0,0,1 \otimes w_{r^2s}^+,0,0\}
        \\
        \{1\otimes w_{rs}^+, r\otimes w_{rs}^+, r^2 \otimes w_{rs}^+, 1 \otimes w_{r^2s}^+, r \otimes w_{r^2s}^+,
        r^2 \otimes w_{r^2s}^+\} \cdot \delta_{r\Z_2}
        &\cong \{0,r\otimes w_{rs}^+, 0,0, r \otimes w_{r^2s}^+,0\}
        \\
        \{1\otimes w_{rs}^+, r\otimes w_{rs}^+, r^2 \otimes w_{rs}^+, 1 \otimes w_{r^2s}^+, r \otimes w_{r^2s}^+, r^2 \otimes w_{r^2s}^+\} \cdot \delta_{r^2\Z_2}
        &\cong \{0,0,r^2 \otimes w_{rs}^+,0,0,r^2 \otimes w_{r^2s}^+\}.
    \end{split}
\end{equation*}
By comparing grading and $A(\mathbb Z_2)$-module structure, we  see that $T_{[r]}(\bm m)$ is indeed isomorphic to $\bm{c_1}$. In some cases, $A(\mathbb Z_2)$-modules are further distinguished by their $\mathbb D_6$-equivariant structure.

\bigskip \noindent
\paragraph{Equivariantisation:} Given the $T$-action defined above, let us now show explicitly that $(\Vect_{\mathbb D_6}^{\mathbb Z_2})^T \cong \Vect_{\mathbb D_6}^{\mathbb D_6}$. Recall that a $T$-module consists of a pair $(M,\varrho \colon T(M)\rightarrow M)$. The morphism $\varrho$ needs to satisfy the conditions 
\begin{equation}\label{eq:rho_conditions}
    \varrho \circ \eta_M = 1_M \q \text{and} \q
    \varrho \circ T(\varrho) = \varrho \circ \mu_M . 
\end{equation}
Remember $\eta_M \colon w \mapsto 1_{\mathbb D_6} \otimes w$ for $w \in M$, while $\mu_M$ is the monad multiplication defined via $g_1 \otimes (g_2 \otimes w) \mapsto g_1g_2 \otimes w$.
The first condition in \eqref{eq:rho_conditions} is simply satisfied via $\varrho(1_G \otimes w) = w$ for all $w \in M$, while the second one requires us to evaluate $g_1 \otimes (g_2 \otimes w)$ for every $g_1,g_2 \in \mathbb D_6$ and $w \in M$.

Solving for this, we find that  $\bm 1 = \mathbb C\{w_1^+\}$ gives a $T$-module $(\bm 1,\varrho)$ we denote $(1)_1$ via $\varrho(s \otimes w_1^+) = w_1^+$ and $\varrho(r \otimes w_1^+) = w_1^+$. Similarly, $\bm e = \mathbb C\{w_1^-\}$ gives a $T$-module $(\bm e,\varrho)$ we denote $(1)_e$ via $\varrho(s \otimes w_1^-) = -w_1^-$ and $\varrho(r \otimes w_1^-) = w_1^-$. Moreover, $\bm 1\oplus \bm e =\mathbb C\{w_1^+,w_1^-\}$ also forms an indecomposable $T$-module $(\bm 1\oplus \bm e,\varrho)$, which we denote $(1)_\pi$, via $\varrho(s \otimes w_1^+) = w_1^+$, $\varrho(s \otimes w_1^-) = -w_1^-$, $\varrho(r \otimes w_1^+) = -\frac{1}{2} w_1^+ - \frac{\sqrt 3}{2} w_1^-$ and $\varrho(r \otimes w_1^-) = \frac{\sqrt{3}}{2} w_1^+ - \frac{1}{2} w_1^-$.  The $T$-modules $(1)_1,(1)_e,(1)_\pi $ recover $\Rep(\mathbb D_6) \subset \Vect_{\mathbb D_6}^{\mathbb D_6}$. 

Now let's construct the other $T$-modules. We start by $\bm m = \mathbb C\{ w_s^+\}$. Notice that e.g.\ $|r \otimes w_s^+ | = r^2 s$, so clearly we cannot find a map $\varrho :T(\bm m) \rightarrow \bm m$ and $\bm m$ alone does not form a $T$-module.  We then consider $\bm m \oplus \bm{c_1} = \mathbb \{w_s^+, w_{rs}^+, w_{r^2s}^+ \}$. This forms a $T$-module $(\bm m \oplus \bm{c_1},\varrho)$, which we denote $(s)_+$, via  
\begin{equation}
\begin{aligned}
    &\varrho (s \otimes w_s^+) =w_s^+,\q \varrho (s \otimes w_{rs}^+) =w_{r^2s}^+ ,\q \varrho (s \otimes w_{r^2s}^+) =w_{rs}^+ ,\\ 
    &\varrho (r \otimes w_s^+) = w_{r^2 s}^+ ,\q \varrho (r \otimes w_{rs}^+) = w_{s}^+ ,\q \varrho (r \otimes w_{r^2s}^+) = w_{rs}^+, \\ 
    &\varrho (r^2 \otimes w_s^+) = w_{r s}^+ ,\q \varrho (r^2 \otimes w_{rs}^+) = w_{r^2s}^+ ,\q \varrho (r^2 \otimes w_{r^2s}^+) = w_{s}^+ .
\end{aligned}    
\end{equation}
Similarly, $\bm f \oplus \bm{c_1} = \mathbb C \{ w_s^-, w_{rs}^+, w_{r^2s}^+ \}$ forms another $T$-module $(\bm f \oplus \bm{c_1},\varrho) $ we denote $(s)_-$ via 
\begin{equation}
\begin{aligned}
    &\varrho (s \otimes w_s^-) =-w_s^-,\q \varrho (s \otimes w_{rs}^+) =w_{r^2s}^+ ,\q \varrho (s \otimes w_{r^2s}^+) =w_{rs}^+ ,
    \\ 
    &\varrho (r \otimes v_s^-) = w_{r^2 s}^+ ,\q \varrho (r \otimes w_{rs}^+) = -w_{s}^- ,\q \varrho (r \otimes w_{r^2s}^+) = -w_{rs}^+, 
    \\ 
    &\varrho (r^2 \otimes w_s^-) = -w_{r s}^+ ,\q \varrho (r^2 \otimes w_{rs}^+) = -w_{r^2s}^+ ,\q \varrho (r^2 \otimes w_{r^2s}^+) = w_{s}^- .
\end{aligned}    
\end{equation}
Finally, let us consider $\bm{c_2} = \mathbb C \{ w^+_r, w^+_{r^2}\}$. This gives another $T$-module $(\bm{c_2},\varrho)$, which we denote $(r)_\gamma$, via the action 
\begin{equation}
\begin{aligned}
    &\varrho (s \otimes w^+_r) = w^+_{r^2},\q \varrho (s \otimes w_{r^2}^+) = w^+_{r},\q  \varrho (r \otimes w^+_r) = \gamma \,w^+_r , \\ 
    &\varrho (r \otimes w^+_{r^2}) = \gamma^2 \,w^+_{r^2} ,\q \varrho (r^2 \otimes w^+_{r}) = \gamma^2 \,w^+_{r} ,\q \varrho (r^2 \otimes w^+_{r^2}) = \gamma \,w^+_{r^2} ,
\end{aligned}    
\end{equation}
with $\gamma$ required to satisfy $\gamma^3=1$. So from $\bm {c_2}$ we actually get three modules $(r)_1,(r)_\omega,(r)_{\omega^2}$, where $\omega = e^{2\pi i /3}$. 

Finally, to confirm that we obtain $\Vect_{\mathbb D_6}^{\mathbb D_6}$, as a fusion category, we need to compute the monoidal structure of $(\Vect_{\mathbb D_6}^{\mathbb Z_2})^T$. To do this, recall that the co-multiplication $\Delta_{W_1,W_2} \colon T(W_1 \otimes W_2) \to T(W_1) \otimes T(W_2)$ of the $T$ bimonad is given by $g \otimes (w_1 \otimes w_2) \mapsto (g \otimes w_1) \otimes (g \otimes w_2)$. Given two $T$-modules $(M_1,\varrho_1)$ and $(M_2,\varrho_2)$, we can define a $T$-action for the tensor product $M_1\otimes M_2$ by composing the comultiplication of $T$ with the individual module actions. That is, $\varrho$ is given by $(\varrho_1 \otimes \varrho_2) \circ \Delta_{M_1,M_2}: T(M\otimes N) \rightarrow T(M) \otimes T(N) \rightarrow M \otimes N$. For example, applying this to the indecomposable $T$-module $(1)_\pi \equiv (\mathbb C \{w_1^+,w_1^-\},\varrho)$, it is immediate to verify 
\begin{equation}
\begin{aligned}
    \mathbb C \{w_1^+,w_1^-\} \otimes \mathbb C \{w_1^+,w_1^-\} = \; &\mathbb C\{ w_1^+ \otimes w_1^+, w_1^+ \otimes w_1^-, w_1^- \otimes w_1^+, w_1^- \otimes w_1^-\} \\ 
    \cong \; &\mathbb C \{ w_1^+ \otimes w_1^+ + w_1^- \otimes w_1^-\} \oplus \mathbb C \{ w_1^+ \otimes w_1^- - w_1^- \otimes w_1^+\}   \\ 
    &\oplus \mathbb C \{ w_1^+ \otimes w_1^+ - w_1^- \otimes w_1^-, -w_1^+ \otimes w_1^- - w_1^- \otimes w_1^+\} ,
\end{aligned}    
\end{equation}
which recovers $(1)_\pi \otimes (1)_\pi \cong (1)_1 \oplus (1)_e \oplus (1)_\pi$ in $\VDD$. 

\bigskip \noindent
\paragraph{Twisted quantum double:} Following the general construction of sec.~\ref{sec:Maths_frac}, let us now consider starting from the twisted quantum double $\Vect_{\mathbb D_6}^{\mathbb D_6, \alpha_p}$, with $\alpha_p$ a normalised representative of a class $[\alpha_p]$ in $H^3(\mathbb D_6,{\rm U}(1)) \cong \mathbb Z_6$, with $p\in \{0,\dots,5\}$. This example was already discussed in ref.~\cite{KNBalasubramanian:2025vum}, but we reformulate it here from the Hopf monad perspective. One can check that the 3-cocycle $\alpha_p$ has no effect on the simple objects and fusion rules in $\Vect_{\mathbb D_6}^{\mathbb D_6, \alpha_p}$ compared to the untwisted case. Moreover, $A(\mathbb  Z_2) \cong (1)_1 \oplus (1)_\pi$ remains a condensable algebra such that $(\Vect_{\mathbb D_6}^{\mathbb D_6, \alpha_p})_{A(\mathbb Z_2)} \simeq \Vect_{\mathbb D_6}^{\mathbb Z_2, \alpha_p}$. On the one hand, for $p \in \{0,2,4\}$ even, one can further check $(\Vect_{\mathbb D_6}^{\mathbb D_6, \alpha_p})_{A(\mathbb Z_2)}^\text{loc} \simeq \Vect_{\mathbb Z_2}^{\mathbb Z_2}$ so that the resulting topological topological order is still that of the toric code. On the other hand, for $p=1,3,5$ odd, we find $(\Vect_{\mathbb D_6}^{\mathbb D_6, \alpha_p})_{A(\mathbb Z_2)}^\text{loc} \simeq \Vect_{\mathbb Z_2}^{\mathbb Z_2,{\alpha}}$, where $\alpha$ is a normalised representative of the unique non-trivial class in $\in H^3(\mathbb Z_2, {\rm U}(1)) \cong \mathbb Z_2$, so that the resulting topological order is that of the so-called double semion model. Focusing on the even case, for every $p \in \{0,2,4\}$, $\Vect_{\mathbb D_6}^{\mathbb Z_2, \alpha_p}$ describes a hypergroup $\DZ$-extension of the topological order $\Vect_{\mathbb Z_2}^{\mathbb Z_2}$. This could in principle give rise to a scenario where two enrichments by the same hypergroup differ by the choice of symmetry fractionalisation, as encoded in the $T$ monad multiplication $\mu_W \colon T(T(W)) \to T(W)$ sending $ g_1 \otimes (g_2 \otimes w)\mapsto \alpha(g_1,g_2|x) \, g_1g_2 \otimes w$ for $w \in W_x$. However, $\alpha(g_1,g_2|x)$ turns out to be trivial in this case. Indeed, notice that the equation 
\begin{equation}
     \alpha(g_1,g_2g_3|x) \, \alpha(g_2,g_3|x) = \alpha(g_1g_2,g_3|x) \, \alpha(g_1,g_2|g_3 x g_3^{-1}),
\end{equation}
can be expressed as ${\rm d}^{(2)} \alpha(g_1,g_2,g_3|x)=1$, where ${\rm d}^{(2)}$ is the coboundary operator associated to the cohomology groups $H^2(\mathbb D_6,\text{Fun}(\mathbb D_6,{\rm U}(1)))$, with $\text{Fun}(\mathbb D_6,{\rm U}(1))$ the abelian group of functions $\phi \colon \mathbb D_6 \rightarrow {\rm U}(1)$ with a right $\mathbb D_6$-module action by conjugation: $(\phi \cdot g)(-)=\phi(g - g^{-1})$. The above equation then states $[\alpha(-,-|-)] \in H^2(
\mathbb D_6,A)$. Such a 2-cocycle is trivial if there exists a 1-cochain $\epsilon(g|x)$ such that 
\begin{equation}
    \alpha(g_1,g_2|x) = {\rm d}^{(2)} \epsilon(g_1,g_2|x) = \frac{\epsilon(g_1|{g_2}xg_2^{-1}) \, \epsilon(g_2|x)}{\epsilon(g_1g_2|x)} \,.
\end{equation}
It follows from \emph{Shapiro's lemma} (see e.g. ref.~\cite{deWildPropitius:1995cf}) that 
\begin{equation}\label{eq:conj_cohomology}
    H^2(\mathbb D_6, A) \cong \bigoplus_{\text{cl}(f) \in \cl(\mathbb D_6)} H^2(Z_{\mathbb D_6}(f),{\rm U}(1)) .
\end{equation}
Since all the cohomology groups on the r.h.s.\ of \eqref{eq:conj_cohomology} are trivial, we learn there must exist a $\epsilon(-|-)$ trivialising $\alpha(-,-|-)$. Therefore, in this case we do not have access to different symmetry fractionalisations. This is also confirmed by the fact that $\Vect_{\mathbb D_6}^{\mathbb Z_2, \alpha_p}$ and $\Vect_{\mathbb D_6}^{\mathbb Z_2}$ have the same fusion ring. The two $\DZ$-enriched topological orders remain distinguished by the choice of discrete torsion, which is encoded in the associator. 
\section{Generalisation and outlook\label{sec:outlook}}

\emph{After summarising our findings, we sketch a generalisation of our construction beyond the topological lattice gauge theory setting. We then illustrate this general setting with several preliminary examples.} 

\subsection{Beyond topological lattice gauge theory}

In this manuscript, we have studied the topological order resulting from condensing an arbitrary condensable algebra of charges in a topological lattice gauge theory. In general, whenever a non-degenerate braided fusion category contains $\Rep(G)$ as a \emph{Tannakian fusion} subcategory, there exists a canonical connected étale algebra whose condensation produces a topological order enriched by an invertible $G$-symmetry \cite{Davydov2013Witt}. This symmetry enriched topological order is encoded in a $G$-crossed braided fusion category. For topological lattice gauge theory, the relevant algebras are of the form $A(N)$ with $N \cat G$ a normal subgroup. We considered instead the case of an arbitrary subgroup $H \leq G$, and showed that the resulting theory $\VGH$ is a hypergroup $\GH$-graded extension of $\VHH$ equipped with a compatible hypergroup $\GH$-action. We then explored several properties of this structure in detail. Specialising to topological lattice gauge theory allowed us to reduce most statements to group theoretic ones that could be proven very explicitly.

The results of this manuscript should be regarded as a guiding principle towards a genuine non-invertible generalisation of the $G$-crossed braided fusion category formalism. Indeed, we expect both our results and the techniques used to derive them to extend beyond the topological lattice gauge theory setting. We sketch such a generalisation below.

\bigskip \noindent
Condensing the algebra $A(H)$ in the topological order $\VGG$ resulted in a theory whose localised excitations are encoded in $\VGH$. We showed that an equivalent description of $\VGH$ is given by the relative centre $\mc Z_{\Vect_H}(\Vect_G)$ of $\Vect_G$ over $\Vect_H$. Recall that, in general, the relative centre construction takes as input a choice of fusion category together with a bimodule category over it \cite{Majid1991}. In our example, $\Vect_G$ is a $\Vect_H$-bimodule category by virtue of $\Vect_H$ being a fusion subcategory of $\Vect_G$. This leads us to consider a generalisation of our framework in which a category $\mc D$ is a spherical fusion subcategory of another spherical fusion category $\mc C$. 

The lattice construction introduced in sec.~\ref{sec:lattice} naturally generalises to this broader context. The starting point is the \emph{string-net model} associated with the input datum $\mc C$ \cite{Levin:2004mi}. In the same vein as our reinterpretation in sec.~\ref{sec:Maths_tube} of the coloured graphs in sec.~\ref{sec:lattice} as morphisms in $\Vect_G$, the degrees of freedom in this string-net model are provided by morphisms in $\mc C$. As with the quantum double model, the dynamics is governed by two families of mutually commuting local projectors. Vertex projectors enforce the fusion rules of $\mc C$ so as to yield a non-trivial hom-space, while plaquette projectors average over a local action of $\mc C$. We then modify this model, mirroring the procedure used to condense the algebra $A(H)$ in $\VGG$. Plaquette projectors are restricted to average over the local action of the fusion subcategory $\mc D$, and new edge projectors are introduced to dynamically penalise edge assignments valued in $\Irr(\mc C) \setminus \Irr(\mc D)$, which effectively reduces the space of allowed morphisms in $\mc C$. Equipped with the general formalism of \cite{KirillovJr2002} and following the same steps as in sec.~\ref{sec:condensation}, one expects the localised excitations of this model to be encoded in the relative centre $\mc Z_\mc D(\mc C)$ of $\mc C$ over $\mc D$, realised as the category of modules over some tube algebra $\Tu_\mc C^\mc D$. Just as in the quantum double example, the relative Drinfel’d centre $\mc Z_\mc D(\mc C) \simeq \Fun_{\mc D| \mc D}(\mc D,\mc C)$ naturally describes localised excitations living at the endpoint of the composite domain wall associated with $\mc D$ \cite{kongBdries}, the different types of domains walls corresponding to indecomposable $\mc D$-bimodules categories in $\mc C$. 

We observe that many of the features of $\VGH \simeq \mc Z_{\Vect_H}(\Vect_G)$ that we identified in this manuscript generalises to $\mc Z_{\mc D}(\mc C)$. First of all, $\mc Z(\mc D)$ is a fusion subcategory of $\mc Z_{\mc D}(\mc C)$, which equips $\mc Z_\mc D(\mc C)$ with the structure of a $\mc Z(\mc D)$-bimodule category. Therefore, one can decompose $\mc Z_\mc D(\mc C)$ into a direct sum of $\mc Z(\mc D)$-bimodule categories \cite{Bruguieres2015}. For every $\mc Z(\mc D)$-bimodule category in this decomposition, one can consider the corresponding \emph{$\mc Z(\mc D)$-regular object}.\footnote{Recall that such regular objects appear in our compatibility condition \eqref{eq:compAdHyperGr}.} It was shown in  ref.~\cite{hempel2023hypergroups} that the span of these $\mc Z(\mc D)$-regular objects form a hypergroup. By analogy with our construction, it would be desirable to express this hypergroup as some \emph{double quotient}. First of all, every \emph{fusion ring} induces a hypergroup \cite{Bischoff2020}, which is simply obtained by rescaling its basis elements. This brings us to the main advantage of hypergroups over fusion rings, namely that we can construct quotients of hypergroups \cite{BloomHeyer+1995}. Concretely, consider the hypergroups associated with the \emph{Grothendieck rings} $\mc K^0(\mc Z_\mc D(\mc C))$ and $\mc K^0(\mc Z(\mc D))$ of $\mc Z_\mc D(\mc C)$ and $\mc Z(\mc D)$, respectively. Following remarks in ref.~\cite{Riesen2025}, the double quotient hypergroup $\mc K^0(\mc Z_\mc D(\mc C))/\!/\mc K^0(\mc Z(\mc D))$ should coincide with the hypergroup of $\mc Z(\mc D)$-regular objects.    
Moreover, it follows from $\mc Z(\mc Z_{\mc D}(\mc C)) \simeq \mc Z(\mc D) \boxtimes \overline{\mc Z(\mc C)}$ that $\mc Z(\mc D)$ is central in $\mc Z_\mc D(\mc C)$. This guarantees that an analogue of eq.~\eqref{eq:compMonHyperGr} will hold \cite{hempel2023hypergroups}. We therefore expect $\mc Z_\mc D(\mc C)$ to be a hypergroup $\mc K^0(\mc Z_\mc D(\mc C))/\!/\mc K^0(\mc Z(\mc D))$-extension of $\mc Z(\mc D)$ the same way $\VGH$ is a hypergroup $\GH$-extension of $\VHH$.

We now turn to the hypergroup action. Suppose we are given a condensable algebra $A$ in $\mc Z(\mc C)$ such that $\mc Z(\mc C)_A \simeq \mc Z_\mc D(\mc C)$ and $\mc Z(\mc C)_A^{\rm loc} \simeq \mc Z(\mc D)$ \cite{DAVYDOV2017149}. Let us consider the composition $\mc Z(\mc C)_A \to \mc Z(\mc C) \to \mc Z(\mc C)_A$ of the forgetful and induction functors, mapping an $A$-module $(M,\rho)$ in $\mc Z(\mc C)_A$ to $(M \otimes A,{\id}_M \otimes \mu)$. According to ref.~\cite{Cui:2018hxz}, this construction should produce a Hopf monad on $\mc Z(\mc C)_A$. By construction, this functor is provided by $- \otimes_A (A \otimes A)$, where $A \otimes A$ is an $(A,A)$-bimodule. It follows from results of ref.~\cite{Riesen2025} that the $A$-bimodule $A \otimes A$ should decompose over $\mc K^0(\mc Z_\mc D(\mc C))/\!/\mc K^0(\mc Z(\mc D))$---which agrees with what we found in eq.~\eqref{eq:decompAA}---in such a way that the Hopf monad defines a categorical action of $\mc K^0(\mc Z_\mc D(\mc C))/\!/\mc K^0(\mc Z(\mc D))$ on $\mc Z_\mc D(\mc C)$. We conjecture that this hypergroup action is compatible with the hypergroup grading in such a way that the analogue of eq.~\eqref{eq:compAdHyperGr} holds true. As in our example, gauging this non-invertible symmetry should amount to computing the Eilbenberg--Moore category of modules over this Hopf monad in $\mc Z_\mc D(\mc C)$. As we illustrate below with further examples, this hypergroup action can be conveniently computed within the tube algebra picture by converting hypergroup elements into tube algebra bimodules.

\subsection{Further examples}

Following the outline above, we now consider two examples that go beyond the strict formalism of our manuscript. The first one, which starts from the string-net model with input datum the spherical fusion category $\msf{Fib}$, is somewhat tautological, as the condensed theory is the trivial topological order. Nonetheless, it showcases several non-trivial features which can be understood as properties of a non-invertible symmetry-protected topological order. The second example starts from the same topological order as the quantum double for the group $\mathbb{D}_6$, but formulated this time as a string-net model with input $\Rep(\mathbb D_6)$. 

\bigskip \noindent
Let $\Fib$ be the spherical fusion category with $\Irr(\Fib) = \{1,\tau\}$, such that $1 \otimes \tau \cong \tau \cong \tau \otimes 1$ and $\tau \otimes \tau \cong 1 \oplus \tau$, with non-trivial $F$-symbols
\begin{equation}
    F^{\tau \tau \tau}_{\tau} = \frac{1}{\varphi} 
    \begin{pmatrix}
        1 & \sqrt \varphi 
        \\
        \sqrt \varphi & -1
    \end{pmatrix},
\end{equation}
where $\varphi := \frac{1}{2}(1 +\sqrt 5)$ is the golden ration. Quantum dimensions of $1$ and $\tau$ are $1$ and $\varphi$, respectively. It follows from $\varphi^2 = 1 + \varphi$ that the global quantum dimension is given by $D^2 := \varphi \sqrt 5$. Moreover, since $\Fib$ is itself a non-degenerate braided fusion category, we have $\mc Z(\Fib) \simeq \Fib \boxtimes \overline{\Fib}$ and thus $\Irr(\mc Z(\Fib))= \{1 \boxtimes \overline 1, 1 \boxtimes \overline \tau, \tau \boxtimes \overline 1, \tau \boxtimes \overline \tau\} \equiv \{1 \overline 1, 1 \overline \tau, \tau \overline 1, \tau \overline \tau\}$. As for any fusion category, the Drinfel'd centre $\mc Z(\mc \Fib)$ of $\Fib$ can be realised as the category of modules over its corresponding tube algebra $\Tu_{\Fib}^{\Fib}$ \cite{ocneanu1994chirality,ocneanu2001operator,Izumi2000,MUGER2003159,Neshveyev_Yamashita_2018}. Briefly, the vector space underlying this tube algebra is given by 
\begin{equation}
    \bigoplus_{X_1,\ldots,X_4 \in \Irr(\Fib)}  \Hom_{\Fib}(X_4,X_1 \otimes X_2) \otimes \Hom_{\Fib}(X_3 \otimes X_1 , X_4),
\end{equation}
which we identify with the span of $\Irr(\Fib)$-coloured graphs on the cylinder of the form
\begin{equation}
    \label{eq:tubeExp}
    \mc T^{X_1}_{X_2X_4X_3}\equiv \tubeTExp{X_1}{X_2}{X_4}{X_3},
\end{equation}
where $X_1,\ldots,X_4 \in \Irr(\Fib)$ are such that $\Hom_{\Fib}(X_3 \otimes X_1,X_4)$ and $\Hom_{\Fib}(X_4,X_1 \otimes X_3)$ are non-trivial. 
The tube algebra multiplication proceeds as before by gluing coloured graphs on the cylinder along circular boundary components, but it now involves the non-trivial $F$-symbols pasted above (see e.g. \cite{Koenig:2010uua,kongBdries,PhysRevB.90.115119,Aasen:2017ubm,Bultinck:2015bot,Bullivant:2019fmk} for details). We omit the explicit expression here and directly list the corresponding minimal central idempotents below
\begin{equation}
\begin{split}
    \mc E_1 &= \frac{1}{D^2} \big( \tub{111}{1} + \varphi \tub{1 \tau 1}{\tau} \big),
    \\
    \mc E_{\tau \overline \tau} &= \frac{1}{D^2} \big( \varphi^2 \tub{111}{1} - \varphi \tub{1 \tau 1}{\tau}  + \varphi \tub{\tau \tau \tau}{1} + \varphi \tub{\tau 1 \tau}{\tau} + \frac{1}{\sqrt \varphi} \tub{\tau \tau \tau}{\tau}\big),
    \\
    \mc E_{\tau \overline 1} &= \frac{1}{D^2} \big(\tub{\tau \tau \tau}{1} + e^\frac{4i \pi}{5} \tub{\tau 1 \tau}{\tau} + \sqrt \varphi e^{-\frac{3 i \pi}{5}} \tub{\tau \tau \tau}{\tau}\big),
    \\
    \mc E_{1 \overline \tau} &= \frac{1}{D^2} \big(\tub{\tau \tau \tau}{1} + e^{-\frac{4i \pi}{5}} \tub{\tau 1 \tau}{\tau} + \sqrt \varphi e^{\frac{3 i \pi}{5}} \tub{\tau \tau \tau}{\tau}\big).
\end{split}
\end{equation}
The primitive idempotents of the only two-dimensional module $\tau \overline \tau$ are
\begin{equation}
    (\mc E_{\tau \overline \tau})_1^1 = \frac{1}{D^2} \big( \varphi^2 \tub{111}{1} - \varphi \tub{1 \tau 1}{\tau} \big) \q \text{and} \q 
    (\mc E_{\tau \overline \tau})_2^2 = \frac{1}{D^2} \big( \varphi \tub{\tau \tau \tau}{1} + \varphi \tub{\tau 1 \tau}{\tau} + \frac{1}{\sqrt \varphi} \tub{\tau \tau \tau}{\tau} \big). 
\end{equation}
Up to isomorphism, there is a unique condensable algebra in $\mc Z(\Fib)$, namely $A = 1 \overline 1 \oplus \tau \overline \tau$, such that $\mc Z(\Fib)_A \simeq \Fib$ and $\mc Z(\Fib)_A^{\rm loc} \simeq \Vect$. Therefore, the condensed theory has trivial topological order, i.e., it does not host any deconfined excitations aside from the trivial one. Here, the relevant hypergroup is provided by the Grothendieck ring of $\Fib$ itself with basis $\{1,\frac{\tau}{\varphi}\}$. Therefore, there are two sectors: the untwisted one labelled by $1$ and the twisted one labelled by $\tau$. Let ${\bm 1}$ be the unique quasi-particle of the untwisted sector and ${\bm \tau}$ the unique confined excitation of the twisted sector. The corresponding minimal central idempotents of the tube algebra $\Tu^{\Vect}_{\Fib}$ are given by
\begin{equation}
    \mc E_{\bm 1} = \tub{111}{1}   \q \text{and} \q \mc E_{\bm \tau} = \tub{\tau \tau \tau}{1}.
\end{equation}
In terms of the minimal central idempotents, condensation is essentially accomplished by restricting them to tube elements $\mc T^{X_1}_{X_2X_3X_2}$ satisfying the rules spelt out above such that $X_1 = \mathbb C \in \Irr(\Vect)$. In symbols,
\begin{equation}
\begin{alignedat}{2}
    \mc E_{1 \overline 1} &\mapsto \frac{1}{D^2} \tub{111}{1},
    &\q\q 
    \mc E_{\tau \overline \tau} &\mapsto \frac{1}{D^2}\big( \varphi^2 \tub{111}{1} + \varphi \tub{\tau \tau \tau}{1} \big),
    \\
    \mc E_{\tau \overline 1} &\mapsto \frac{1}{D^2} \tub{\tau \tau \tau }{1},
    &\q \q
    \mc E_{1 \overline \tau} &\mapsto \frac{1}{D^2}\tub{\tau \tau \tau}{1},
\end{alignedat}
\end{equation}
from which we deduce the image of the initial quasi-particles in the condensed theory: $1 \overline 1 \mapsto \bm 1$, $\tau \overline \tau \mapsto {\bm 1} \oplus {\bm \tau}$, $\tau \overline 1 \mapsto {\bm \tau}$ and $1 \overline \tau \mapsto {\bm \tau}$. Naturally, the fusion rules of ${\bm 1}$ and ${\bm \tau}$ are identical to those of $\Fib$.

Let us now compute the action of the non-invertible symmetry encoded in $\Fib$ itself. The simple object $\tau$---here regarded as a representative of the corresponding hypergroup element---acts via its corresponding $\Tu_{\Fib}^{\Vect}$-bimodule $M_\tau$. We denote this action by $T_\tau$. As a vector space, $M_\tau$ is spanned by tubes of the form $\tub{X_1X_2X_3}{\tau}$, where $X_1,X_2,X_3 \in \Irr(\Fib)$. Using the minimal central idempotents written above, we find
\begin{equation}
\begin{split}
    \mc E_{\bm 1} \cdot M_\tau \cdot \mc E_{\bm 1} &= \mathbb C\big\{\tub{1\tau 1}{\tau}\big\},
    \\
    \mc E_{\bm 1} \cdot M_\tau \cdot \mc E_{\bm \tau} &= \mathbb C\big\{ \tub{\tau \tau 1}{\tau} \big\},
    \\
    \mc E_{\bm \tau} \cdot M_\tau \cdot \mc E_{\bm 1} &= \mathbb C\big\{ \tub{1 \tau \tau }{\tau} \big\},
    \\
    \mc E_{\bm \tau} \cdot M_\tau \cdot \mc E_{\bm \tau} &= \mathbb C\big\{ \tub{\tau \tau \tau}{\tau}, \tub{\tau 1 \tau}{\tau}\big\},
\end{split}
\end{equation}
which implies that $T_\tau(\bm 1) \cong \bm 1 \oplus \tau$ and $T_\tau(\bm \tau) \cong \bm 1 \oplus 2 \cdot \bm \tau$. We leave it to the reader to verify that the analogue of eq.~\eqref{eq:compAdHyperGr} holds.

\bigskip \noindent
For our second example, we revisit the condensation in $\VDD$ encoded in  $A(\mathbb Z_2)$. However, invoking $\VDD \simeq \mc Z(\Rep(\mathbb D_6))$, we now formulate it as a string-net model with input $\Rep(\mathbb D_6)$. In this formulation, the condensed theory is provided by the relative centre $\mc Z_{\Rep(\mathbb Z_2)}(\Rep(\mathbb D_6))$, where the $\Rep(\mathbb Z_2)$-bimodule structure of $\Rep(\mathbb D_6)$ is provided by the inclusion of $\Rep(\mathbb Z_2)$ in $\Rep(\mathbb D_6)$, as a fusion subcategory. This statement may seem surprising in light of the results of sec.~\ref{sec:Maths_ZVGH}. Indeed, we argued there that $\VGH$ is equivalent to the relative centre $\mc Z_{\Rep(G)}(\Rep(H))$ with the bimodule structure provided by the restriction functor. But $\mathbb Z_2$ is both a subgroup and a quotient group of $\mathbb D_6$, which explains why there is yet another formulation. We can also trace it back to the fact that $\VDZ \simeq \Rep(\mathbb Z_2) \boxtimes \Rep(\mathbb D_6)$ as a monoidal category.   

Since we have already treated this condensation process from various perspectives, we keep the exposition brief. The eight simple objects in $\VDD$ are in one-to-one correspondence with minimal central idempotents of a tube algebra spanned by $\Irr(\Rep(\mathbb D_6))$-coloured graphs of the form \eqref{eq:tubeExp}. The multiplication rule involves $F$-symbols of $\Rep(\mathbb D_6)$, which are a subset of those we listed in sec.~\ref{sec:Maths_example}. For illustration, we list below the minimal central idempotents associated with three simple objects:
\begin{equation}
\begin{split}
    \mc E_{(\cl(1),e)} &= \frac{1}{6}\big(\tub{eee}{1} + \tub{e1e}{e}-2\tub{e\pi e}{\pi} \big),
    \\
    \mc E_{(\cl(s),+)} &= \frac{1}{2}\big(\tub{111}{1} - \tub{1e1}{e}+\frac{1}{2}\tub{\pi \pi \pi}{1} +\frac{1}{2}\tub{\pi \pi \pi }{e} + \frac{1}{2}\tub{\pi 1 \pi}{\pi} +\frac{1}{2}\tub{\pi e \pi}{\pi} \big),
    \\
    \mc E_{(\cl(r),1)} &= \frac{1}{6}\big( \tub{\pi \pi \pi}{1} - \tub{\pi \pi \pi}{e}+\tub{\pi 1 \pi}{\pi} -\tub{\pi e \pi }{\pi} + \sqrt 2 \tub{\pi \pi \pi}{\pi} \big).
\end{split}
\end{equation}
Let us also list three minimal central idempotents of the condensed theory $\VDZ \simeq \mc Z_{\Rep(\Z_2)}(\Rep(\mathbb D_6)) \simeq \Mod(\Tu_{\Rep(\mathbb D_6)}^{\Rep(\mathbb Z_2)})$:
\begin{equation}
\begin{split}
    \mc E_{\bm e} &= \frac{1}{2} \big( \tub{eee}{1} + \tub{e1e}{e} \big),
    \\
    \mc E_{\bm m} &= \frac{1}{2} \big( \tub{111}{1} - \tub{1e1}{e}  \big),
    \\
    \mc E_{\bm{c_1}} &= \frac{1}{2} \big( \tub{\pi \pi \pi}{1} + \tub{\pi \pi \pi}{e} \big). 
\end{split}
\end{equation} 
Condensation amounts to restricting the minimal central idempotents to tube elements of the form $\tub{X_2X_3X_2}{X_1}$, where $X_1 \in \Irr(\Rep(\mathbb Z_2))$ and $X_2,X_3 \in \Irr(\Rep(\mathbb D_6))$. Under this operation, we have for instance 
\begin{equation}
    \mc E_{(\cl(s),+)} \mapsto \frac{1}{2} \big(\tub{111}{1} - \tub{1e1}{e}  +\frac{1}{2} \tub{\pi \pi \pi}{1} + \frac{1}{2}\tub{\pi \pi \pi}{e} \big) = \mc E_{\bm m} + \frac{1}{2} \mc E_{\bm{c_1}},
\end{equation}
from which we recover the fact that $(\cl(s),+) \mapsto \bm{m} + \bm{c_1}$. 

The grading is now provided by the double quotient $\mc K^0(\Rep(\mathbb D_6))/\!/\mc K^0(\Rep(\mathbb Z_2))$ hypergroup, which contains two elements, namely $[1] = \mc K^0(\Rep(\mathbb Z_2))$ and $[\pi ] = \{\pi\}$. Let us compute the action of the non-trivial hypergroup element $[\pi]$. It acts via its corresponding $\Tu_{\Rep(\mathbb D_6)}^{\Rep(\mathbb Z_2)}$-bimodule $M_\pi$. We denote this action by $T_\pi$. As a vector space, $M_\pi$ is spanned by tubes of the form $\tub{X_1X_2X_3}{\pi}$, where $X_1,X_2,X_3 \in \Irr(\Rep(\mathbb D_6))$. For instance, we find

\begin{equation}
\begin{split}
    \mc E_{\bm{1}} \cdot M_\pi \cdot \mc E_{\bm{1}} \cong
    \mathbb C\big\{ \tub{1\pi 1}{\pi}\big\},
    \\
    \mc E_{\bm{e}} \cdot M_\pi \cdot \mc E_{\bm{1}} \cong
    \mathbb C\big\{\tub{1\pi e}{\pi}\big\},
    \\
    \mc E_{\bm{c_1}} \cdot M_\pi \cdot \mc E_{\bm m} \cong \mathbb C\big\{\tub{1\pi \pi}{\pi} \big\},
\end{split}
\end{equation}
from which we deduce that $T_\pi(\bm{1}) \cong \bm{1} \oplus \bm{e}$ and
$T_\pi(\bm m) \cong \bm{c_1}$. Naturally, we recover the same result we found before, but from a different realisation of the hypergroup.

\newpage

\renewcommand*\refname{\hfill References \hfill}
\bibliographystyle{alpha}
\bibliography{ref}

\end{document}